%% file: qgp4_v2chapter.tex
\documentclass{ws-ijmpe}
\usepackage{graphicx}         % standard LaTeX graphics tool

\begin{document}

\markboth{P. Sorensen}{Review of elliptic flow $v_2$}

%%%%%%%%%%%%%%%%%%%%% Publisher's Area please ignore %%%%%%%%%%%%%%%
\catchline{}{}{}{}{}
%%%%%%%%%%%%%%%%%%%%%%%%%%%%%%%%%%%%%%%%%%%%%%%%%%%%%%%%%%%%%%%%%%%%

\title{ELLIPTIC FLOW: A STUDY OF SPACE-MOMENTUM CORRELATIONS IN RELATIVISTIC NUCLEAR COLLISIONS}

\author{\footnotesize PAUL SORENSEN}

\address{Physics Department, Brookhaven National Laboratory, Building 510\\
Upton, NY 11973,U.S.A.\\
prsorensen@bnl.gov}

\maketitle

\begin{history}
\received{(received date)}
\revised{(revised date)}
%\accepted{(Day Month Year)}
%\comby{(xxxxxxxxxx)}
\end{history}

\begin{abstract}
  Here I review measurements of $v_2$, the second component in a
  Fourier decomposition of the azimuthal dependence of particle
  production relative to the reaction plane in heavy-ion
  collisions. $v_2$ is an observable central to the interpretation of
  the subsequent expansion of heavy-ion collisions. Its large value
  indicates significant space-momentum correlations, consistent with
  the rapid expansion of a strongly interacting Quark Gluon
  Plasma. Data is reviewed for collision energies from
  $\sqrt{s_{_{NN}}}=2$ to 200 GeV. Scaling observations and
  comparisons to hydrodynamic models are discussed.
\end{abstract}

\input{./introduction}

\input{./rhicdata}

\input{./hydro}

\section{Summary}

In this review, $v_2$ measurements were presented as a method for
studying space-momentum correlations in heavy-ion collisions. The
measurements of $v_2$ indicate the eccentricity in the initial overlap
region is transferred efficiently to momentum-space. At top RHIC
energy, the conversion is near that expectated from zero
mean-free-path hydrodynamic predictions. The comparisons of data to
hydrodynamics, however, depends on model calculations of the initial
eccentricity. Several models for the initial eccentricity have been
discussed. The mass, and $p_T$ dependence of $v_2$ at $p_T<1$~GeV/c is
found to be consistent with emission from a boosted source. Above
that, the particle type dependence of $v_2$ exhibits a dependence on
the number of constituent quarks in the hadron, with baryons obtaining
$v_2$ values larger than mesons.

The relationship between two-particle correlations, $v_2$, and $v_2$
fluctuations has also been discussed. Calculations showing that some
of the structures in two-particle correlations can be ascribed to
fluctuations in the initial conditions, have been
reviewed. Measurements of correlations and $v_2$ fluctuations can
therefore be used to constrain models for the initial
conditions. These constraints, along with improved measurements of the
shape of $v_2(p_T)$ as a function of system-size, improved
measurements of $\phi$ and $\Omega$ $v_2$, measurements of $v_2$ for
heavy-flavor hadrons, measurements at LHC energies, and a beam-energy
scan at RHIC will further improve our understanding of the properties
of the matter created in heavy-ion collisions.

\section*{Acknowledgements}

I'm grateful to Yan Lu and Alexander Wetzler for providing the code to
produce several figures. My thanks go to Tuomas Lappi for providing
Fig.~\ref{fig:gluondensity} and to Paul Romatschke, Peter Filip,
Arthur Poskanzer, and Anton Andronic for comments provided on the
text. I also want to thank Arthur Poskanzer for the advice he gives
me. I always appreciate it even though sometimes I fail to follow it
(usually to my own disadvantage). I also appreciate the many
discussions held with the STAR Flow Group and the STAR
Bulk-Correlations Physics Working Group. I want to thank all the
people who have contributed to the study of elliptic flow and its
interpretation, some of whom, I have failed to recognize in this
review. I refer the reader to a presentation on the history of flow
measurements\cite{histflow} for more context.

\end{document}

%% file: introduction.tex
%=======================================================================
\section{Introduction}

\begin{figure}[hbt]
\begin{center}
\centerline{  \includegraphics[width=.98\textwidth]{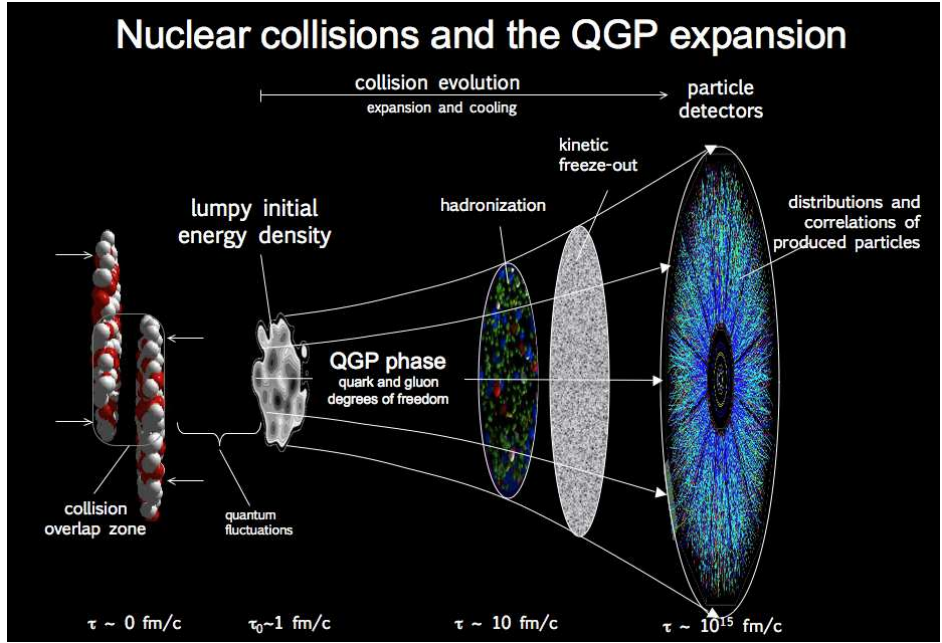}}
\caption{ A schematic diagram of the expansion after an
  ultra-relativistic heavy-ion collision. }
  \label{fig:expansion}
\end{center}
\end{figure}

Collisions of heavy nuclei have been exploited for decades to search
for and study the transition of hadronic matter to quark gluon
plasma\cite{Reisdorf:1997fx,Herrmann:1999wu}. In these collisions,
the extended overlap area, where the nuclei intersect and initial
interactions occur, does not possess sphrerical symmetry in the
transverse plane. Rather, for non-central collisions, the overlap area
is roughly elliptic in shape. If individual nucleon-nucleon collisions
within the interaction region are independent of each other
(\textit{e.g.}  point-like) and no subsequent interactions occur, this
spatial anisotropy will not be reflected in the momentum distribution
of particles emitted from the interaction region. On the other hand,
if the initial interactions are not independent, or if there are
subsequent interactions after the initial collisions, then the spatial
anisotropy can be converted into an anisotropy in momentum-space. The
extent to which this conversion takes place allows one to study how
the system created in the collision of heavy nuclei deviates from a
point-like, non-interacting system. The existence and nature of {\bf
  space-momentum} correlations is therefore an interesting subject in
the study of heavy ion collisions and the nature of the matter created
in those collisions\cite{Ollitrault:1992bk,Ollitrault:1995dy}.
Fig.~\ref{fig:expansion} shows an illustration of the possible stages
of a heavy-ion collisions starting with some initial energy density
deposited at mid-rapidity, followed by a QGP expansion, a
hadronization phase boundary, a kinetic freeze-out boundary and
finally the observation of particle trajectories in a detector.

\begin{figure}[ht]
\vspace{1.0cm}
\begin{minipage}[b]{0.49\linewidth}
\centering
\includegraphics[width=1.\textwidth]{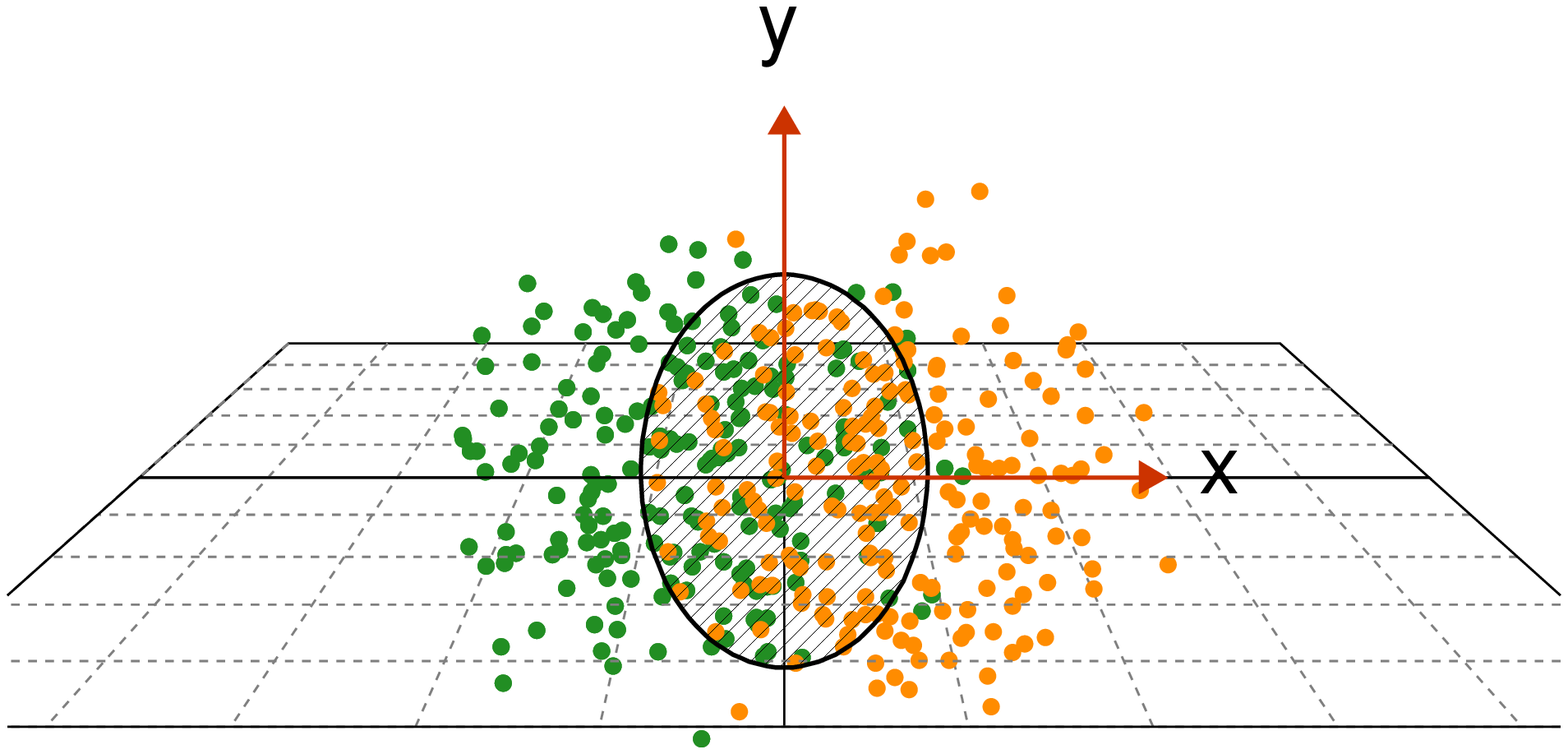}
\end{minipage}
\begin{minipage}[b]{0.49\linewidth}
\centering
\includegraphics[width=1.\textwidth]{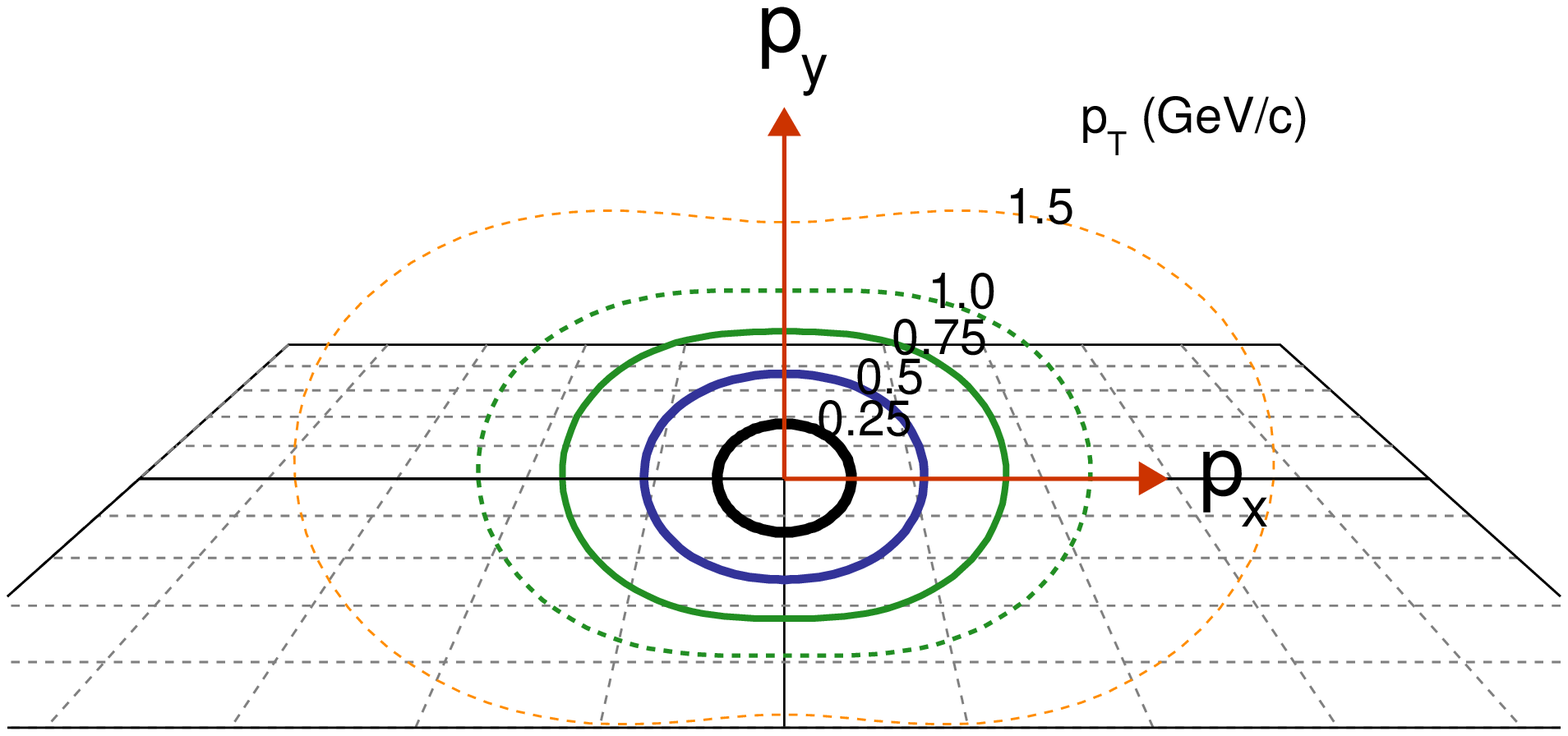}
\end{minipage}

\caption{Schematic illustrations of a $\sqrt{s_{_{NN}}}=200$ GeV Au+Au
  collision with a 6 fm impact parameter. The left panel shows the
  nucleons of the two colliding nuclei with an ellipse outlining the
  approximate interaction region. The right panel shows a
  momentum-space representation of $v_2$. The average radius of each
  successive ring represents the $p_T$ of the particles while the
  anisotropy of the ring represents the magnitude of $v_2$. The highest
  $p_T$ particles (outer-ring) exhibit the strongest $v_2$ while the
  lowest $p_T$ particles (inner-ring) exhibit a vanishingly small
  $v_2$. }
\label{fig:collision2}
\end{figure}

One can consider a number of ways to study space-momentum
correlations: \textit{e.g.} two-particle
correlations\cite{Voloshin:2003ud} and
HBT\cite{Lisa:2005dd,Voloshin:1995mc1996ch}. In this review we
discuss elliptic flow $v_2$; an observable that has been central in
the interpretation of heavy-ion data and QGP
formation\cite{whitepapers}. Given the predominantly elliptic shape
of the initial overlap region, it is natural to ask whether this shape
also shows up in the distribution of particles in
momentum-space. Fig.~\ref{fig:collision2} shows a schematic
illustration of the conversion of coordinate-space anisotropy to
anisotropy in momentum-space. The left panel shows the position of
nucleons in two colliding nuclei at the moment of impact. The overlap
region is outlined and shaded. A Fourier decomposition can be used to
describe the azimuthal dependence of the final triple momentum-space
distributions\cite{Voloshin:1994mz}:
\begin{equation}
 \frac{d^3N}{p_Tdp_Tdyd(\phi-\Psi)} = \frac{1}{2\pi}\frac{dN}{p_Tdp_Tdy}\times
\left[1 + 2v_1\cos( \phi - \Psi) + 2v_2\cos( 2 (\phi - \Psi)) + ...\right],
\end{equation}
where $\phi$ is the azimuth angle of the particle, $y$ the
longitudinal rapidity variable, $p_T$ the transverse momentum, and
$\Psi$ is the reaction plane angle defined by the vector connecting
the centers of the two colliding nuclei. Positive $v_2$ implies that
more particles are emitted along the short axis of the overlap region.
To study the extent to which space-momentum correlations develop in
heavy-ion collisions, one can measure the second component $v_2$ and
compare it to the initial spatial
eccentricity\cite{Ollitrault:1992bk,Voloshin:1999gs}.  The right
panel of Fig.~\ref{fig:collision2} shows the final azimuthal
distribution of particles in momentum-space. The curves represent the
anisotropy at different $p_T$ values measured in 200 GeV Au+Au
collisions\cite{Adams:2004bi}: \textit{i.e.} $f(p_T, \phi)=p_T*(1 +
2v_2(p_T)\cos(atan2(p_{y},p_{x})))$. The goal of $v_2$ measurements is
to study how the initial spatial anisotropy in the left panel is
converted to the momentum-space anisotropy in the right panel. In this
review, a summary of $v_2$ data for different colliding systems,
different center-of-mass energies, and different centralities is
given.

This review will focus on results from the first four years of
operation of the Relativistic Heavy Ion Collider (RHIC). We start with
a brief discussion of the beam energy dependence of $v_2$ and some
ideas about what physics might be relevant. Even before considering
physics scenarios to explain how a space-momentum correlation
develops, one can see that to interpret $v_2$ it is important to
understand the initial geometry and how it varies with the collision
centrality and system-size.  Since, the concept of the reaction-plane
is so central to the definition of $v_2$ and eccentricity is so
central to it's interpretation, I discuss the two in a sub-section
below.  Then a review of RHIC data is provided. This will include the
dependence of $v_2$ on center-of-mass energy, centrality, colliding
system, pseudo-rapidity, $p_T$, particle mass, constituent quark number
and various scaling laws. In the following section, I will discuss
comparisons to models and the emergence of the hydrodynamic paradigm
at RHIC. Particular emphasis is given to uncertainties in the model
comparisons. In that section I will also discuss current attempts to
extract viscosity and future directions of investigation.

Voloshin, Poskanzer, and Snellings recently wrote a review
article\cite{psv} on collective phenomena in non-central nuclear
collisions that deals with a similar subject matter. That article
provides valuable detail on technical aspects of measuring $v_2$. In
this review I will attempt to avoid duplicating that work by
discussing interpretations of $v_2$ more extensively and refer the
reader to that review where appropriate.

\subsection{Two Decades in Time and Five Decades in Beam Energy}

\begin{figure}[hbt]
\begin{center}
%FOPI data are for Z=1 (~=p for E/A > 1 GeV, but quite larger flow 
%than p for lower energies), the E895 are for protons,
%E877 all charged, NA49 pions, RHIC: all charged (I believe).
\centerline{  \includegraphics[width=.8\textwidth]{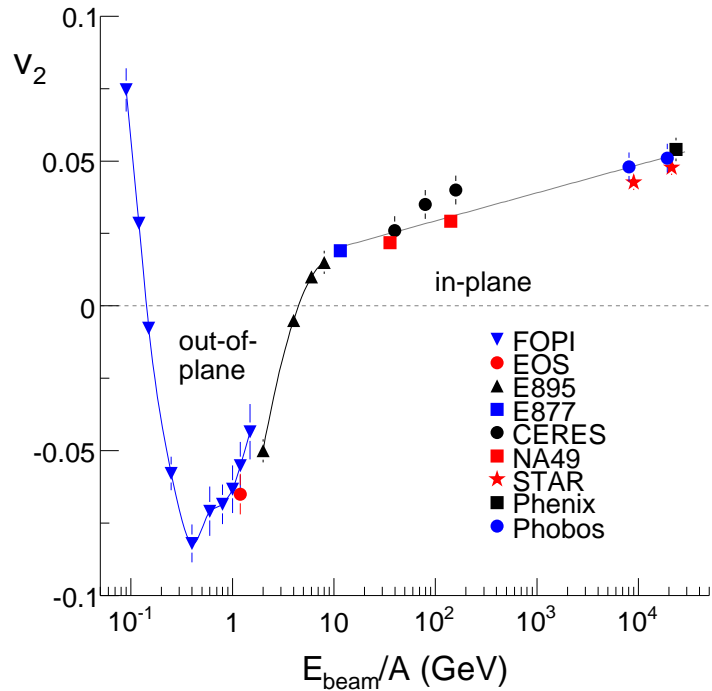}}
\caption{The beam energy dependence of elliptic flow measurements.
  %Data are shown for 0\%--20\% most central A+A collisions. 
  The RHIC and E877 data are for charged hadrons, independent of
  species. The NA49 data is for charged pions. The E895 data are for
  protons and the FOPI data are for atomic number $Z$=1. At each
  energy, the sample of particles is close to the total
  charge. Positive $v_2$ values indicate that particles tend to be
  more aligned with the reaction-plane (in-plane). RHIC and SPS data
  suggest a smooth trend of in-plane $v_2$ growing with
  $\log(\sqrt{s_{_{nn}}})$ above $E_{beam}/A \approx 20$~GeV or
  $\sqrt{s_{_{NN}}}\approx 6$ GeV.}
  \label{fig:v2excitation}
\end{center}
\end{figure}

Positive values of $v_2$ imply that particles tend to be produced more
abundantly in the $x-$direction than in the $y-$direction. This is
referred to as in-plane flow. Fig.~\ref{fig:v2excitation} shows $v_2$
measured in an interval of beam energies covering five orders of
magnitude\cite{v2ex,Barrette:1994xr,Barrette:1996rs,Pinkenburg:1999ya,Chung:2001qr,Ackermann:2000tr,Manly:QM2002,CERES:QM2002,Alt:2003ab,Andronic:2004cp}.
For $E_{beam}/A$ ranging from approximately $0.12-5$ GeV
($1.92<\sqrt{s_{_{nn}}}<3.3$ GeV), $v_2$ is negative. For this energy
range, spectator protons and neutrons are still passing the
interaction region while particles are being produced. Their presence
inhibits particle emission in the in-plane direction leading to the
phenomenon termed squeeze-out. At still lower energies, $v_2$ is
positive as the rotation of the matter leads to fragments being
emmitted in-plane. At this energy beam rapidity and mid-rapidity are
essentially indistinguishable with $y_{beam}<0.41$ units.

{\bf In-plane flow:} As the beam energy is increased, the nuclei
become more Lorentz contracted and the time it takes the spectators to
pass each other decreases. It was predicted by
Ollitrault\cite{Ollitrault:1992bk} that at high enough beam energy,
the squeeze-out phenomena would cease and $v_2$ would take on positive
values. Positive $v_2$ values were measured at the AGS for energies
above $E_{beam}/A=5$ GeV ($\sqrt{s_{_{nn}}}=3.3$~GeV). For energies
above $E_{beam}/A \approx 20$ GeV ($\sqrt{s_{_{nn}}}=6.3$~GeV), $v_2$
exhibits a steady log-linear increase: $v_{2}\approx 0.01 + 0.0042
\log(E_{beam}/A)$ or $v_{2}\approx 0.008 + 0.0084
\log(\sqrt{s_{_{nn}}})$ where the data represented are from
intermediate impact parameter A+A collsions.  It appears therefore
that RHIC $v_2$ data may be part of a smooth trend that began at SPS
energies. This trend was noted previously at least
once\cite{Andronic:2004tx}. Understanding the physics that underlies
that trend is one of the challenges of heavy-ion physics.

One class of models that has provided an illustrative reference for
heavy-ion collisions are hydrodynamic models which are used to model
the expansion the matter remaining in the fireball after the initial
collisions\cite{Landau:1953gs,Kolb:1999it,Teaney:1999gr,Kolb:2000sd,Kolb:2000fha,Teaney:2000cw,Huovinen:2001cy,Teaney:2001av,Teaney:2001gc,Heinz:2001xi}. This
model can be used to determine how matter with a vanishingly small
mean free path would convert the initial eccentricity into
$v_2$. These models typically treat all elliptic flow as arising from
the final state expansion rather than from some initial state
effects\cite{Boreskov:2008uy,Kopeliovich:2008nx,Teaney:2002kn,Krasnitz:2002ng}. In
the hydrodynamic models, large pressure gradients in the in-plane
direction lead to a preferential flow of matter in the in-plane
direction. In this review, we will use hydrodynamic models as a
convenient reference. Other models providing a valuable reference for
measurements include hadronic and partonic cascades and transport
models\cite{Pang:1992sk,Sorge:1996pc,Bleicher:1999xi,Bass:1998ca,Burau:2004ev,Geiger:1991nj,Geiger:1997pf,Zhang:1997ej,Molnar:2001ux}. Additionally,
the blast-wave model provides a successful parametrization of low
$p_T$ heavy-ion data, including $v_2$, HBT, and spectra in terms of
several freeze-out parameters\cite{Huovinen:2001cy,Retiere:2003kf}.

\subsection{Initial Geometry: The Reaction Plane and Eccentricity}

\begin{figure}[ht]

\vspace{1.0cm}
\begin{minipage}[b]{0.49\linewidth}
\centering
\includegraphics[width=1.\textwidth]{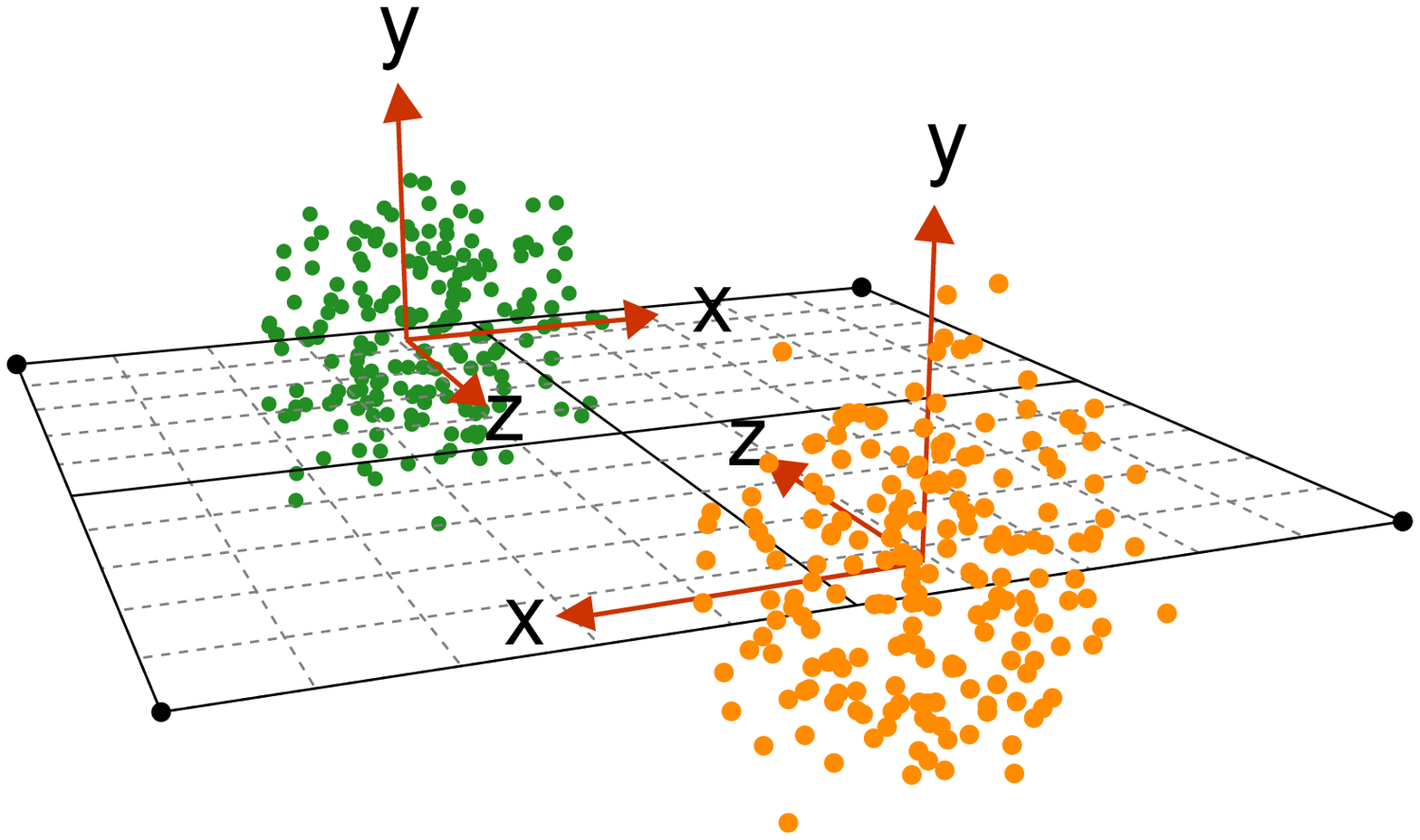}
\end{minipage}
\begin{minipage}[b]{0.49\linewidth}
\centering
\includegraphics[width=1.\textwidth]{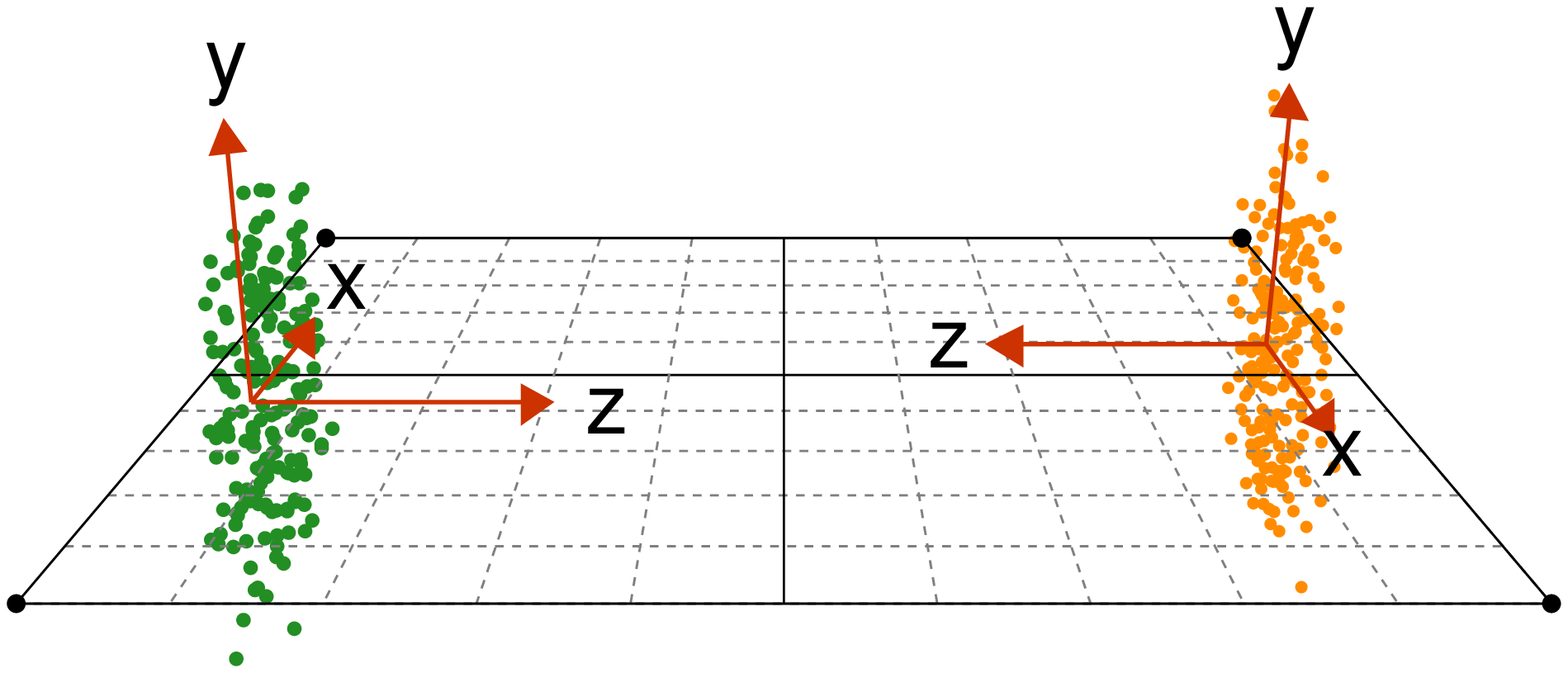}
\end{minipage}

\begin{minipage}[b]{0.49\linewidth}
\centering
\vspace{1.0cm}
\includegraphics[width=1.\textwidth]{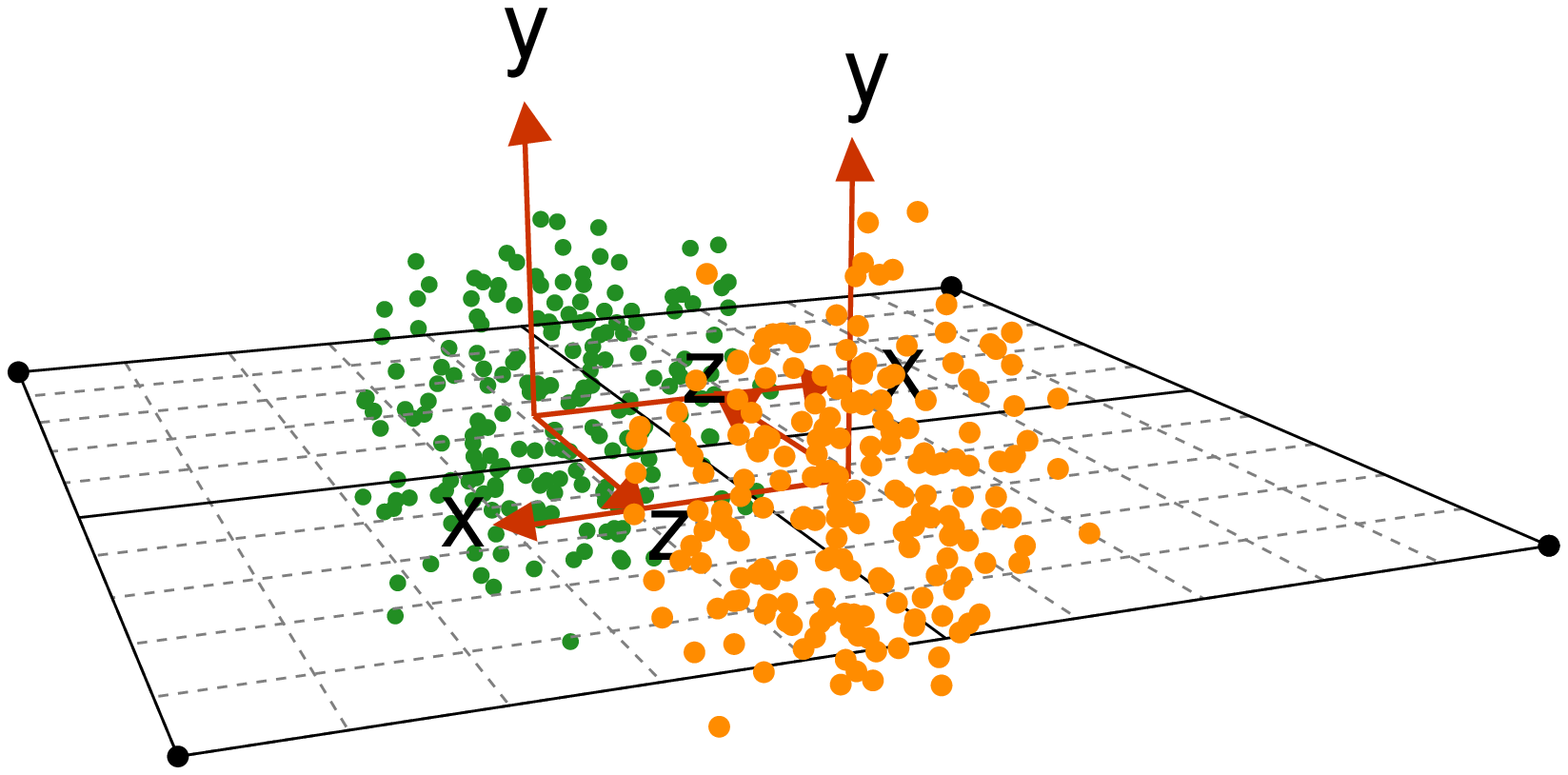}
\end{minipage}
\begin{minipage}[b]{0.49\linewidth}
\vspace{1.0cm}
\centering
\includegraphics[width=1.\textwidth]{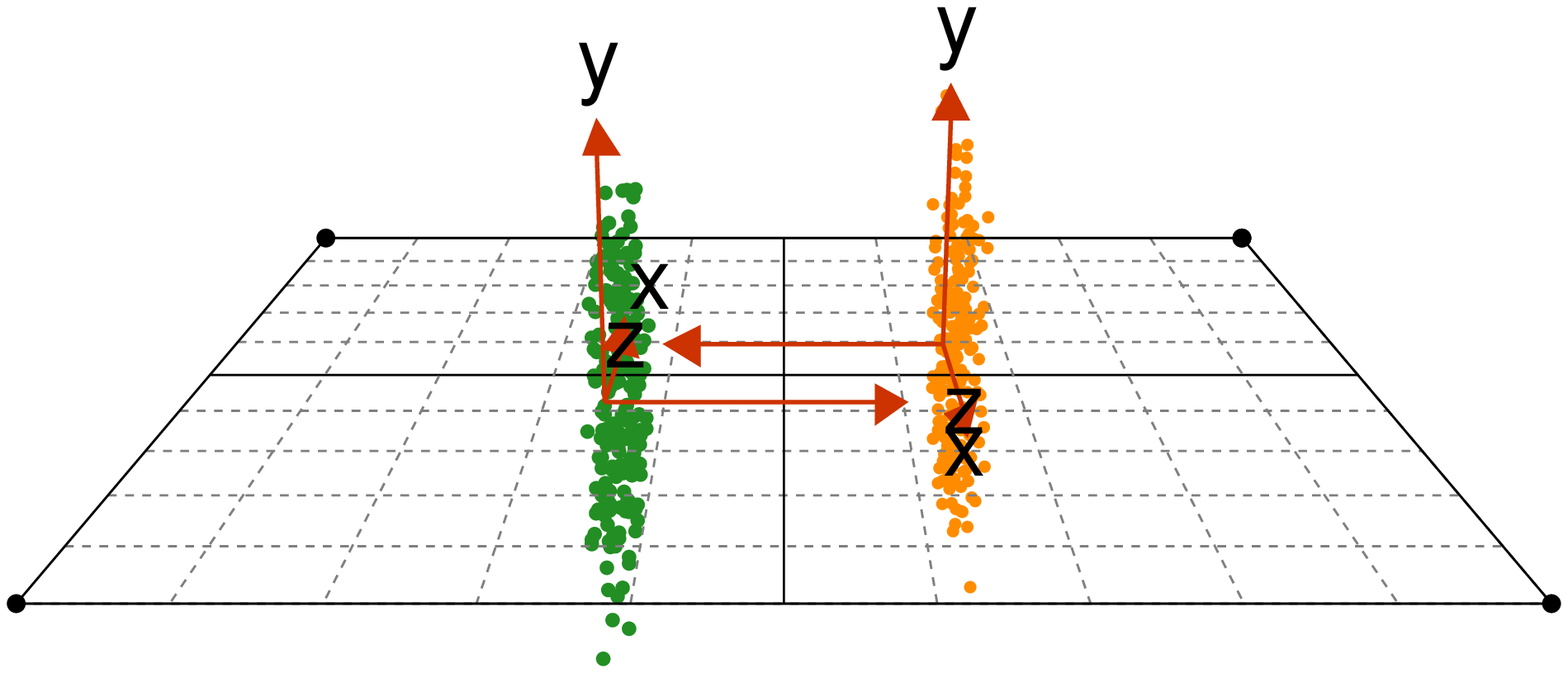}
\end{minipage}

\caption{Schematic illustrations of a $\sqrt{s_{_{NN}}}=200$ GeV
  Au+Au collision with a 6 fm impact parameter. The right-handed
  coordinate systems defined by the momentum of the nucleus (here the
  z-axis) crossed with the vector pointing to the center of the
  approaching nucleus (the x-axis) is shown for each nucleus. The
  reaction-plane is the plane normal to the y-axis, containing the
  centers of the two colliding nuclei. The nucleons are distributed
  inside the nucleus according to a Woods-Saxon distribution. The
  nuclei are Lorentz contracted in the z-direction. }
\label{fig:collision}
\end{figure}

In the collision of two symmetric nuclei, a unique vector (the
$y$-axis) can be defined by applying the right-hand-rule to the
momentum vector of one nucleus and the vector pointing to the center
of the other nucleus. The $y$-axis is a pseudovector. The
reaction-plane is then the plane perpendicular to the $y$-axis
containing the points at the center of the two nuclei. The
reaction-plane and the right-handed coordinate system are illustrated
in Fig.~\ref{fig:collision}. The figure contains perspective
illustrations of two nuclei approaching with an impact parameter of 6
fm. The impact parameter is the distance between the centers of the
two nuclei at the moment of their closest approach. The two nuclei in
this illustration are Lorentz contracted by a Lorentz gamma factor of
10 which roughly corresponds to the appropriate gamma for top SPS
energies.
%Using the vector
%from the right-hand-rule as the y-axis, the eccentricity usually
%calculated as
%\begin{equation}
%\varepsilon = \frac{\langle y^2-x^2\rangle}{\langle x^2 + y^2\rangle}.
%\end{equation}

\begin{figure}[hbt]
\begin{center}
\centerline{  \includegraphics[width=.8\textwidth]{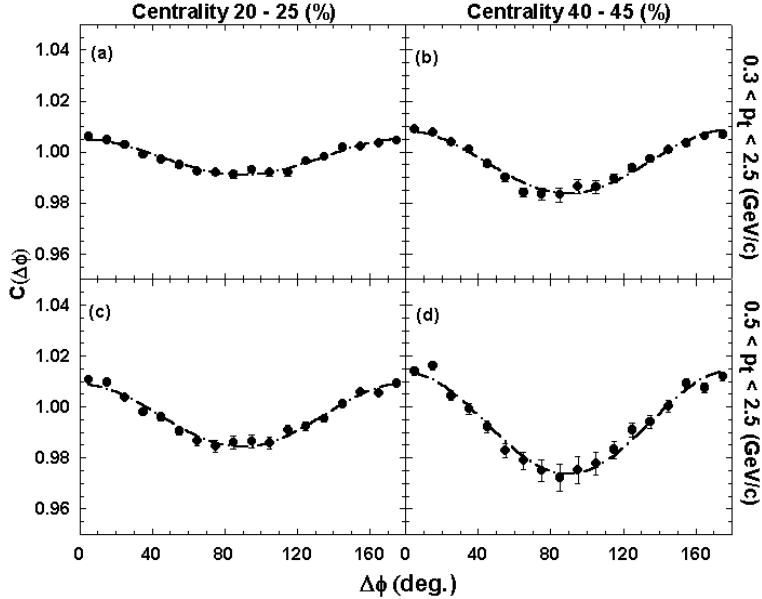}}
\caption{ Charged hadron correlation functions in Au+Au collisions at
  $\sqrt{s_{_{nn}}}=200$~GeV for two centrality intervals and two
  $p_T$ ranges. }
  \label{fig:v2corr}
\end{center}
\end{figure}

The reaction-plane is not directly observed in experiments, however,
and this introduces a systematic uncertainty into the measurement of
$v_2$. One often relies instead on indirect observations to estimate
$v_2$\cite{Ollitrault:1997di,Ollitrault:1997vz,Poskanzer:1998yz,Danielewicz:1985hn,Ollitrault:1993ba}. For
example, when forming two particle azimuthal correlations such as
$\frac{dN}{d(\phi_1-\phi_2)}$, a non-zero $v_2$ value will lead to a
modulation in $\Delta\phi=\phi_1-\phi_2$ of the form $1+2\langle v_2^2
\rangle\cos(2\Delta\phi)$. Fig.~\ref{fig:v2corr} shows the correlation
function for hadrons produced at mid-rapidity at
RHIC\cite{Wang:1991qh,Adcox:2002ms}. The panels show different
centralities. The area normalized correlation function is
\begin{equation}
C(\Delta \phi) \equiv 
\frac{Y^{AB}_{\rm Same}(\Delta\phi)}
                            {Y^{AB}_{\rm Mixed}(\Delta\phi)}
                            \times
\frac{\int Y^{AB}_{\rm Mixed}(\Delta\phi)}
                            {\int Y^{AB}_{\rm Same}(\Delta\phi)}
\propto  \frac{dN^{AB}}{d(\Delta\phi)}                         
\label{Eq:CF_defined}
\end{equation}
\noindent
where $Y^{AB}_{\rm Same}(\Delta\phi)$ and $Y^{AB}_{\rm
  Mixed}(\Delta\phi)$ are, respectively, the uncorrected yields of
pairs in the same and in mixed events within each data sample.
$C(\Delta\phi)$ shows a clear $\cos(2\Delta\phi)$ dependence. We note
here that what is measured in these correlation functions is $\langle
v_2^2\rangle = \langle v_2\rangle^2 + \sigma_{v_{2}}^2$ in
anticipation of a discussion of $v_2$ fluctuations.

\begin{figure}[hbt]
\begin{center}
\centerline{  \includegraphics[width=.8\textwidth]{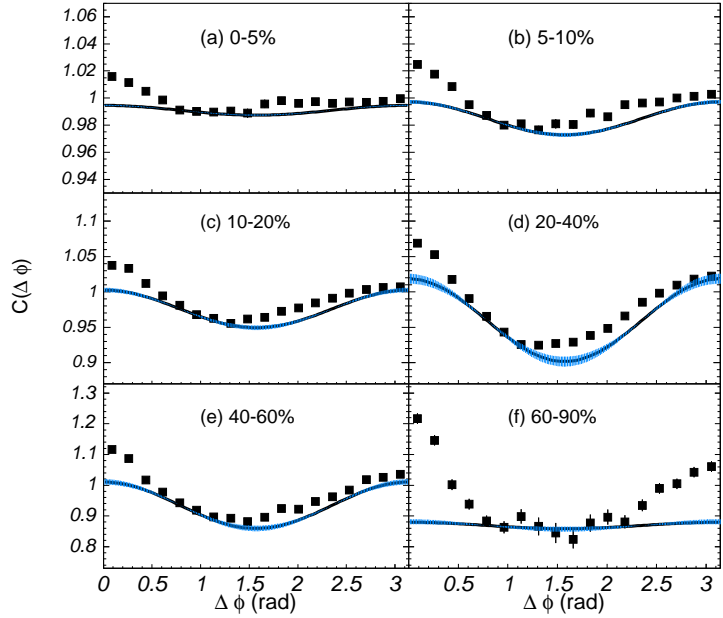}}
\caption{ Charged hadron correlations for a variety of centrality
  intervals. The correlation function is formed between two samples of
  hadrons based on their $p_T$: the ``trigger'' particle sample is
  selected with $2.5$ GeV/c $<p_T<4.0$~GeV/c while the ``associated''
  particle sample is selected with $1$ GeV/c $<p_T<2.5$~GeV/c.  }
  \label{fig:jetcorr}
\end{center}
\end{figure}

$v_2$ will not be the only contribution to the azimuthal dependence of
the two-particle azimuthal correlations. Other processes that are not
related to the reaction-plane can give rise to structures in the shape
of the two-particle $\Delta\phi$ distribution as well. These
non-reaction plane contributions are commonly called "non-flow". The
subject of non-flow is an important one and will be discussed
throughout this review. The contribution of non-flow can be seen more
clearly by looking at very peripheral collisions or by selecting high
momentum particles to increase the chance that a particular pair of
hadrons are correlated to a hard scattered parton
(jet). Fig.~\ref{fig:jetcorr} shows the correlation function for
higher momentum particles\cite{Adler:2005ee}. The solid line shows
what a pure $v_2$ correlation would look
like\cite{Bielcikova:2003ku}. The difference between those curves and
the data are often taken as a measurement of jet
correlations\cite{Adler:2005ee,Adler:2002tq,Adler:2002ct}.

\begin{figure}[ht]

\vspace{1.0cm}
\begin{minipage}[b]{0.49\linewidth}
\centering
\includegraphics[width=50mm]{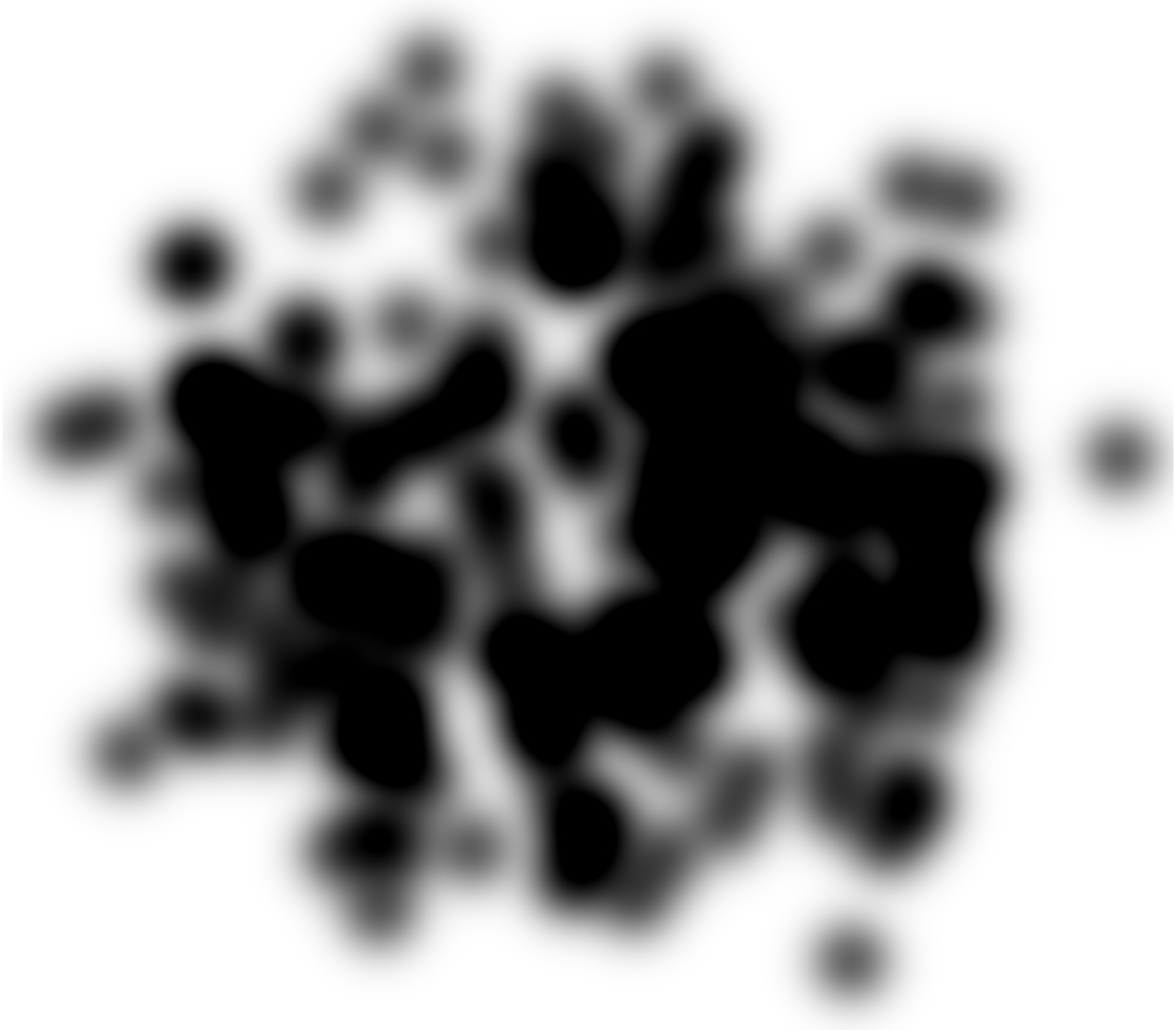}
\end{minipage}
\begin{minipage}[b]{0.49\linewidth}
\centering
\includegraphics[width=50mm]{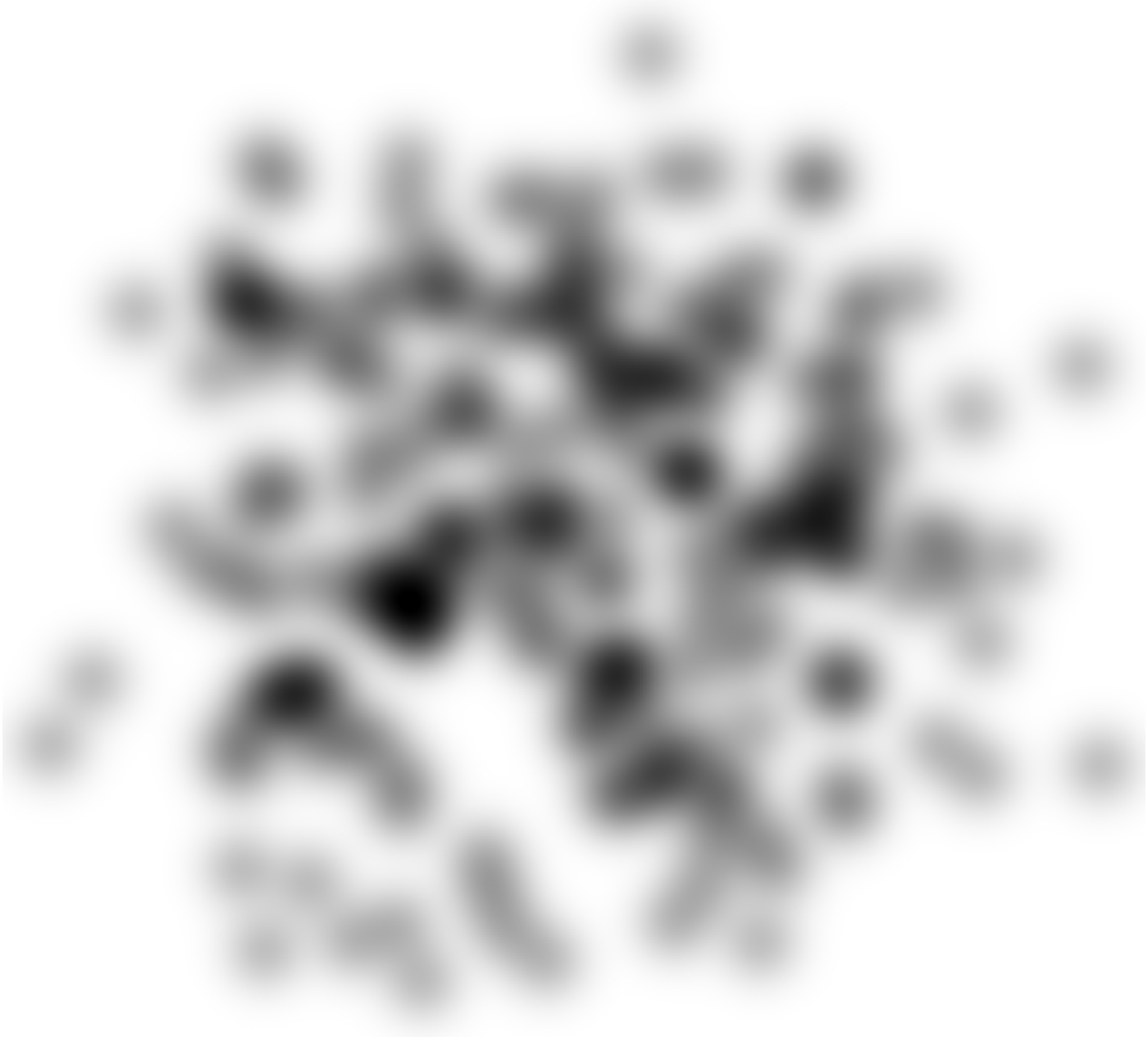}
\end{minipage}

\caption{ Gluon density in the transverse plane when the nucleus is
  probed at different $x$ values by a 0.2 fm quark-antiquark dipole in
  the IPsat CGC model. }
\label{fig:gluondensity}
\end{figure}

Even if the reaction-plane were known with precision, there is no
first principles calculation of the initial matter distribution in the
overlap region, so the eccentricity is uncertain. Various models can
be used to calculate the initial spatial eccentricity which can then
be compared to $v_2$. Defining the $y-$axis according to the
right-hand-rule, the eccentricity $\varepsilon_s$ is traditionally
calculated as:
\begin{equation}
\varepsilon_s = \frac{\langle y^2-x^2\rangle}{\langle x^2 + y^2\rangle},
\end{equation}
where the average represents a weighted mean. Other eccentricity
definitions have also been considered\cite{Bhalerao:2006tp}. The
weights can be some physical quantity in a model such as energy or
entropy density, or simply the position of nucleons participating in
the collision. One popular method for calculating the eccentricity is
to use a Monte Carlo Glauber model. Details can be found in a recent
Review\cite{Miller:2007ri}. In that model, a finite number of
nucleons are distributed in a nucleus according to a Woods-Saxon
distribution. Then two nuclei are overlaid with a fixed impact
parameter and the $x$ and $y$ positions of the participating nucleons
is determined based on whether the nucleons overlap in the transverse
plane; each nucleon is considered to be a disk with an area determined
by the $\sqrt{s}$ dependent nucleon-nucleon cross-section. The $x$ and
$y$ coordinates of the participating nucleons are then used to
calculate the eccentricity. Those nucleons that do not participate in
this initial interaction are called spectators. One can anticipate
that due to the finite number of nucleons in this model, the initial
geometry will fluctuate. Other models used to determine the initial
matter distribution including HIJING\cite{Wang:1991hta},
NEXUS\cite{Drescher:2000ec}, and Color Glass Condensate
models\cite{Kowalski:2007rw,Kowalski:2003hm,Lappi:2006xc,Drescher:2006pi}
also reach the same conclusions; the initial overlap region is
expected to be lumpy rather than smooth. Fig.~\ref{fig:gluondensity}
shows the gluon density in the transverse plane which is probed by a
0.2 fm quark-antiquark dipole at two different $x$ values in the IPsat
CGC model\cite{Kowalski:2007rw} ($x = 2p_T/\sqrt{s_{_{NN}}}$ is
$10^{-5}$ in the left panel and $10^{-3}$ in the right panel). The
lumpiness is immediately apparent.

\begin{figure}[ht]

\vspace{1.0cm}
\begin{minipage}[b]{0.49\linewidth}
\centering
\includegraphics[width=1.\textwidth]{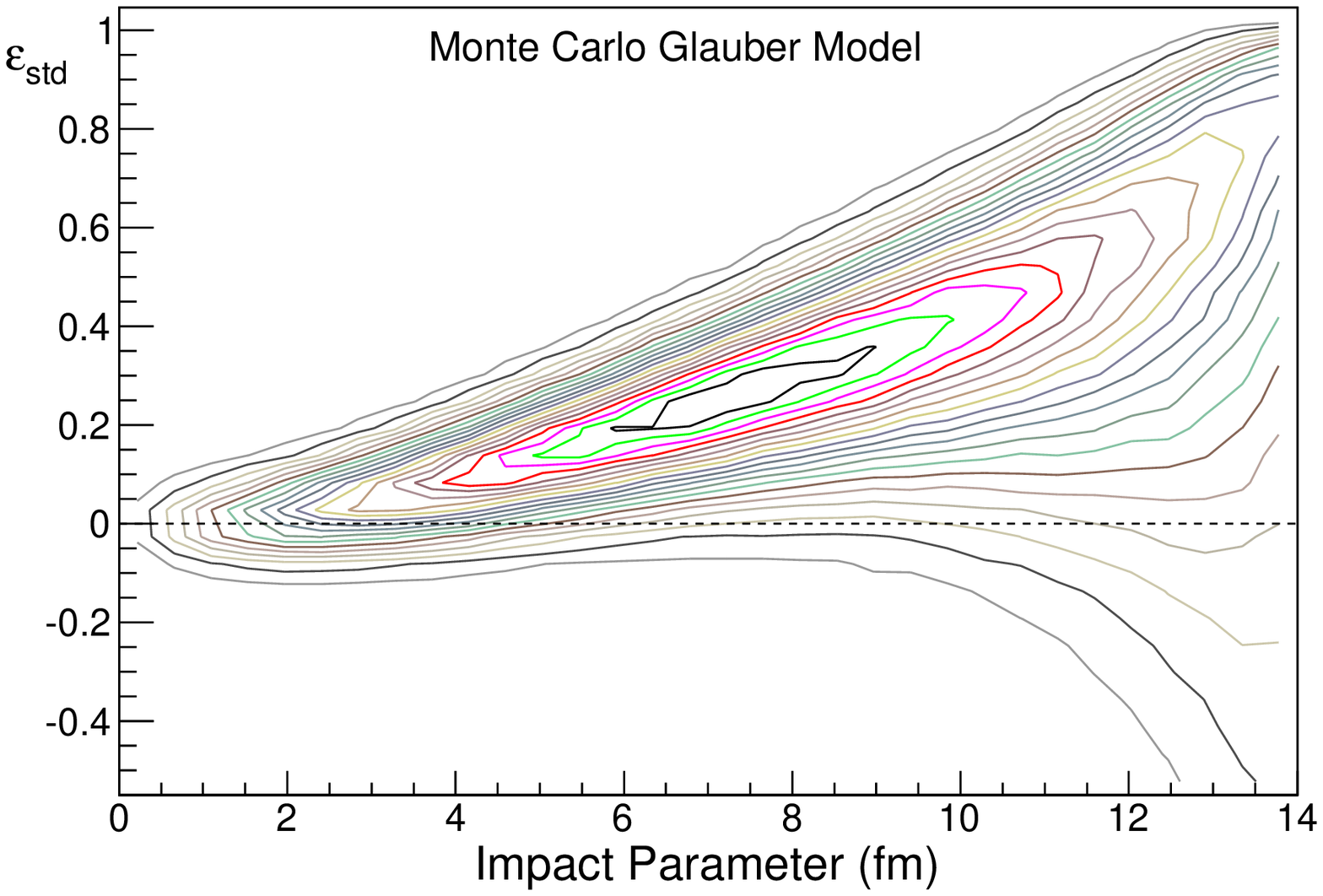}
\end{minipage}
\begin{minipage}[b]{0.49\linewidth}
\centering
\includegraphics[width=1.\textwidth]{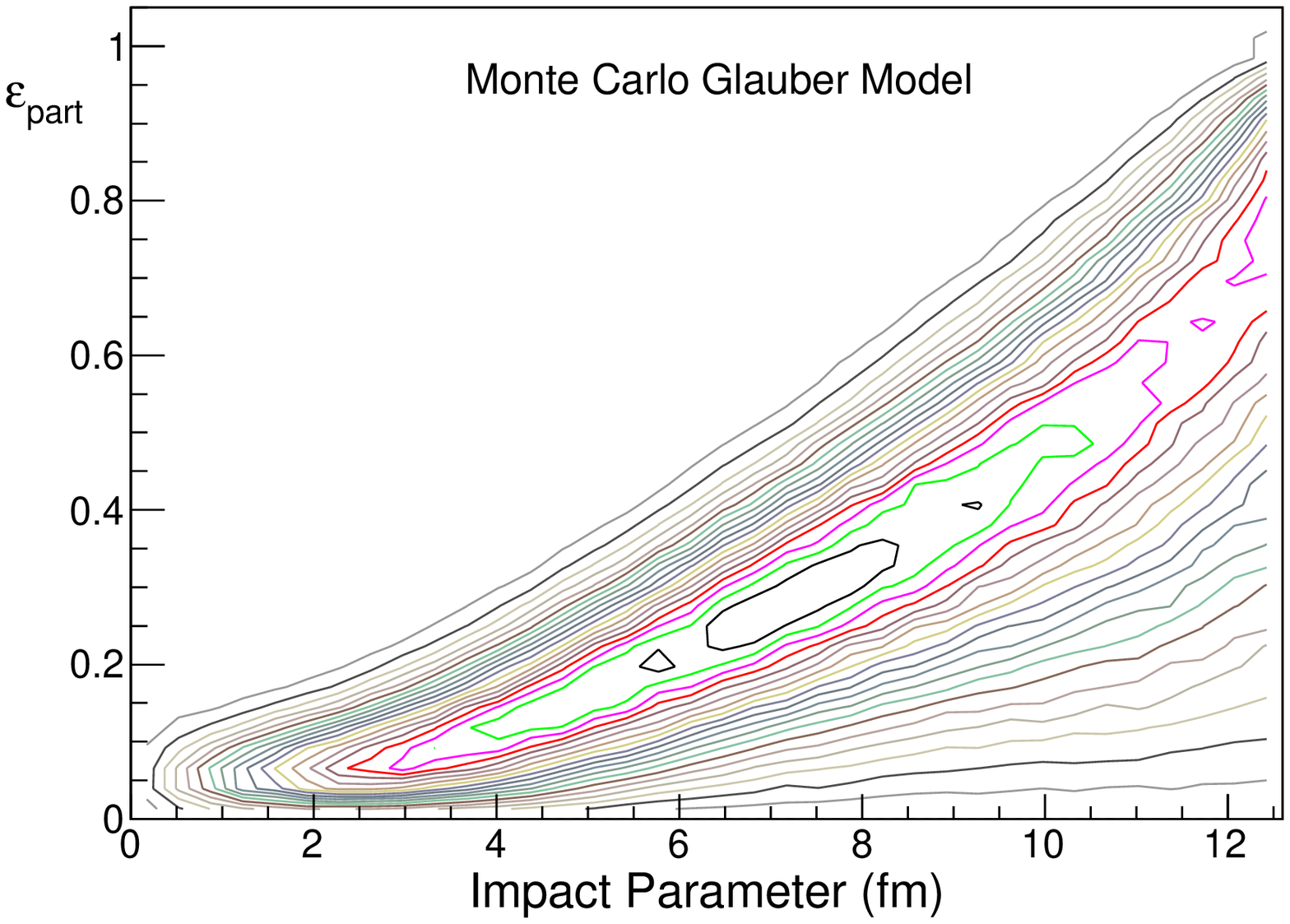}
\end{minipage}

\caption{ The distribution of eccentricity on an event-by-event basis
  when calculated relative to the reaction-plane (left) and the
  participant plane (right). }
\label{fig:ecc}
\end{figure}

Until recently however\cite{Aguiar:2001ac,Miller:2003kd}, $v_2$ data
was compared almost exclusively to calculations assuming an infinitely
smooth initial matter distribution (for example in initializing a
hydrodynamic expansion). Improving on that approximation may be
important for understanding the shape expected for the
$\frac{d^2N}{d\Delta\eta d\Delta\phi}$
distribution\cite{Voloshin:2003ud,Sorensen:2008dm}. This distribution
is also investigated in heavy ion collisions in order to search for
jets. In any scenario where space-momentum correlations develop, the
correlations and fluctuations in the initial geometry can be
manifested in the $\frac{dN}{d\Delta\phi}$ distribution and
understanding these correlations is important for interpreting
heavy-ion collisions. Fluctuations in the initial geometry have also
led to the idea of measuring particle distributions relative to the
participant-plane rather than the
reaction-plane\cite{Manly:2005zy,Bhalerao:2006tp}. The
participant-plane is defined by the major axis of the eccentricity
which, due to fluctuations, can deviate from the reaction-plane. The
eccentricity relative to the participant-plane is a positive definite
quantity and is always larger than the eccentricity relative to the
reaction-plane; the participant-plane is defined by rotating to the
axis that maximizes the eccentricity. %In accounting for possible
%fluctuations, one defines two quantities:
%\begin{equation}
%  \varepsilon_{std}^2 = \langle\varepsilon\rangle^2 - \sigma_{\varepsilon}^2 \\
%  \varepsilon_{part}^2 = \langle\varepsilon\rangle^2 + \sigma_{\varepsilon}^2
%\end{equation}
%where $\varepsilon_{std}$ is the eccentricity defined relative to the
%x-axis and $\varepsilon_{part}$ is the eccentricity defined relative
%to the major axis of the eccentricity. The left panel of
Fig.~\ref{fig:ecc} shows the event-by-event distribution of the
standard eccentricity (left panel) and the participant eccentricity
(right panel) as a function of impact parameter determined from a
Monte-Carlo Glauber calculation. The fluctuations in this model are
large as illustrated by the widths of the distributions. The
relationship between the different definitions of eccentricity and
their fluctuations are explained clearly in two recent
papers\cite{Bhalerao:2006tp,Voloshin:2007pc}.

Different models for the initial matter distribution yield different
estimates of $\langle\varepsilon\rangle$ and
$\sqrt{\langle\varepsilon^2\rangle}$. The deviations in the
$\langle\varepsilon\rangle$ for different models can be of the order
of 30\% and strongly centrality dependent\cite{Hirano:2005xf}. That
uncertainty in $\langle\varepsilon\rangle$ leads to an inherent
uncertainty when comparing models to $v_2/\varepsilon$. This level of
uncertainty becomes important when attempting to estimate transport
properties of the matter based on comparisons of the observed $v_2$ to
the initial $\varepsilon$.

%% file: rhicdata.tex
\section{Review of Recent Data}

\begin{figure}[hbt]
\begin{center}
\centerline{  \includegraphics[width=.8\textwidth]{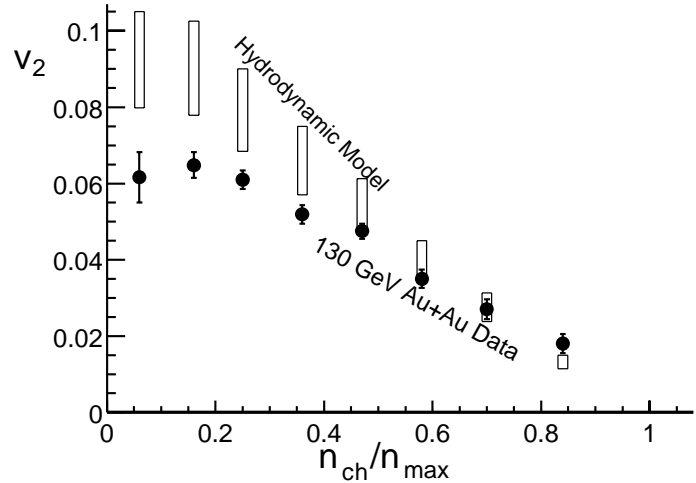}}
\caption{ First measurements of $v_2$ versus centrality at
  RHIC. Positive values are observed which are largest for events with
  the largest eccentricity and decrease for more central, symmetric
  collisions. The trend with centrality clearly indicates a
  space-momentum correlation driven by the eccentricity of the initial
  overlap zone. Centrality is expressed in terms of the observed
  multiplicity of a give event relative to the highest multiplicity
  observed $n_{ch}/n_{max}$. Data are compared to eccentricity scaled
  by 0.19 (bottom edge of the boxes) and 0.25 (top edge of the
  boxes). The values are chosen to represent the typical conversion of
  eccentricity to $v_2$ in a hydrodynamic model.}
  \label{fig:v2cent}
\end{center}
\end{figure}

The first paper published on RHIC data was on elliptic flow in
$\sqrt{s_{NN}}=130$~GeV Au+Au
collisions\cite{Ackermann:2000tr}. Fig.~\ref{fig:v2cent} shows that
data on the centrality dependence of $v_2$. The values of $v_2$ reach
a maximum of approximately 6\% for peripheral collisions where the
initial eccentricity of the system is largest. That value is 50\%
larger than the values reached at SPS energies\cite{Alt:2003ab} and
the $v_2$ values are a factor of two larger than the those predicted
by the RQMD transport-cascade model\cite{Bleicher:1999xi}. For
central collisions, the measurements approach the zero mean-free-path
limit estimated from the eccentricity shown in the figure as open
boxes. The boxes represent the eccentricity scaled by 0.19 (bottom
edge of the boxes) and 0.25 (top edge of the boxes). Those values are
chosen to represent the typical conversion of eccentricity to $v_2$ in
hydrodynamic models. At lower $\sqrt{s_{NN}}$ energies, the RQMD model
provided a better description of the data, while hydrodynamic models
significantly over-predicted the data. The conclusion based on this
early comparison, therefore, was that heavy-ion collisions
approximately satisfy the assumptions made in the hydrodynamic models:
1) zero mean-free-path between interactions, and 2) early local
thermal equilibrium\cite{Heinz:2001xi}. These conclusions remain at
the center of scientific debate in the heavy-ion community.

\begin{figure}[hbt]
\begin{center}
\centerline{  \includegraphics[width=.8\textwidth]{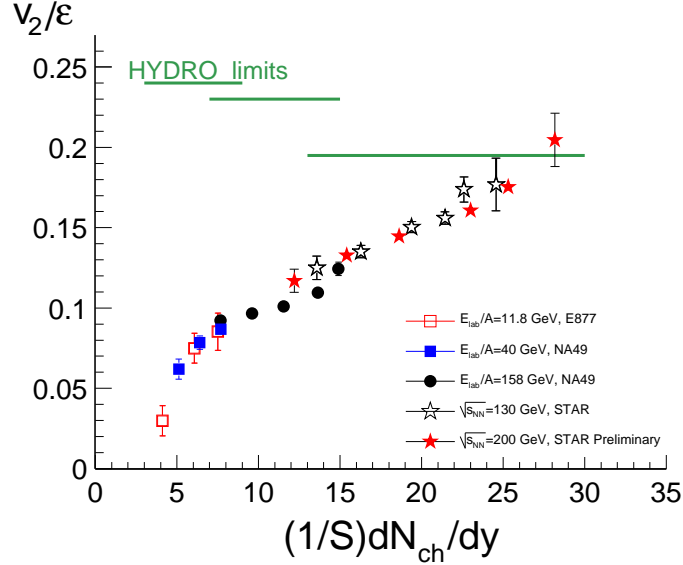}}
\caption{ $v_2$ scaled by a Monte Carlo Glauber model calculation of
  the initial overlap eccentricity. The ratio is plotted
  versus transverse particle density $(1/S)dN_{ch}/dy$, where
  $S=\pi\sqrt{\langle x^2\rangle\langle y^2\rangle}$ is a weighted
  average area calculated with the same model as the
  eccentricity. Data are taken from different $\sqrt{s_{_{NN}}}$
  values and different centralities. Plotted in this format, the data
  suggest $v_2/\varepsilon$ for different energies and overlap
  geometries is determined by the transverse particle density, and
  approaches a zero mean-free-path hydrodynamic limit for most central
  top energy collisions at RHIC. This conclusion is not universally
  accepted and is still being investigated. }
  \label{fig:v2ecc}
\end{center}
\end{figure}

In Fig.~\ref{fig:v2ecc}, $v_2$ is scaled by model calculations of the
initial eccentricity and plotted versus transverse particle density
$\frac{1}{S}\frac{dN}{dy}$\cite{Alt:2003ab}. This facilitates
comparisons of $v_2$ across different $\sqrt{s_{_{NN}}}$ energies,
collision centralities and
system-sizes\cite{Sorge:1996pc,Voloshin:1999gs}.

For the case of ballistic expansion of the system -- that is an
expansion for which the produced particles escape the initial overlap
zone without interactions -- $v_2$ should only reflect the
space-momentum correlations that arise from the initial conditions.
Those can exist in the case that the initial interactions are not
point-like\cite{Boreskov:2008uy} but rather involve cross-talk
between different $N+N$ interactions within the overlap zone.
%a large enough area to be
%sensitive to the geometry of the overlap region. 
%The exact value of
%the ballistic expansion limit is not known but is thought to be small
%and is often taken as zero. 
The opposite extreme from the ballistic expansion limit is the zero
mean-free-path limit represented by ideal hydrodynamic models. Lacking
a length scale, the zero mean-free-path models should not depend on
system-size and instead should be a function of density.

The measurements of $v_2$ are expected to rise from values near the
ballistic expansion limit and asymptotically approach the zero
mean-free-path limit as the density of the system is increased. Data
in Fig.~\ref{fig:v2ecc} exhibit such a behavior with the most central
collisions at full RHIC energy apparently becoming consistent with the
hydrodynamic model. This conclusion however depends on the model
calculations for the initial eccentricity and on the assumption that
the observed $v_2$ dominantly arises from an expansion phase where
anisotropic pressure gradients are the origin of the space-momentum
correlations. Different models for the eccentricity yield
$\varepsilon$ results that deviate both in their centrality dependence
and in their overall magnitude. Reasonable models for the eccentricity
can easily give magnitudes $30\%$ larger than those used in
Fig.~\ref{fig:v2ecc} with a stronger centrality dependence. The ratio
$v_2/\varepsilon$ can therefore be smaller than what is shown and have
a different shape\cite{Hirano:2005xf}. Given this level of
uncertainty, the conclusion that heavy-ion collisions at
$\sqrt{s_{_{NN}}}=200$~GeV approximately satisfy the assumptions made
in the hydrodynamic models \textit{i.e.} early local thermal
equilibrium and interactions near the zero mean-free path limit, would
be more convincing if an asymptotic approach to a limiting value were
observed. Rather, for the eccentricity calculation used in
Fig.~\ref{fig:v2ecc}, the data suggest a nearly linear rise with no
indication of asymptotic behavior.

\begin{figure}[hbt]
\begin{center}
\centerline{  \includegraphics[width=.7\textwidth]{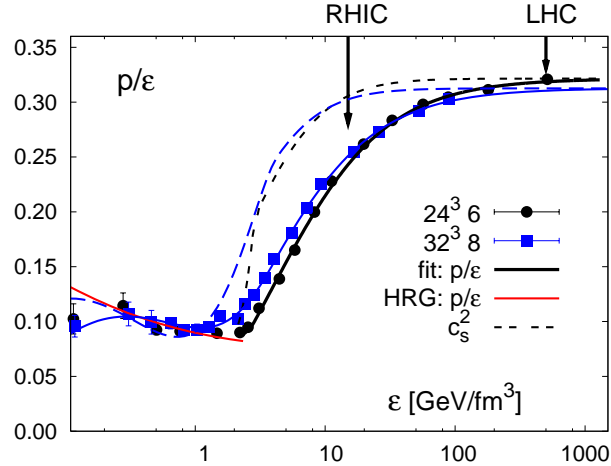}}
  \caption{ The QCD equation-of-state (pressure over energy density
    versus the fourth power of the energy density) as determined in
    Lattice calculations. }
  \label{fig:eos}
\end{center}
\end{figure}

In the hydrodynamic picture, one might also expect that
$v_{2}/\varepsilon$ versus $\frac{1}{S}\frac{dN}{dy}$ will be
sensitive to the equation-of-state of the matter formed during the
expansion phase. Since $v_2$ is expected to reflect space-momentum
correlation developed due to pressure gradients and
$\frac{1}{S}\frac{dN}{dy}$ is a measure of the transverse particle
density, $v_{2}/\varepsilon$ versus $\frac{1}{S}\frac{dN}{dy}$ could
be considered as a proxy for the pressure versus energy density or the
equation-of-state. It's difficult to identify in the data the features
that are expected in the equation-of-state. Fig.~\ref{fig:eos} shows
the equation-of-state calculated in a recent lattice QCD
calculation\cite{eos}. The onset of the QGP phase is seen to lead to an
increase in the pressure as the energy density is increased above a
critical value. The energy density in heavy-ion collisions is often
estimated from the Bjorken formula\cite{bjorken}:
\begin{equation}
  \epsilon_{Bj}=\frac{1}{A\tau}\frac{dE_{T}}{dy}
\end{equation}
which depends only on the experimentally accessible quantity
$\frac{dE_{T}}{dy}$, the overlap area of the nuclei and $\tau$, the
unknown formation time which is often assumed to be 1 fm. The Bjorken
estimate for the energy density is closely related to the transverse
particle density $(1/S)(dN/dy)$.

\subsection{Differential Elliptic Flow}

\begin{figure}[hbt]
\begin{center}
\centerline{  \includegraphics[width=.7\textwidth]{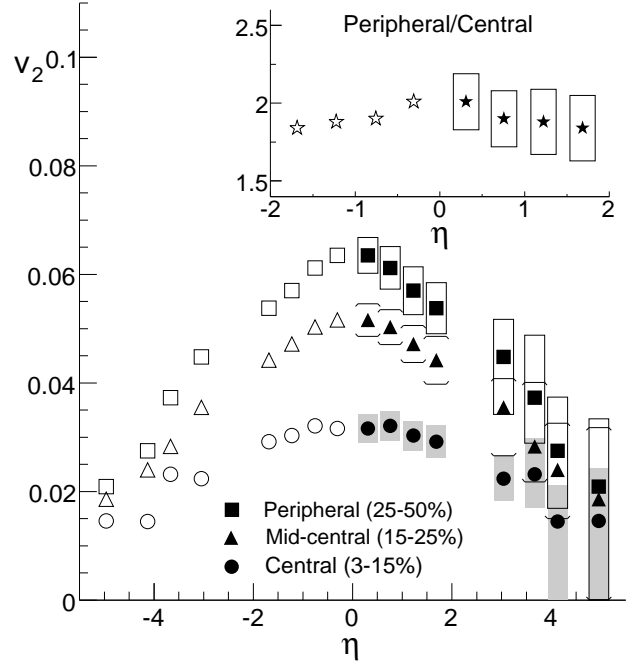}}
\caption{ $v_{2}(\eta)$ for 200 GeV Au+Au collisions for three
  centrality intervals. The inset shows the ratio of $v_{2}(\eta)$ for
  central collisions over peripheral collisions. Open symbols are data
  reflected to negative $\eta$.}
  \label{fig:v2eta}
\end{center}
\end{figure}

In addition to studying how $v_2$ integrated over all particles
depends on the centrality or $\sqrt{s_{_{NN}}}$ of the collision, one
can study how $v_2$ depends on the kinematics of the produced
particles (differential elliptic flow). Fig.~\ref{fig:v2eta} shows the
centrality and pseudo-rapidity dependence of $v_2$ for 200 GeV Au+Au
collisions\cite{Back:2004mh}. $v_2$ is largest at mid-rapidity where
the transverse particle density is largest and then falls off at
larger $|\eta|$ values. This behavior is therefore consistent with the
trends seen in integrated $v_2$ where $v_2/\varepsilon$ appears to
increase with increasing transverse particle density.

The fall off of $v_2(\eta)$ with increasing $|\eta|$ is common to the
three centrality intervals studied. The inset of the figure shows the
ratio of $v_2$ in peripheral over central collisions. Within errors
the ratio is flat indicating a similar shape for all centralities with
$v_2(\eta)$ only changing by a scale factor. Scaling of $v_2(\eta)$
for different energies and system sizes will be discussed in a later
section.

\begin{figure}[hbt]
\begin{center}
\centerline{  \includegraphics[width=.95\textwidth]{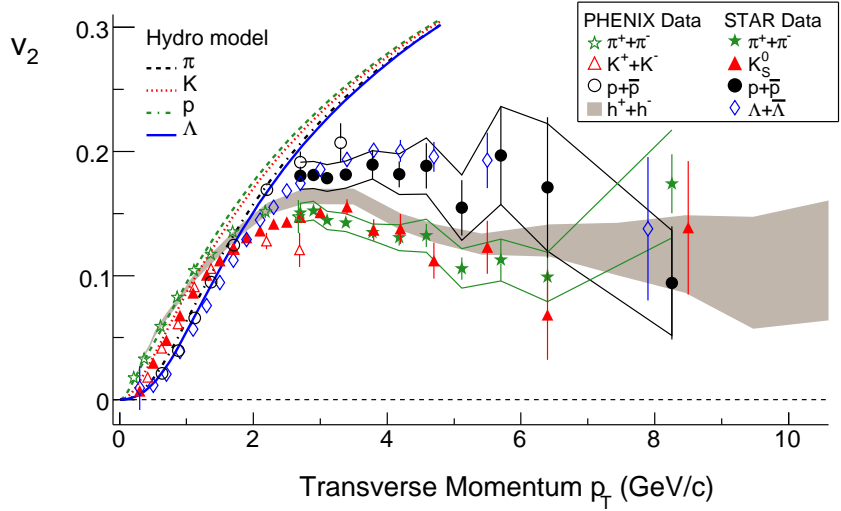}}
\caption{ $v_2(p_T)$ for a variety of identified particle species and
  inclusive charged hadrons. Data are for Au+Au collisions at 200 GeV
  averaged over all centralities. At low momentum $v_2(p_T)$ exhibits
  mass ordering while at larger $p_T$ the identified particle $v_2$
  appears to be grouped according to constituent quark number. The
  mass ordering at low $p_T$ is approximately reproduced by the
  hydrodynamic calculation.}
  \label{fig:v2pid}
\end{center}
\end{figure}

Fig.~\ref{fig:v2pid} shows $v_2$ for a variety of particle species as
a function of their transverse momentum
$p_T$\cite{Adler:2001nb,Adler:2002pb,Adams:2003am,Adler:2003kt,Abelev:2008ed,Abelev:2007qg}. In
the region below $p_T \sim 2$~GeV/c, $v_2$ follows mass ordering with
heavier particles having smaller $v_2$ at a given $p_T$. Above this
range, the mass ordering is broken and the heavier baryons take on
larger $v_2$ values.

\begin{figure}[hbt]
\begin{center}
\centerline{  \includegraphics[width=.7\textwidth]{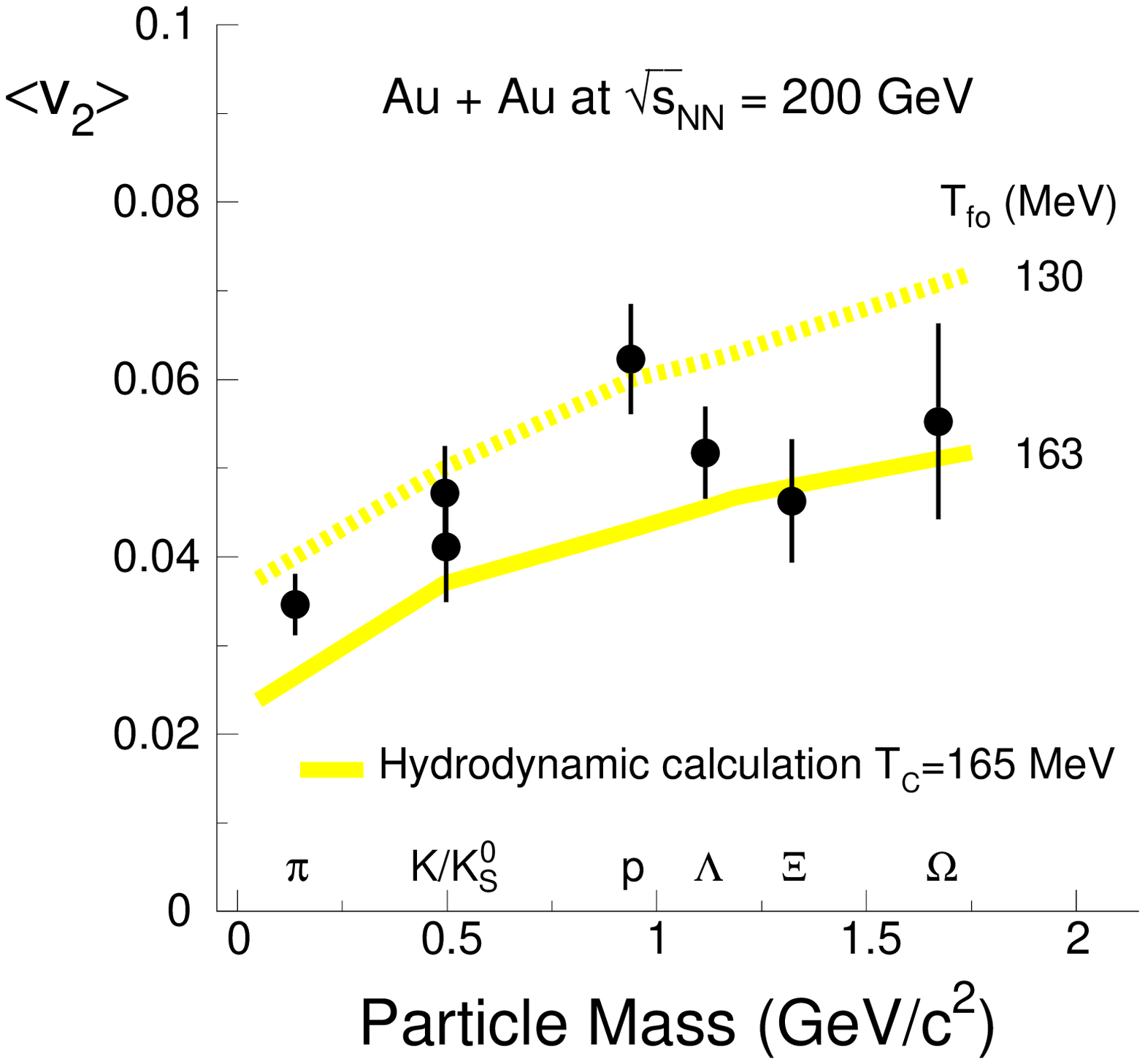}}
\caption{ $p_T$ integrated $v_2$ versus mass for a variety of identified
  particle species. Yellow bands show hydrodynamic calculations with
  different freeze-out temperatures.}
  \label{fig:v2mass}
\end{center}
\end{figure}

A hydrodynamic model for $v_2(p_T)$ is also shown which describes the
$v_2$ in the lower $p_T$ region well. This mass ordering is a feature
expected for particle emission from a boosted source. In the case that
particles move with a collective velocity, more massive particles will
receive a larger $p_T$ kick. As the particles are shifted to higher
$p_T$, the lower momentum regions become depopulated with a larger
reduction in the direction with the largest boost (in-plane). This
reduction reduces $v_2$ at a given $p_T$, with the reduction largest
for more massive particles. Note that this does not imply that the
more massive particles have a smaller integrated $v_2$ value, and in
fact the opposite is true. Fig.~\ref{fig:v2mass} shows $v_2$ for
identified particles integrated over all
$p_T$\cite{Sorensen:2003kp,Adams:2005zg}. The integration shows that
$v_2$ increases with particle mass. This is because the more massive
particles have a larger $\langle p_T\rangle$ and $v_2$ is generally
increasing with $p_T$ in the $p_T$ region where the bulk of the
particles are produced. The hydrodynamic model also exhibits this
trend.

\subsubsection{Identified Particle $v_2(p_T)$: RHIC versus SPS}

\begin{figure}[hbt]
\begin{center}
\centerline{  \includegraphics[width=.95\textwidth]{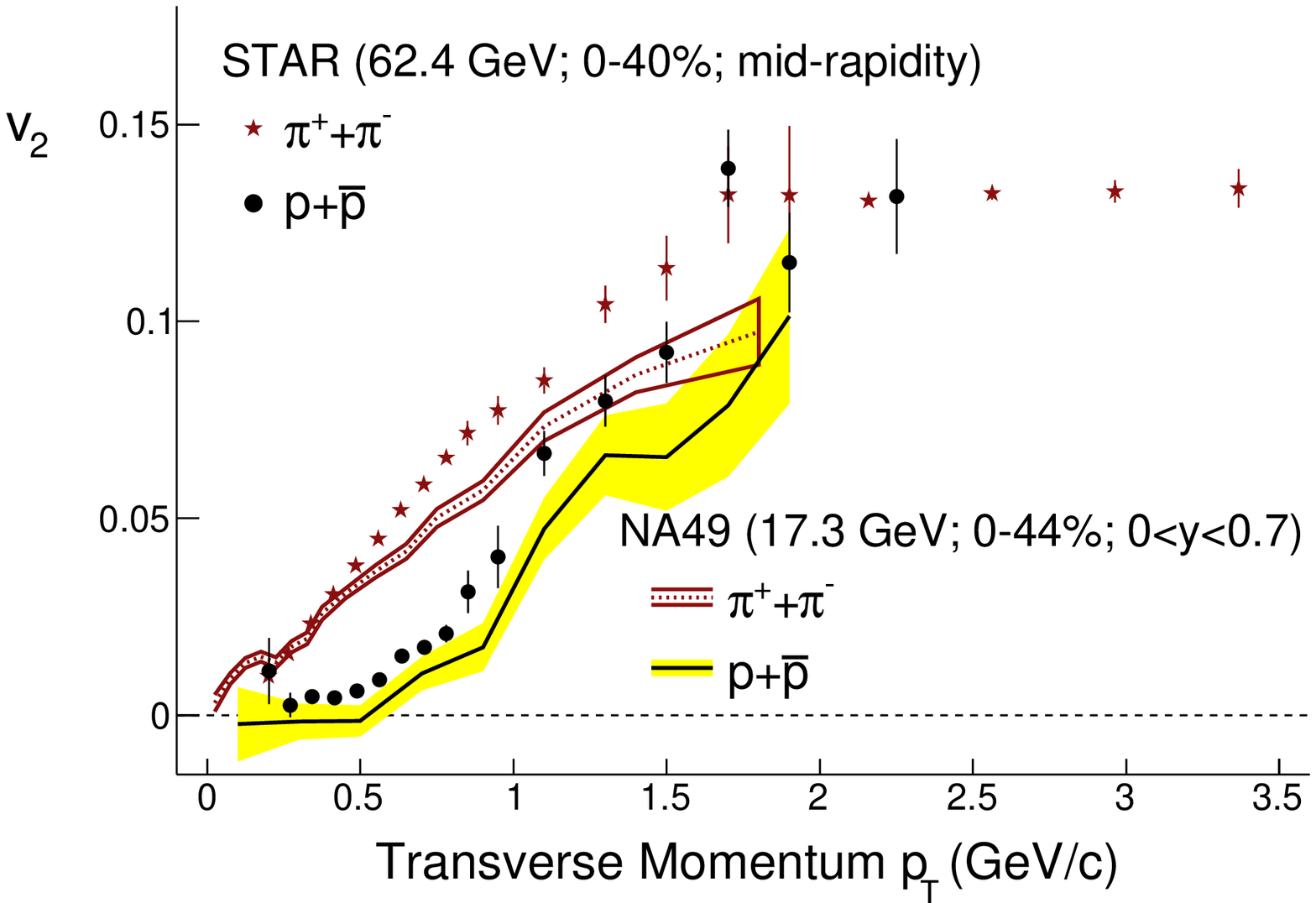}}
\caption{ $v_2(p_T)$ for pions and protons at $\sqrt{s_{_{NN}}}=$ 62.4
  and 17.3 GeV.  }
  \label{fig:v2spsrhic}
\end{center}
\end{figure}

Fig.~\ref{fig:v2spsrhic} shows pion and proton $v_2$ from
$\sqrt{s_{_{NN}}}=62.4$ Au+Au\cite{Abelev:2007qg} and 17.3~GeV Pb+Pb
collisions\cite{Alt:2003ab}. The centrality intervals have been
chosen similarly for the 17.3 GeV and 62.4 GeV data. The STAR data at
62.4 GeV are measured within the pseudo-rapidity interval $|\eta|<1.0$
and the 17.3 GeV data are from the rapidity interval $0<y<0.7$. These
intervals represent similar $y/y_{beam}$ intervals. It has been shown
that $v_2$ data for pions and kaons at 62.4 GeV are similar to 200 GeV
data; the 62.4 GeV data only tending to be about 5\% smaller than the
200 GeV data.

Appreciable differences are seen between the 17.3 GeV and 62.4 GeV
data. At $p_T>0.5$~GeV/c, for both pions and protons, the $v_2$ values
measured at 62.4 GeV are approximately 10\%--25\% larger than those
measured at 17.3~GeV. Although the magnitude of $v_2$ is different at
the lower energy, the systematics of the particle-type dependencies
are similar. In particular, pion $v_2$ and proton $v_2$ cross over
each other at $p_T$ near $1.7$~GeV/c for $\sqrt{s_{_{NN}}}=$ 17.3,
62.4 and 200 GeV data. Due to the limited kinematic range covered by
the 17.3 GeV data, it's not possible to determine if the $v_2$ of
baryons at $p_T>2$~GeV becomes larger than that for the lighter
mesons.

The increase in the magnitude of $v_2$ from 17.3 GeV to 62.4 GeV and
the similarity of 62.4 GeV $v_2$ to 200 GeV $v_2$ has been taken as a
possible indication for the onset of a limiting
behavior\cite{ph62v2}. In a collisional picture, a saturation of
$v_2$ could indicate that for $\sqrt{s_{_{NN}}}$ at and above 62.4~GeV
the number of collisions the system constituents experience in a given
time scale can be considered large and that hydrodynamic equations can
therefore be applied. Hydrodynamic model calculations of $v_2$ depend
on the model initialization and the poorly understood freeze-out
assumptions. As such, rather than comparing the predicted and measured
values at one energy, the most convincing way to demonstrate that a
hydrodynamic limit has been reached may be to observe the onset of
limiting behavior with $\sqrt{s_{_{NN}}}$. For this reason, $v_2$
measurements at a variety of center-of-mass energies are of
interest. Fig.~\ref{fig:v2spsrhic} shows that when the 17.3 and 62.4
GeV $v_2(p_T)$ data are compared within similar $|y|/y_{beam}$
intervals, the differences between $v_2(p_T)$ within the data sets may
be as small as 10\%--15\%. As such, a large fraction of the deviation
between the SPS data and hydrodynamic models arises due to the wide
rapidity range covered by those measurements ($v_2$ approaches zero as
beam rapidity is approached\cite{Back:2004mh}), increased $\langle
p_T \rangle$ values at RHIC and the larger $v_2$ values predicted for
the lower colliding energy by hydrodynamic models.

\subsection{High $p_T$}

\begin{figure}[hbt]
\begin{center}
\centerline{  \includegraphics[width=0.7\textwidth]{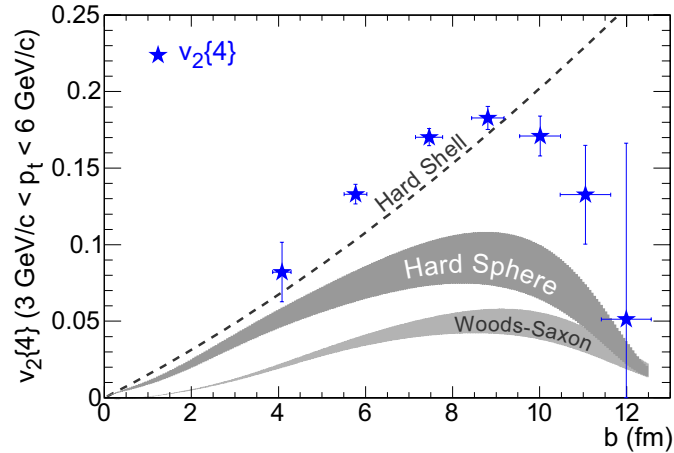}}
\caption{ The four-particle cumulant $v_2$ at $3<p_T<6$~GeV/c versus
  impact parameter, b, compared to models of $v_2$ based on geometry
  alone. Data in this region does not agree well with these surface
  emission pictures.  }
  \label{fig:highpt}
\end{center}
\end{figure}

At higher $p_T$, $v_2$ no longer rises with $p_T$ and the mass
ordering is broken. Above $p_T\sim 2$~GeV/c the more massive baryons
exhibit a larger $v_2$ than the mesons. While the pion and kaon $v_2$
reach a similar maximum of $v_2 \approx 0.14$ at $p_T \approx
2.5~GeV/c$, the baryon $v_2$ continues to rise until it reaches a
maximum of $v_2 \approx 0.20$ at $p_T \approx 4.0$~GeV/c. For still
larger $p_T$, the $v_2$ values exhibit a gradual decline until $v_2$
for all particles is consistent with $v_2 \approx 0.10$ at $p_T
\approx 7$~GeV/c. Fine detail cannot yet be discerned at $p_T>7$ due
to statistical and systematic uncertainties. At these higher $p_T$
values one expects that the dominant process giving rise to $v_2$ is
jet-quenching\cite{Wang:1991xy} where hadron suppression is larger
along the long axis of the overlap region than along the short
axis\cite{Wang:2000fq,Gyulassy:2001kr,Gyulassy:2000gk}. For very
large energy loss, the value of $v_2$ should be dominated by the
geometry of the collision region. Fig.~\ref{fig:highpt} shows a
comparison of $v_2$ data\cite{Adams:2004wz} for $3<p_T<6$~GeV/c
compared to several geometric models\cite{Drees:2003zh}. This
comparison seems to indicate that $v_2$ in this intermediate $p_T$
range is still too large to be related exclusively to quenching.

\begin{figure}[hbt]
\begin{center}
\centerline{  \includegraphics[width=0.9\textwidth]{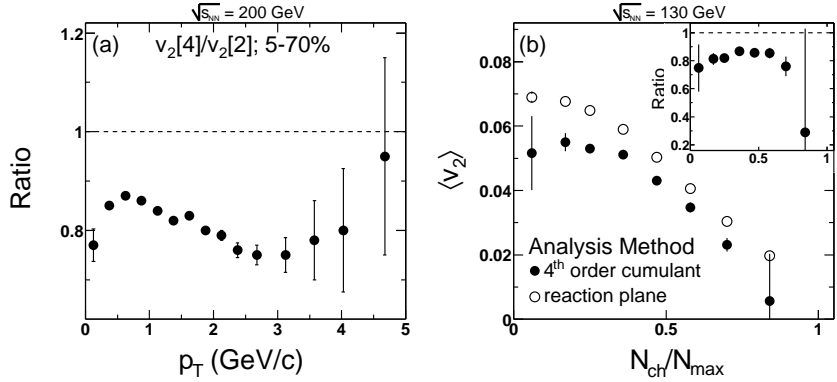}}
\caption{ The left panel shows the ratio $v_2\{4\}/v_2\{2\}$ for the
  0\%-70\% centrality interval in 200 GeV Au+Au collisions. The ratio
  falling below unity indicates the importance of non-flow and $v_2$
  fluctuations. The larger reduction at high $p_T$ seems indicative of
  an increase in non-flow due to jets. The right panel shows the
  centrality dependence of the $v_2\{4\}$ and $v_2$ measured with
  respect to the event plane in 130 GeV Au+Au collisions. The inset
  shows the ratio of the two. }
  \label{fig:nonflow}
\end{center}
\end{figure}

In the higher $p_T$ regions, significant azimuthal structure will
arise from jets, so non-flow correlations are thought to be
significant in this region\cite{Adler:2002ct}. These effects have
been studied in several ways. The four-particle cumulant $v_2$ has
been studied as a function of $p_T$ and the ratio of the four- and
two-particle cumulants $v_2\{4\}/v_2\{2\}$ is found to decrease with
increasing $p_T$\cite{Borghini:2001zr,Adler:2002pu}. This decrease
is identified with a gradual increase in the contribution of jets to
$v_2\{2\}$ (see Fig.~\ref{fig:nonflow} left panel). The four-particle
cumulant suppresses contributions due to intra-jet correlations but
the statistical errors of the measurement are larger. One can also
suppress jet structure in the $v_2$ measurement by implementing a
$\Delta\eta$ cut in the pairs of particles being used in the
analysis\cite{Adams:2004bi}. In this case, a high $p_T$ particle is
correlated with other particles in the event that are separated by a
minimum $\Delta\eta$. This method relies on the assumption that jet
correlations do not extend beyond a given $\Delta\eta$
range. Interactions of jets with the medium in nuclear collisions
however can change the structure of jets and extend the correlations
in $\Delta\eta$ beyond the widths observed in $p+p$
collisions\cite{Putschke:2007mi}. This method therefore is not guaranteed to
eliminate non-flow from jets. The problem of measuring $v_2$ without
non-flow and of measuring modifications of jet structure by the medium
are entirely coupled. If one is known, the other is trivial.

Other methods for suppressing non-flow include measuring correlations
between particles at mid-rapidity and and an event-plane determined
from particles observed at forward rapidity\cite{Voloshin:2007af}. In
the extreme and the most effective case, the event-plane was
reconstructed from spectator neutrons in a Zero-Degree Calorimeter to
measure $v_2$ of produced particles near $\eta=0$. An extension of
analyses based on the change in correlations across various rapidity
intervals is the analysis of the two dimensional correlation landscape
for two-particle correlations \textit{e.g.}
$d^2N/\Delta\phi\Delta\eta$\cite{Trainor:2007fu}. After unfolding the
two particle correlations one can attempt to identify various
structures with known physics such as jets, resonance decay, or HBT
based on their width in $\eta$ and $\phi$. The remaining
$\cos(2\Delta\phi)$ structure can then be used to estimate $\langle
v_2\rangle^2 + \sigma_{v_2}^2$. This method will be discussed below.

\begin{figure}[hbt]
  \begin{center}
    \centerline{  \includegraphics[width=0.9\textwidth]{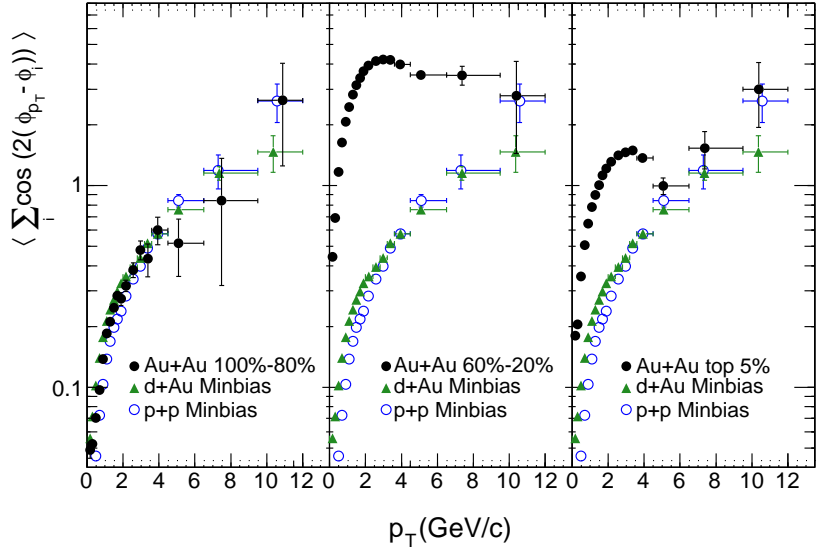}}
    \caption{ Second harmonic azimuthal correlations in $p+p$, $d+$Au, and
      Au+Au collisions. The quantity
      $\langle\Sigma_{i}\cos(n(\phi_{p_{T}}-\Phi_{i}))\rangle = M\langle
      v_{2}v_{2}(p_{T})\rangle + M\delta_{2}$ facilitates comparisons
      between different systems particularly where a reaction-plane may
      not be well defined. }
    \label{fig:scalar}
  \end{center}
\end{figure}

As we study progressively smaller systems the connection between the
nucleus-nucleus reaction-plane and the azimuthal structure breaks
down. In the limit that one proton from each nucleus participates in
the interaction, the reaction-plane defined by the colliding protons
will not necessarily be related to the reaction-plane defined by the
vector connecting the centers of the colliding nuclei. In order to
facilitate a comparison between the $p_T$ dependence of azimuthal
correlations in large systems and small systems, the scalar product
$\langle uQ^*\rangle$ is used where $u=e^{i2\phi}$ and
$Q^*=e^{-i2\Psi}$\cite{Adams:2004wz}. The mean of $uQ^*$ therefore
yields a quantity that depends on $\langle v_2v_2(p_T)\rangle$ and
non-flow as follows:
\begin{equation}
  \langle uQ^*\rangle =  \langle\Sigma_{i}\cos(n(\phi_{p_{T}}-\Phi_{i}))\rangle = M\langle v_{2}v_{2}(p_{T})\rangle + M\delta_{2},
\end{equation}
where $M$ is the multiplicity used in the sum. Fig.~\ref{fig:scalar}
shows this quantity for $p+p$, $d+$Au, and three Au+Au centrality
intervals. The $p+p$ and $d+$Au data are repeated in each of the
panels. The most peripheral Au+Au collisions are shown in the left
panel. The centrality bin shown is not usually presented since trigger
inefficiencies for low multiplicity events makes it difficult to
define the actual centrality range sampled. In this case, the data has
been published in order to compare $uQ^*$ between the most peripheral
sample of events and $p+p$ collisions. The data in Au+Au has a similar
shape and magnitude as the data in $p+p$. This suggests that
peripheral collisions are dominated by the same azimuthal structure as
$p+p$ collisions; an observation consistent with two-particle
$\Delta\eta$, $\Delta\phi$ correlations\cite{Adams:2004pa}. The data
from mid-central Au+Au collisions shown in the middle panel however,
exhibit a magnitude and shape clearly different than $p+p$
collisions. While $uQ^*$ for $p+p$, $d+$Au and very peripheral Au+Au
collisions rises monotonically with $p_T$, for mid central Au+Au
collisions, the data rises to a maximum at $p_T=3$~GeV/c and then
falls.  For central collisions shown in the right panel, a similar
feature is seen with data rising to a maximum at $p_T=3$~GeV/c, then
falling until $p_T=6$~GeV/c, where it begins rising again. This second
rise is presumably a manifestation of non-flow at high $p_T$ in
central collisions. These data suggest that azimuthal structure in
Au+Au collisions above $p_T=6$~GeV/c is dominated by jets. This is
also consistent with the conclusions reached by examining the particle
type dependence of $v_2$ and $R_{CP}$\cite{Adams:2003am}.

\begin{figure}[hbt]
\begin{center}
\centerline{  \includegraphics[width=0.9\textwidth]{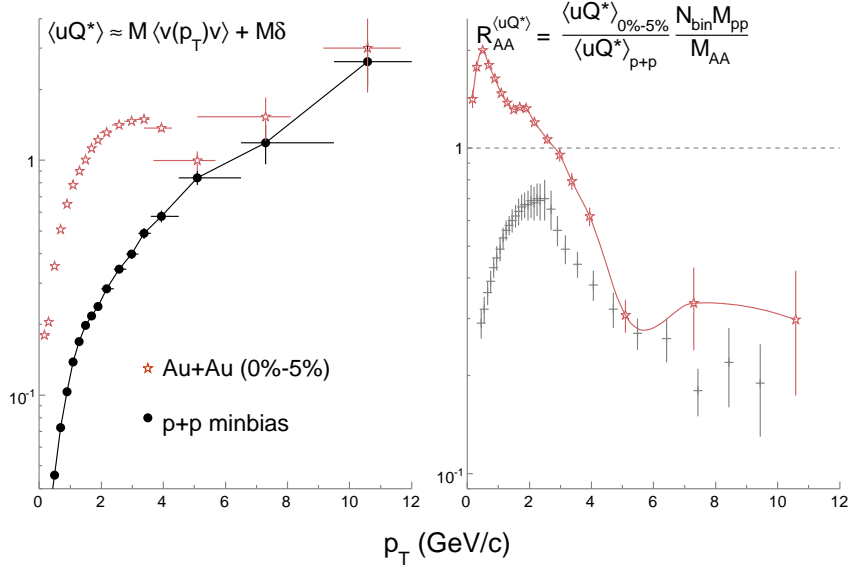}}
\caption{ The left panel reproduces the data in
  Fig.~\ref{fig:scalar}. The right panel shows $R_{AA}^{\langle
    uQ*\rangle}$. Binary scaling of $\delta_{2}$ from $p+p$ collisions
  corresponds to $R_{AA}^{\langle uQ*\rangle}=1$. In comparison to
  binary scaling, non-flow above $p_T \sim 5$~GeV/c in Au+Au
  collisions is suppressed by a factor of 5. Also shown for comparison
  is $R_{AA}$ for single particle spectra of charged hadrons.}
  \label{fig:raav2}
\end{center}
\end{figure}

In Fig.~\ref{fig:scalar} the $p+p$ data is replotted in each panel to
facilitate a comparison between the shape and magnitude in $p+p$ to
that in Au+Au. In the absence of jet-quenching however, non-flow at
high $p_T$ is expected to scale with the number binary nucleon-nucleon
collisions $N_{\mathrm{binary}}$. The plotting format in Fig.~\ref{fig:scalar}
on the other hand, assumes that $\delta_2 \propto 1/M$ rather than
$N_{bin}/M^2$ as would be expected for hard scattering. The
multiplicity has been shown to scale as $(1-x_{hard})N_{\mathrm{part}} +
x_{hard}N_{\mathrm{binary}}$ with $x_{hard}\approx
0.11$\cite{Kharzeev:2000ph,Back:2001xy}. This is referred to as the
two-component model. In order to compare azimuthal structure in Au+Au
collisions to $N_{\mathrm{binary}}$ scaling of $p+p$ collisions we can form a
ratio in analogy with $R_{AA}$ for single hadrons:
\begin{equation}
R_{AA}^{uQ^*}=\frac{\langle uQ^*\rangle_{AA}/M_{AA}}{N_{\mathrm{binary}}\langle uQ^*\rangle_{pp}/M_{pp}}
\end{equation}
where $M_{AA}$ and $M_{pp}$ are the multiplicities in $A+A$ and $p+p$
collisions with $M_{AA}$ taken according to the two component
model. In the case that jet production in Au+Au collisions scales
with the number of binary collisions, as hard processes are expected
to, $R_{AA}^{uQ^*}$ should be unity. The right panel of
Fig.~\ref{fig:raav2} shows $R_{AA}^{uQ^*}$ for charged hadrons in
$0\%-5\%$ central Au+Au collisions. For comparison, $R_{AA}$ from
single particle charged hadron spectra is also shown in the figure.
$R_{AA}^{uQ^*}$ first rises abruptly with $p_T$ to a maximum of 2 at
$p_T \approx 0.5$~GeV/c and then falls to a value of 0.25 at $p_T
\approx 5$~GeV/c. At $p_T>5$~GeV/c $R_{AA}^{uQ^*}$ is similar to
$R_{AA}$. This shows that jet-quenching suppresses the charged hadron
spectra, and the azimuthal structure by a similar amount; confirming
that the single hadron suppression is indeed related to
jet-quenching. $R_{AA}^{uQ^*}$ is complimentary to studies of
$I_{AA}$, the ratio of dihadron correlations in Au+Au and $p+p$
collisions\cite{Adler:2002tq}.

We note the presence of what appears to be a local minimum and local
maximum $p_T \approx 1.5$ and 2.0~GeV/c respectively. It is not clear
if this is a real feature or simply an artifact largely caused by the
shape of the $p+p$ data. In the case that it is a real feature, it is
possibly related to the changing particle composition in Au+Au
collisions where baryons with larger $v_2$ values become more
prominent. At $p_T=3$~GeV/c, baryons and mesons in $p+p$ collisions are
created in the proportion 1:3 while at the same $p_T$ in central Au+Au
collisions the proportion is approximately 1:1. $R_{AA}^{uQ^*}$ will
be an interesting quantity to investigate for identified
particles. One can anticipate a quark number dependence at
intermediate $p_T$ as seen in $R_{CP}$ and $v_2$.

\subsection{Multiply strange hadrons and heavy flavor}

The build-up of space-momentum correlations throughout the collision
evolution is cumulative. Information about space-momentum correlations
developed during a Quark-Gluon-Plasma phase can be masked by
interactions during a later hadronic phase. For studying a QGP phase,
it is useful to use a probe that is less sensitive to the hadronic
phase. Multi-strange hadrons have hadronic cross-sections smaller than
the equivalent non-strange hadrons, and the $v_2$ values measured for
hadrons such as $\phi$-mesons ($s\overline{s}$) and $\Omega$-baryons
($sss$) are therefore thought to be more sensitive to a
quark-gluon-plasma phase than to a hadronic
phase\cite{Teaney:2001gc}.

\begin{figure}[hbt]
\begin{center}
\centerline{  \includegraphics[width=.8\textwidth]{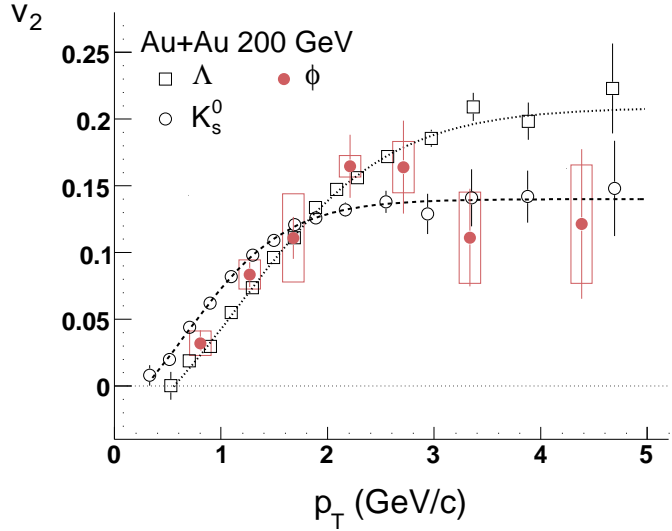}}
\caption{ $v_2(p_T)$ for the $\phi$ meson. The $\phi$ is composed of
  an $s\overline{s}$ pair and is expected to have a smaller hadronic
  cross-section than non-strange, or singly-strange hadrons. The
  $\phi$ $v_2$ is compared to $K_S^0$ and $\Lambda$ $v_2$.}
  \label{fig:v2phi}
\end{center}
\end{figure}

Fig.~\ref{fig:v2phi} shows $v_2(p_T)$ for the
$\phi$-meson\cite{Abelev:2007rw,Afanasiev:2007tv}. The $v_2$ rises
with $p_T$ and reaches a maximum of approximately $15\%$ at $p_T$ near
2 GeV/c. At intermediate $p_T$, the $\phi$-meson $v_2$ appears to
follow a trend similar to the other meson $K_S^0$. This observation
suggests that either the $\phi$-meson cross section is larger than
anticipated or re-scattering during the hadronic phase does not
contribute significantly to $v_2$. The latter possibility requires
that $v_2$ is established prior to a hadronic phase, suggestive of
development of $v_2$ during a QGP phase.

\begin{figure}[hbt]
\begin{center}
\centerline{  \includegraphics[width=.8\textwidth]{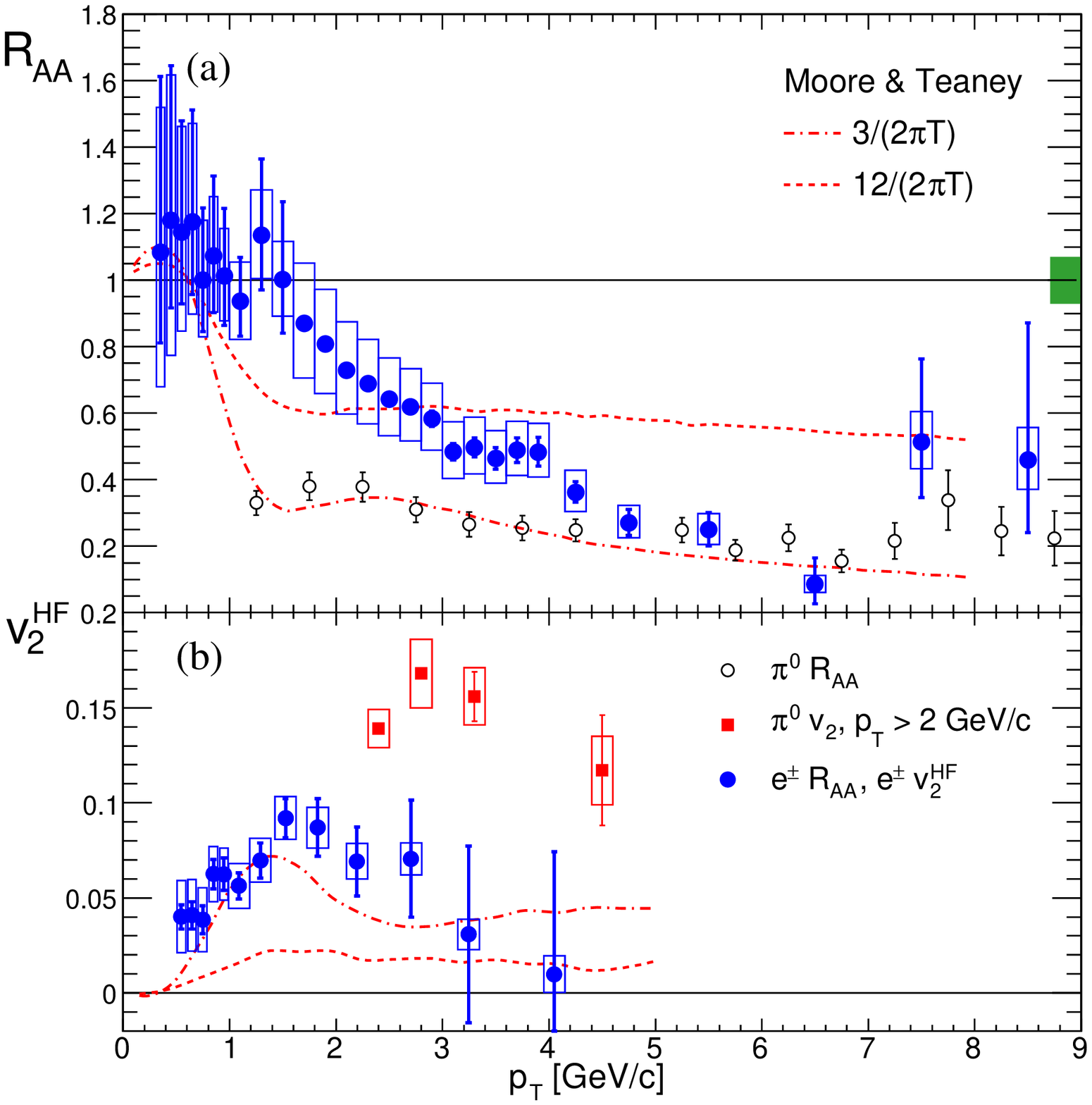}}
\caption{ $R_{AA}$ and $v_2$ for non-photonic electrons. The two
  measurements can be used in conjunction to provide constraints on
  models. Data are compared to a model with the conjectured minimum
  viscosity-to-entropy ratio $\eta/s$ and one with $\eta/s$ four times
  that value.}
  \label{fig:v2npe}
\end{center}
\end{figure}

Measurements of multiply strange hadrons are interesting because they
should be less coupled to the matter in a hadronic phase and therefore
a better reflection of the QGP phase. Heavy quarks on the other hand
(\textit{e.g.} charm and bottom quarks) may be less coupled to even
the QGP matter\cite{Dokshitzer:2001zm,Djordjevic:2003zk}. It's not a
priori obvious that heavy quarks will couple significantly to the
medium and be influenced by its apparent expansion. The extent to
which they do couple to the medium should be reflected in how large
$v_2$ for heavy flavor hadrons becomes and how much the nuclear
modification ($R_{AA}$) deviates from unity. Precision measurements of
Heavy Flavor mesons or baryons are not yet available from the RHIC
experiments. As a proxy for identifying D-mesons, the STAR and PHENIX
experiments have measured non-photonic
electrons\cite{Adler:2005xv,Abelev:2006db}. Non-photonic electrons
are generated from the weak-decays of heavy flavor hadrons and after
various backgrounds have been accounted for can, with some
caveats\cite{Sorensen:2005sm}, be used to infer the $R_{AA}$ and
$v_2$ of D-mesons.

The top panel (a) of Fig.~\ref{fig:v2npe} shows $R_{AA}$ for
non-photonic electrons\cite{Adler:2005ab,Adare:2006nq}. Prior to the
measurement of non-photonic electron $R_{AA}$, it was expected that
heavy-flavor hadrons would be significantly less suppressed than light
flavor hadrons. These expectations based on a decrease in the coupling
of charm quarks to the medium because of the dead-cone
effect\cite{Dokshitzer:2001zm}, are contradicted by the data; At $p_T
\approx 5$~GeV/c, non-photonic electrons are as suppressed as
pions. This suppression suggests a stronger than expected coupling of
charm quarks to the medium. This coupling apparently also leads to
significant $v_2$ as seen in Fig.~\ref{fig:v2npe} (b).

Also shown in the figure is a calculation of $v_2$ and $R_{AA}$ based
on a Langevin model\cite{Moore:2004tg}. In that model, the strength of
the energy loss and momentum diffusion of charm quarks is
characterized in terms of a diffusion coefficient (D). $R_{AA}$ and
$v_2$ for charm quarks is then computed for several values of D. Two
of these values are shown in Fig.~\ref{fig:v2npe}. Although neither
curve provides an entirely satisfactory simultaneous description of
$v_2$ and $R_{AA}$, the comparison suggests that the diffusion
coefficient is large. This comparison only achieves rough agreement,
but the calculation illustrates the sensitivity of heavy flavor
hadrons to transport coefficients of the QGP and $v_2$ is an important
quantity to measure for these hadrons. This is also in agreement with
H.~van~Hees, {\it et al.}\cite{vanHees:2005wb} where coalescence at
the hadronization phase boundary is also considered and found to help
improve the agreement with data.

\subsection{Fluctuations and Correlations}

Comparisons between data and models are complicated by uncertainties
in the initial eccentricity and by uncertainties in the
data. Estimating transport quantities from the data may require a
precision comparison between eccentricity and $v_2$ so it is important
to reduce the uncertainties in both. As discussed in the previous
sections, a CGC model of the initial conditions yields eccentricity
values typically $30\%$ larger than a Gluaber model while the
fluctuations ($\sigma_{\varepsilon}$) are still of the same width. The
ratio of $\sigma_{\varepsilon}/\varepsilon$ in a CGC model is
therefore smaller than in a Glauber
model\cite{Lappi:2006xc,Drescher:2006pi}. One can expect the
statistical fluctuations in eccentricity to show up as dynamical
fluctuations in $v_2$ measurements. Measuring the dynamic $v_2$
fluctuations in conjunction with $\langle v_2\rangle$ can therefore
provide an additional constraint on the initial
conditions\cite{Sorensen:2006nw,Alver:2007qw,Sorensen:2008zk,Alver:2008zza}.

Several methods have been employed for measuring $v_2$ and the various
methods have different dependencies on non-flow correlations and $v_2$
fluctuations\cite{Miller:2003kd,Borghini:2001vi,Bhalerao:2003xf,Borghini:2004ke,Voloshin:2007pc,Bhalerao:2006tp,Zhu:2005qa,Bilandzic:2008nx}. The
differences between these measurements give information on non-flow
correlations and $v_2$ fluctuations. If one uses a two particle
correlation to estimate $v_2$, then one finds $v_{2}\{2\}^2 =
\langle\cos(2(\phi_i-\phi_j))\rangle = \langle v_2\rangle^2 +
\sigma_{v_{2}}^2 + \delta_2$ where the average is over all unique
pairs of particles. $v_2$ is the single particle anisotropy with
respect to the reaction plane $v_2=\langle\cos(2(\phi-\Psi))\rangle$
and $\delta_2$ is the non-flow parameter which summarizes the
contributions to $\langle\cos(2(\phi_i-\phi_j))\rangle$ from
correlations not related to the reaction plane. If one uses a
4-particle cumulant $v_2\{4\}$ calculation, then for most cases the
non-flow term will be suppressed by large combinatorial factors and
$v_2$ fluctuations will contribute with the opposite sign. For
Gaussian fluctuations, $v_{2}\{4\}^2 \approx \langle v_2\rangle^2 -
\sigma_{v_{2}}^2$\cite{Voloshin:2007pc}. Without knowing $\delta$ or
$\sigma_{v_{2}}$, one cannot determine the exact value of $\langle
v_{2}\rangle$. Rather, $\langle v_2\rangle^2$ could lie anywhere
between $v_2\{4\}^2$ and $(v_2\{2\}^2+v_2\{4\}^2)/2$.

\begin{figure}[hbt]
\begin{center}
\centerline{  \includegraphics[width=.8\textwidth]{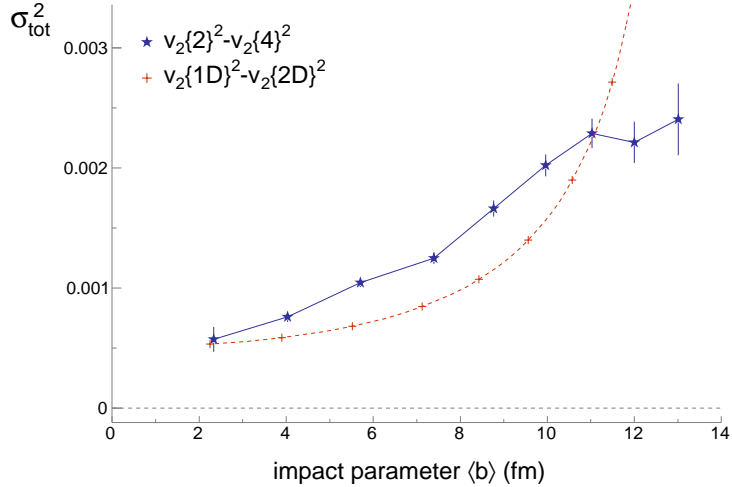}}
\caption{ The quantity $\sigma_{\mathrm{tot}}^2 = \delta_2 + 2\sigma_{v_{2}}^2$
  estimated from the from the difference $v_{2}\{2\}^2 -
  v_{2}\{4\}^2$. Also shown is a parametrization of
  $v_{2}\{1D\}^2-v_{2}\{2D\}^2$; the difference between
  $\langle\cos(2\Delta\phi)\rangle$ extracted for all correlations and
  that extracted by first fitting other structures in the correlations
  identified with various non-flow sources.}
  \label{fig:sigtot}
\end{center}
\end{figure}

It is advantageous to confront various models with the data that is
experimentally accessible. The difference between the two- and
four-particle cumulants in the case of Gaussian $v_2$ fluctuations is:
\begin{equation}
  v_2\{2\}^2 - v_2\{4\}^2 \approx \delta_2 + 2\sigma_{v_{2}}^2.
\end{equation}
The term $\delta+2\sigma_{v_{2}}^2$ is also approximately equivalent
to the non-statistical width of the distribution of the length of the
flow vector distribution $(dN/d|q_2|)$ and is called
$\sigma_{\mathrm{tot}}^2$. The flow vector for the $n^{th}$ harmonic is defined
as $q_{n,x} = \Sigma_i\cos(n\phi_i)$ and $q_{n,y} =
\Sigma_i\sin(n\phi_i)$. Fig.~\ref{fig:sigtot} shows $\sigma_{\mathrm{tot}}^2$
extracted from the difference between the two- and four-particle
cumulants\cite{Adams:2004bi}.
%\begin{equation}
%  v_2\{2\}^2 + v_2\{4\}^2 \approx 2\langle v_2\rangle^2 + \delta_2.
%\end{equation}
In the case of Gaussian fluctuations, higher cumulants such as
$v_{2}\{6\}$ are equal to $v_2\{4\}$. In this case, the quantity
$\sigma_{\mathrm{tot}}^2$ and $v_{2}\{2\}^2$ summarizes the
information available experimentally from the second harmonic flow
vector distribution. No more information can be accessed without
applying more differential techniques or by making assumptions about
the shape or centrality dependence of flow, non-flow, or flow
fluctuations. An example of a more differential analysis is also shown
in Fig.~\ref{fig:sigtot}, where two particle correlations have been
fit in $\Delta\phi$-$\Delta\eta$ space\cite{Trainor:2007ny}. Terms
identified with various non-flow sources have been included with the
fit and the remaining $\cos(2\Delta\phi)$ modulation is then
identified as $v_2\{2D\}^2$. In the case that the sources of non-flow
are correctly parametrized, $\delta = v_2\{1D\}^2 - v_2\{2D\}^2$,
where $v_2\{1D\}^2$ is $\langle\cos(2\Delta\phi)\rangle$ integrated
over all azimuthal structure. Then $v_2\{2D\}^2 = \langle v_2\rangle^2
+ \sigma_{v_2}^2$. This procedure is discussed below.

\begin{figure}[hbt]
\begin{center}
\centerline{  \includegraphics[width=.8\textwidth]{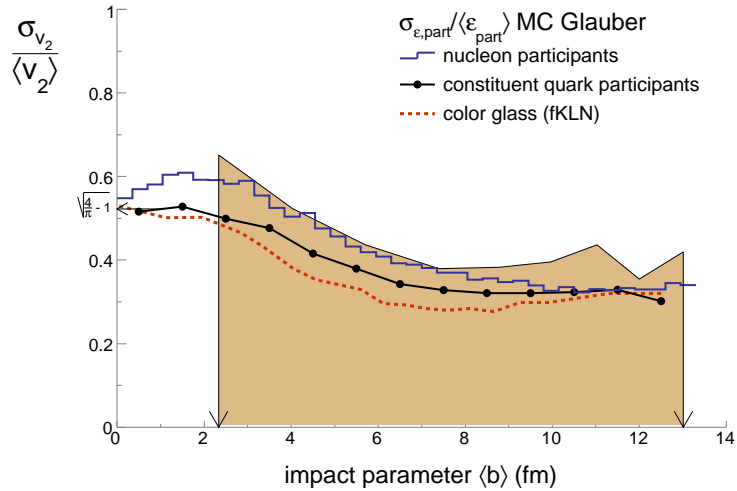}}
\caption{ An upper limit on the ratio $\sigma_{v_{2}}/\langle
  v_{2}\rangle$ is shown. The upper limit can be established from
  $\sigma_{\mathrm{tot}}$ and $v_{2}\{4\}$. The upper limit is
  compared to various models of eccentricity fluctuations. The
  existence of non-flow correlations implies that the true value of
  $\sigma_{v_{2}}/\langle v_2\rangle$ may be significantly below the
  upper limit, potentially challenging the Monte-Carlo Glauber
  calculation of the eccentricity.}
  \label{fig:fluct}
\end{center}
\end{figure}

Even without attempting to disentangle flow fluctuations from non-flow
correlations, the assumption that non-flow is a positive quantity
(consistent with $v_2\{1D\}^2 - v_2\{2D\}^2$) can be used with
$\sigma_{\mathrm{tot}}^2$ to provide an upper limit on $v_2$ fluctuations
\begin{equation}
  \sigma_{v_2}^2 < \frac{\sigma_{\mathrm{tot}}^2}{2}.
\end{equation}
To facilitate a comparison between this limit and models of the
initial eccentricity the upper limit on $\frac{\sigma_{v_{2}}}{\langle
  v_2\rangle}$ is compared to
$\frac{\sigma_{\varepsilon}}{\langle\varepsilon\rangle}$ in the
model. To form this ratio appropriately, the same assumptions should
be made for $\sigma_{v_{2}}$ and $\langle v_2\rangle$ \textit{i.e.}
zero non-flow.
\begin{equation}
\frac{\sigma_{v_2}^2}{\langle v_{2}\rangle^2} < \frac{\sigma_{\mathrm{tot}}^2/2}{(v_{2}\{2\}^2 + v_2\{4\}^2)/2} \\
= \frac{v_{2}\{2\}^2 - v_2\{4\}^2}{v_{2}\{2\}^2 + v_2\{4\}^2}
\end{equation}
This upper limit is shown in Fig.~\ref{fig:fluct} and compared to
several models of $\sigma_{\varepsilon}/\varepsilon$. The models
include two Glauber Monte Carlo models\cite{Broniowski:2007ft}; one
using the coordinates of participating nucleons to calculate the
eccentricity, the other using the coordinates of constituent quarks
confined inside the nucleons. The constituent quark Monte Carlo
Glauber Model (cqMCG)\cite{Sorensen:2008zk} treats the nucleus as
$3\times A$ constituent quarks grouped in clusters of three confined
to the size of a hadron. This increases the number of participants by
roughly a factor of three, reducing the fluctuations in
eccentricity. The correlations between the constituent quarks required
by confinement partially counteract this effect since those
correlations act to broaden the eccentricity distribution. The net
effect, however, is a narrowing of the distribution. Also shown is a
Color Glass Condensate (CGC) based model which yields eccentricity
values 30\% larger than the Glauber models leading to a reduction of
$\sigma_{\varepsilon}/\varepsilon$. The Monte Carlo Glauber model
based on the eccentricity of nucleons already exhausts most of the
width $\sigma_{\mathrm{tot}}^2=\delta_2+2\sigma_{v_{2}}^2$. This shows
that the statistical width of the eccentricity fluctuations in the
Glauber model already accounts for almost all of the non-statistical
width of the flow vector distribution thus leaving little room for
other sources of fluctuations and correlations. This is particularly
challenging since non-flow has been neglected in setting the upper
limit and the only $v_2$ fluctuations considered are those arising
from eccentricity fluctuations. We have therefore neglected
fluctuations that would arise during the expansion
phase\cite{Mrowczynski:2002bw,Vogel:2007yq,Trainor:2007ny}. One can
write the total width including these terms:
\begin{equation}
  \sigma_{\mathrm{tot}}^2 = \delta_2 + 2(v_2\frac{\sigma_{\varepsilon}}{\varepsilon})^2 + 2\sigma_{v_{2},dyn}^2,
\end{equation}
where the middle term in the right-hand-side is the $v_2$ fluctuations
from eccentricity fluctuations and the final term is $v_2$
fluctuations from the expansion phase. The middle term arises from the
approximation that to first order $\sigma_{v_{2}}/v_2 =
\sigma_{\varepsilon}/\varepsilon$. This approximation is prevalent in
the literature. The last term can be related to the Knudsen number of
the matter during the expansion\cite{Vogel:2007yq}. Measurements
demonstrating the existence of non-flow or dynamic $v_2$ fluctuations
($\sigma_{v_{2},dyn}$) therefore would challenge the model. The CGC or
cqMCG models provide a more likely description since
$\sigma_{\varepsilon}/\varepsilon$ is smaller in those models than
the upper limit on $\sigma_{v_2}/\langle v_2\rangle$. The upper limit
on $\sigma_{v_2}$ provided by $\sigma_{\mathrm{tot}}^2$ provides a
valuable test, therefore, for models of the initial eccentricity and
can help to reduce the uncertainty on $\varepsilon$; an essential
component in extracting meaning from the value of $v_2$.

\subsubsection{Two Dimensional Correlations and $v_2\{2D\}$}

\begin{figure}[ht]
\vspace{0.5cm}
\begin{minipage}[b]{0.49\linewidth}
\centering
\includegraphics[width=1.\textwidth]{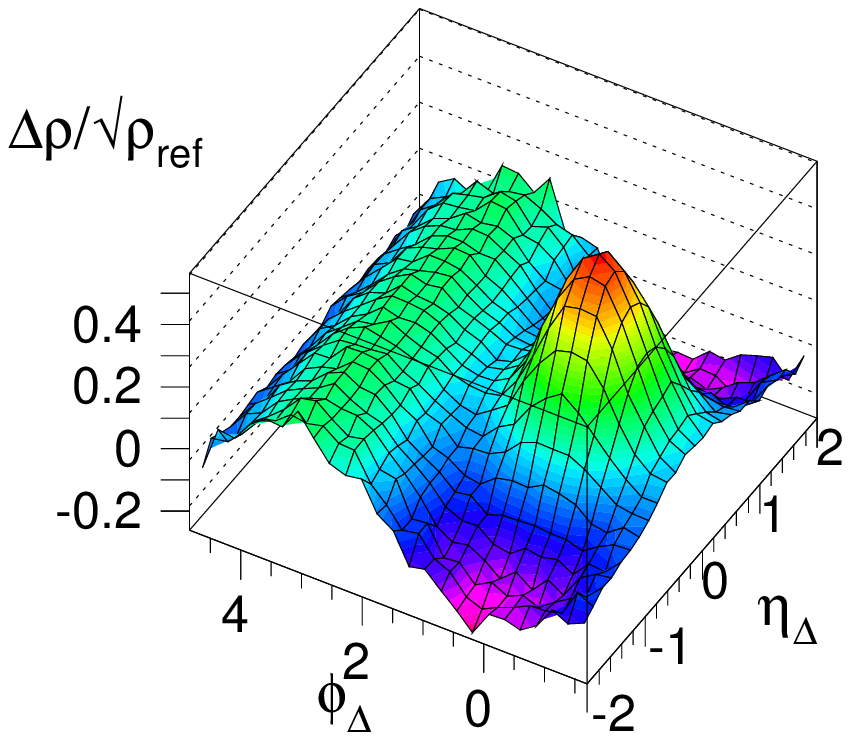}
\end{minipage}
%\hspace{0.01cm}
\begin{minipage}[b]{0.49\linewidth}
\centering
\includegraphics[width=1.\textwidth]{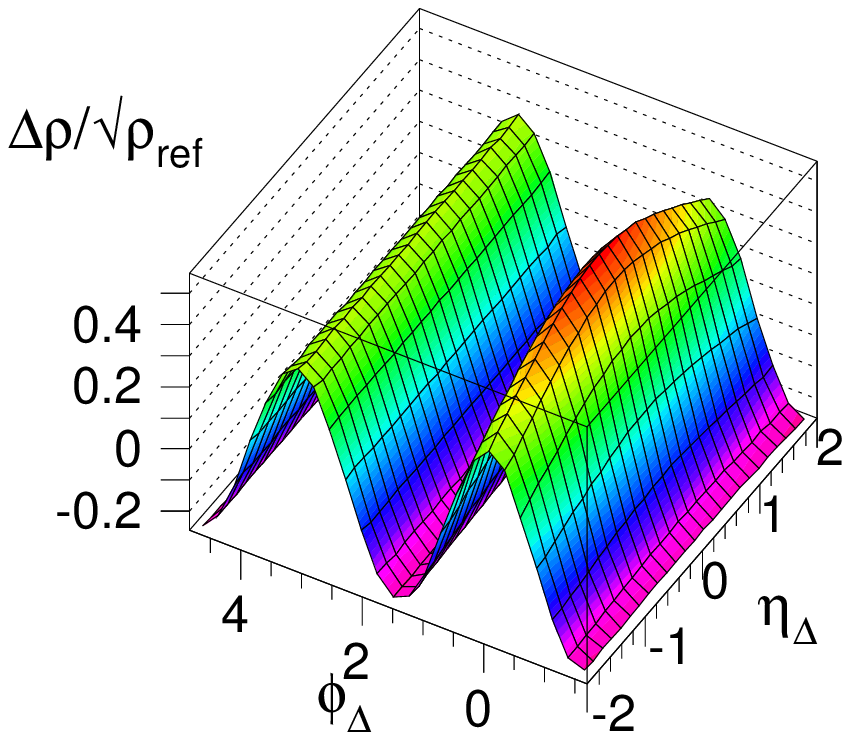}
\end{minipage}
\begin{minipage}[b]{0.49\linewidth}
\centering
\vspace{0.4cm}
\includegraphics[width=0.9\textwidth]{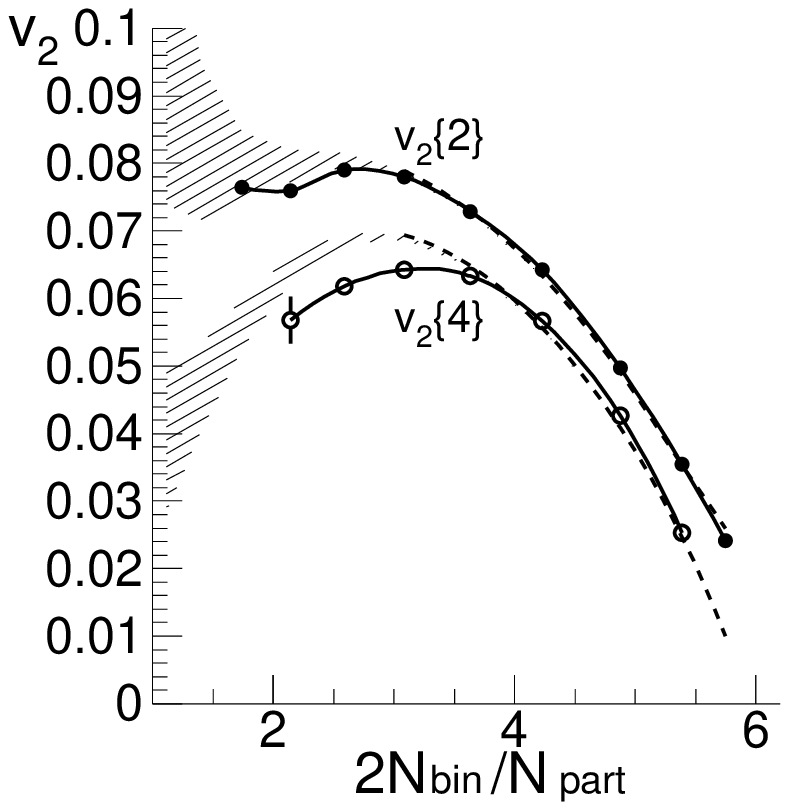}
\end{minipage}
%\hspace{0.01cm}
\begin{minipage}[b]{0.49\linewidth}
\vspace{0.4cm}
\centering
\includegraphics[width=0.95\textwidth]{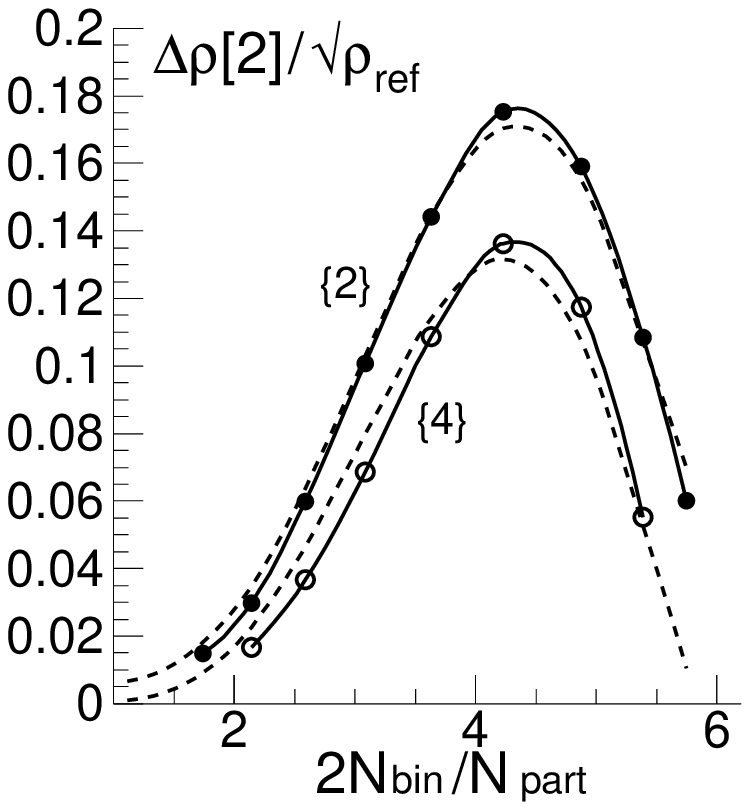}
\end{minipage}
\caption{ The top panels show simulated data for the two-particle
  correlation measure $\Delta\rho/\sqrt{\rho_{\mathrm{ref}}}$. The simulations
  are tuned to yield data similar to $p+p$ collisions (left) and Au+Au
  collisions (right). The bottom left panel shows $v_2$ estimated from
  the 2- and 4-particle cumulants. The shaded regions show the $v_2$
  derived by integrating over all structure in
  $\Delta\rho/\sqrt{\rho_{\mathrm{ref}}}$ (upper shaded region) or by
  performing a 2-Dimensional fit to the surface of
  $\Delta\rho/\sqrt{\rho_{\mathrm{ref}}}$ (lower shaded region). The bottom
  right panel shows the 2- and 4-particle cumulant data transformed to
  compare directly to $\Delta\rho[2]/\sqrt{\rho_{\mathrm{ref}}}$:
  $\{2\}=nv_{2}\{2\}^2$ and $\{4\}=nv_{2}\{4\}^2$ .}
\label{fig:estruct}
\end{figure}

One way to study non-flow contributions to two-particle correlations
is to measure the correlations as a function of $\Delta\eta$ and
$\Delta\phi$\cite{Trainor:2007fu}. This allows different sources of
two particle correlations to be studied where each source is
identified by its characteristic dependence on $\Delta\phi$ and
$\Delta\eta$. Additional information can be obtained by including
information about the charge-sign dependence of the correlations. An
example of such an analysis is shown in Fig.~\ref{fig:estruct}. The
figure displays four panels. The top panels show the correlation
density
\begin{equation}
\frac{\Delta\rho}{\sqrt{\rho_{\mathrm{ref}}}}=\frac{\rho-\rho_{\mathrm{ref}}}{\sqrt{\rho_{\mathrm{ref}}}},
\end{equation}
where $\rho$ is the pair density and $\rho_{\mathrm{ref}}$ is the
product of the single particle densities. This normalization is chosen
to search for deviations of the correlations in large systems from
those in small systems. If Au+Au collision were simply a superposition
of $p+p$ collisions for example,
$\Delta\rho/\sqrt{\rho_{\mathrm{ref}}}$ would be the same in $p+p$ and
Au+Au collisions. This figure was produced based on simulated data
which was tuned to match real data in very peripheral Au+Au or $p+p$
collisions (top left) and 20\%-30\% central Au+Au collisions (top
right)\cite{Trainor:2007ny,Daugherity:2008su}. The correlations in
these two systems have been found to be very different. The $p+p$
collisions exhibit structures characteristic of fragments from string
breaking (a narrow ridge at $\Delta\eta=0$ independent of
$\Delta\phi$) and fragments from semi-hard scattered partons or
mini-jets. These mini-jets yield a two-dimensional Gaussian
correlation at $0, 0$, and a broad ridge at $\Delta\phi=\pi$. The
away-side jet can sweep over a wide $\Delta\eta$ range since the
partons can have a momentum within the proton or Au nucleus. For
semi-central and central Au+Au collisions, the correlation landscape
is drastically different with the most prominent feature being $v_2$
giving rise to a clear $\cos(2\Delta\phi)$ shape. If the shape of the
various non-flow terms structures is well understood, the correlation
landscape can be fit and $\langle v_2^2\rangle$ can be extracted,
independent of the sources of non-flow. This procedure depends on
having an accurate description of the shape of the non-flow
sources. Some of these are easily identified based on their charge
dependence or their characteristic shapes. Other sources may be less
easily identifiable though, particularly if they become modified by
the medium in Au+Au collisions.

The bottom panels of Fig.~\ref{fig:estruct} show a proof-of-principle
extraction of $\langle v_2^2\rangle$ based on the simulated data. The
left panel shows $\sqrt{\langle v_2^2\rangle}$ while the right panel
shows the per-particle measure $\Delta\rho[2]/\sqrt{\rho_{\mathrm{ref}}}$. Also
shown are the two-particle and four-particle cumulant data. The
upper hatched region in the bottom left panel shows 
\begin{equation}
\langle v_{2}^2\rangle =\frac{2\pi}{\overline{n}}\Delta\rho[2]/\sqrt{\rho_{\mathrm{ref}}}
\end{equation}
extracted without separating out non-flow; called
$v_2^2\{1D\}$. $\overline{n}$ is the multiplicity of measured
tracks. The lower hatched region shows the same without including the
non-flow structures in the calculation; called $v_2^2\{2D\}$. Data are
plotted versus $2N_{\mathrm{binary}}/N_{\mathrm{part}}$ so that most
central collisions are on the right. This procedure makes use of the
two-particle correlations landscape to separate different
contributions to the azimuthal structure. The fit procedure does
require some assumptions be made in order to separate non-flow from
$v_2$\cite{Daugherity:2008su}. These include that $v_2$ and
$\sigma_{v_{2}}$ are independent of $\Delta\eta$ and that higher
harmonics of $v_n$ and $\sigma_{v_{n}}$ do not contribute. Relaxing
those assumptions may make it difficult to distinguish between
non-flow and more complicated correlations related to the reaction
plane. More information can be made use of however, by examining the
charge, sign dependence and particle-type dependence of the various
correlation structures. This is therefore a promising method for
disentangling $v_2$ and non-flow.

\subsection{Scaling Observations}

Elliptic flow measurements represent an extensive data set. $v_2$ has
been measured for $0.1 < p_T < 12$~GeV/c, for $-5<\eta<5$, for mesons
from the pion to the $\phi$ and $J/\psi$, for baryons from the proton
to the $\Omega$, and for transverse particle densities
$3<\frac{1}{S}\frac{dN}{dy}<30$. And yet, given the complexity of
heavy-ion collisions and such a large data-set, the measurements
exhibit many surprisingly simple features. These we can summarize in
terms of simple scaling observations where a large amount of data is
found to behave in a regular and simple way when plotted versus the
appropriate variable. The observation of a particular scaling then
motivates the question: why does the data only depend on that
variable? These scaling observations, therefore, not only allow us to
summarize large amounts of data in a simple form, but they also
suggest simple physical explanations for the data with perhaps deeper
implications. In this section I review several observed scaling laws.

\subsubsection{Longitudinal scaling} Fig.~\ref{fig:v2eta} shows the
centrality dependence of $v_2(\eta)$\cite{Back:2004mh}. Although more
detailed measurements at small $\eta$ show that $v_2$ is approximately
independent of $\eta$ for $|\eta|<1$\cite{Adams:2004bi}, the data
extending to larger $\eta$ exhibit a nearly triangular shape: having a
maximum at $\eta=0$ with a nearly linear decrease with $|\eta|$. A
similar shape is seen for $\sqrt{s_{_{NN}}} = $ 200, 130, 62.4, and
19.6 GeV Au+Au collisions\cite{Alver:2006wh}.

\begin{figure}[hbt]
\begin{center}
\centerline{  \includegraphics[width=.8\textwidth]{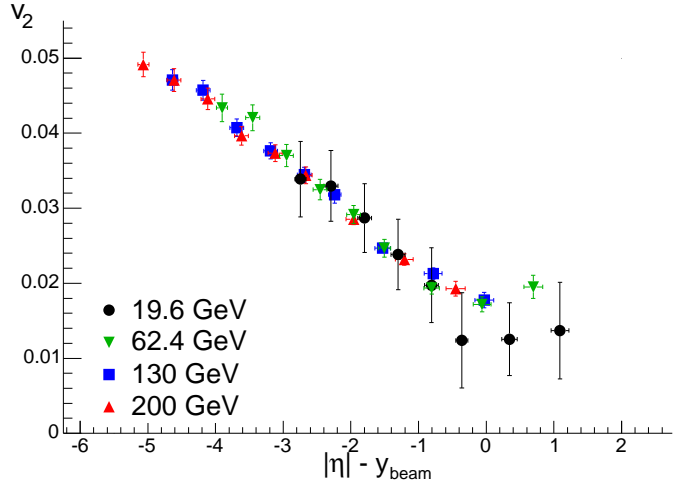}}
\caption{ $v_{2}(\eta-y_{beam})$ for a variety of energies. The
  centrality interval used is $0\%-40\%$. Plotted in this format, data
  at all energies fall on one curve. This scaling is also seen in
  $dN/dy(\eta-y_{beam})$ and is referred to as limiting
  fragmentation. $v_2$ is largest at middle rapidity ($\eta=0$ or
  $\eta-y_{beam}=-y_{beam}$) and vanishes at $\eta=y_{beam}$
  ($\eta-y_{beam}=0$). }
  \label{fig:v2etafold}
\end{center}
\end{figure}

Fig.~\ref{fig:v2etafold} shows $v_2(|\eta|-y_{beam})$ for
$\sqrt{s_{_{NN}}}$ from 19.6 to 200 GeV; one order of magnitude in
$\sqrt{s_{_{NN}}}$. The data are for the 40\% most central
collisions. Ideally the x-axis would display $y-y_{beam}$ but data on
identified particle $v_2$ spanning such a large range of rapidity are
not available. One finds that within errors, all data lie on a single
curve. This suggests a smooth variation of the development of
space-momentum correlations from forward rapidity to
mid-rapidity. This scaling observation also implies that the value of
$v_2$ obtained at mid-rapidity is a smooth function of $y_{beam}$ or
equivalently of $\log(\sqrt{s_{_{NN}}})$; consistent with the smooth
trend seen in Fig.~\ref{fig:v2excitation} for $v_2$ above
$\sqrt{s_{_{NN}}}$ of approximately 5-10 GeV. An energy scan at RHIC
extending down to $\sqrt{s_{_{NN}}}=5$ GeV will make it possible to
investigate this trend with better precision and with a single
detector, eliminating many systematic uncertainties\cite{escan}. This
simple trend may be confirmed with more precision or perhaps
deviations will point to a softest point in the equation-of-state\cite{Sorge:1998mk,Stocker:2007pd}.

\subsubsection{Kinetic Energy and Constituent Quark Number Scaling}

At low $p_T$, $v_2$ is ordered by mass with heavier particles having a
smaller $v_2$ value at a given $p_T$
value\cite{Adler:2001nb,Adler:2002pb}. This ordering is indicative of
particle emission from a boosted source with the boost larger in the
in-plane direction than the out-of-plane direction. Indeed, blast-wave
fits implementing this scenario agree very well with the data in this
region\cite{Retiere:2003kf}. It is also found that in this same $p_T$
region, when $v_2$ is plotted versus $m_T-m_0$ all data fall on a
common line\cite{Sorensen:2003wi}. $m_T-m_0$ is the particles
transverse kinetic energy and sometimes labeled
KE$_T$\cite{Adare:2006ti}

\begin{figure}[hbt]
\begin{center}
\centerline{  \includegraphics[width=.9\textwidth]{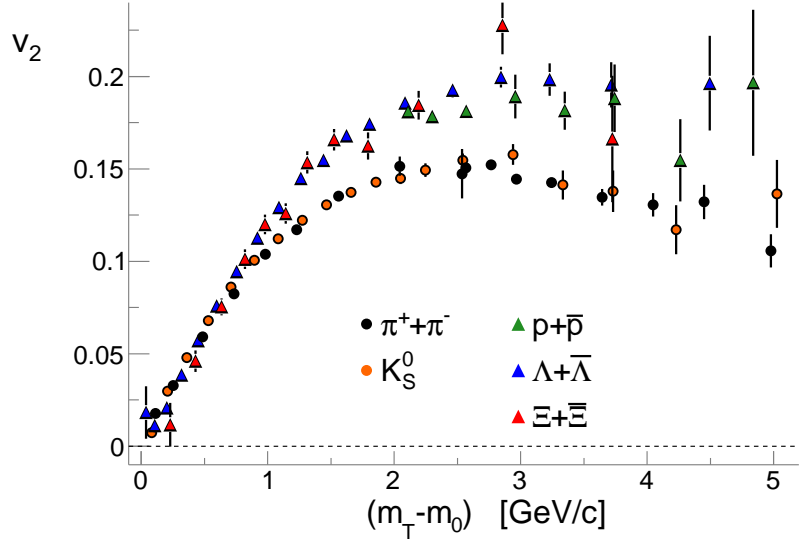}}
\caption{ $v_2$ for a variety of particles plotted versus $m_T-m_0$ where
  $m_T^2 = p_T^2+m_0^2$, and $m_0$ is the rest mass of the
  particle. $m_T-m_0$ is also the transverse kinetic energy of the
  particle $KE_T$.}
  \label{fig:mtscaling}
\end{center}
\end{figure}

Fig.~\ref{fig:mtscaling} shows $v_2$ versus $m_T-m_0$ for particles
ranging in mass from the pion with mass of 0.1396 GeV/c$^2$ to the
$\Xi$ with mass of 1.321 GeV/c$^2$. The measurement is made for the
0-80\% centrality interval in 200 GeV Au+Au collisions. Similar
scaling has also been demonstrated for 62.4 GeV
collisions\cite{Abelev:2007qg}. The data exhibit obvious trends. At
low $m_T-m_0$, $v_2$ values for all particles rise linearly with no
apparent differences between the particles with different masses. Near
$m_T-m_0 = 0.8$ GeV/c$^2$, $v_2(m_T-m_0)$ for mesons and baryons
diverges. The meson $v_2$ begins to saturate, obtaining a maximum
value of 14-15\% near $m_T-m_0 = 2.5$~GeV/c$^2$. The baryon $v_2$
continues to rise, obtaining a maximum value of approximately 19-20\%
at $m_T-m_0 = 3$ to 3.5 ~GeV/c$^2$. The relative masses of the baryons
and mesons do not seem to be relevant, rather the number of
constituent quarks in the hadron determines the $v_2$ values in this
range. The mass dependence can be better checked using the
$\phi$-meson which has a mass slightly larger than that of the
proton. The statistical significance of the $\phi$ $v_2$ is limited
but measurements seem to indicate that the $\phi$ lies closer to the
mesons than to the baryons \textit{i.e.}  closer to the particles with
a common number of constituent quarks than to particles with a common
mass\cite{Abelev:2007rw}.

The observation of the quark-number dependence of $v_2$ at
intermediate $p_T$ led to speculation that hadron formation through
the coalescence of dressed quarks at the hadronization phase boundary
could lead to an amplification of $v_2$ with baryons getting amplified
by a factor of 3 while mesons were amplified by a factor of
2\cite{Voloshin:2002wa,Lin:2001zk,Molnar:2003ff,Fries:2008hs,Hwa:2003bn,Fries:2003kq,Greco:2003mm,Greco:2003xt,Fries:2003vb,Pratt:2004zq,Ravagli:2008rt,Fries:2008hs}. This
picture was subsequently strengthened by the observation that a
similar quark-number dependence arises in
$R_{CP}$\cite{Adams:2003am,Adler:2003kg}: the ratio of the single
particle spectra in central collisions to that in peripheral
collisions. At intermediate $p_T$ the $R_{CP}$ values for various
particle species are also grouped by the number of constituent quarks,
with baryons having a larger $R_{CP}$. The larger $R_{CP}$ for baryons
signifies that baryon production increases with collision centrality
faster than meson production; an observation consistent with the
speculation that hadrons from Au+Au collisions are formed by
coalescence such that baryon production becomes easier as the density
of the system increases. The more general and less model dependent
statement is that the baryon versus meson dependence arises from high
density and therefore most likely from multi-quark or gluon effects or
sometimes called" higher twist" effects. The combination of large
baryon $v_2$ and large baryon $R_{CP}$ also immediately eliminates a
class of explanations attempting to describe one or the other
observation: \textit{e.g.}  originally it was speculated that the
larger $R_{CP}$ for baryons might be related to a smaller
jet-quenching for jets that fragment to baryons than for jets that
fragment to mesons. This explanation would lead to a smaller baryon
$v_2$ and is therefore ruled out by the larger $v_2$ for baryons. The
same can be said for color transparency models\cite{Brodsky:2008qp}
which would account for the larger baryon $R_{CP}$ in this $p_T$
region but would predict a smaller baryon $v_2$. Color transparency
may still be relevant to the particle type dependencies at $p_T>5$
where $R_{CP}$ for protons is slightly larger than $R_{CP}$ for
pions\cite{Abelev:2007ra} and the $v_2$ measurements are not yet
precise enough to conclude whether the baryon $v_2$ is also smaller
than the meson $v_2$. This is a topic that needs to be studied
further.

\begin{figure}[hbt]
\begin{center}
\centerline{  \includegraphics[width=.8\textwidth]{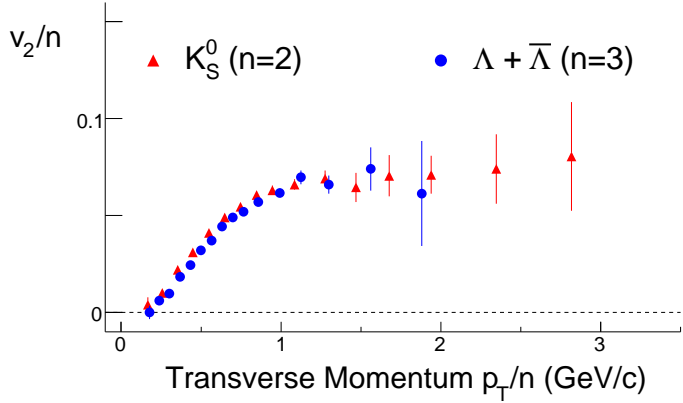}}
\caption{ $v_2$ for $K_S^0$ and $\Lambda$ scaled by the number of
  their constituent quarks ($n$) and plotted versus $p_T/n$. The data
  appear to fall on a universal curve which has been taken as an
  indication of hadron formation via coalescence of quarks from a
  flowing medium.}
  \label{fig:ncqprl}
\end{center}
\end{figure}

In a coalescence picture, the final momentum of the observed hadron
would depend on the momentum of the coalescing constituent quarks. The
exact dependence is not known but a relatively good scaling of $v_2$
for $K_S^0$ and $\Lambda$ was found when $v_2/n$ was plotted as a
function of $p_T/n$. Such a scaling implies that the momentum of the
hadron is simply the sum of the momenta of the coalescing
quarks. Fig.~\ref{fig:ncqprl} shows $v_2/n$ versus $p_T/n$ for
$K_S^0$-mesons and $\Lambda$-baryons. The scaling appears to be good
throughout the whole $p_T$ range but part of this perception is due to
the decrease of $v_2$ for both particles at small $p_T$. When a ratio
is taken between the $v_2/n(p_T/n)$ values, a clear deviation from
scaling is seen in the lower $p_T$ region. A combination of the
$m_T-m_0$ scaling in Fig.~\ref{fig:mtscaling} and the $v_2/n$ scaling
in Fig.~\ref{fig:ncqprl} will lead to a good scaling over the whole
measured momentum range; since $v_2(m_T-m_0)$ for all particles fall
on a single line at low $m_T-m_0$, dividing the x- and y-axis by n
will not destroy that scaling seen in
Fig.~\ref{fig:mtscaling}. Plotting $v_2/n$ versus $(m_T-m_0)/n$ should
therefore provide a good scaling across a large kinematic range.

\begin{figure}[hbt]
\begin{center}
\centerline{  \includegraphics[width=0.99\textwidth]{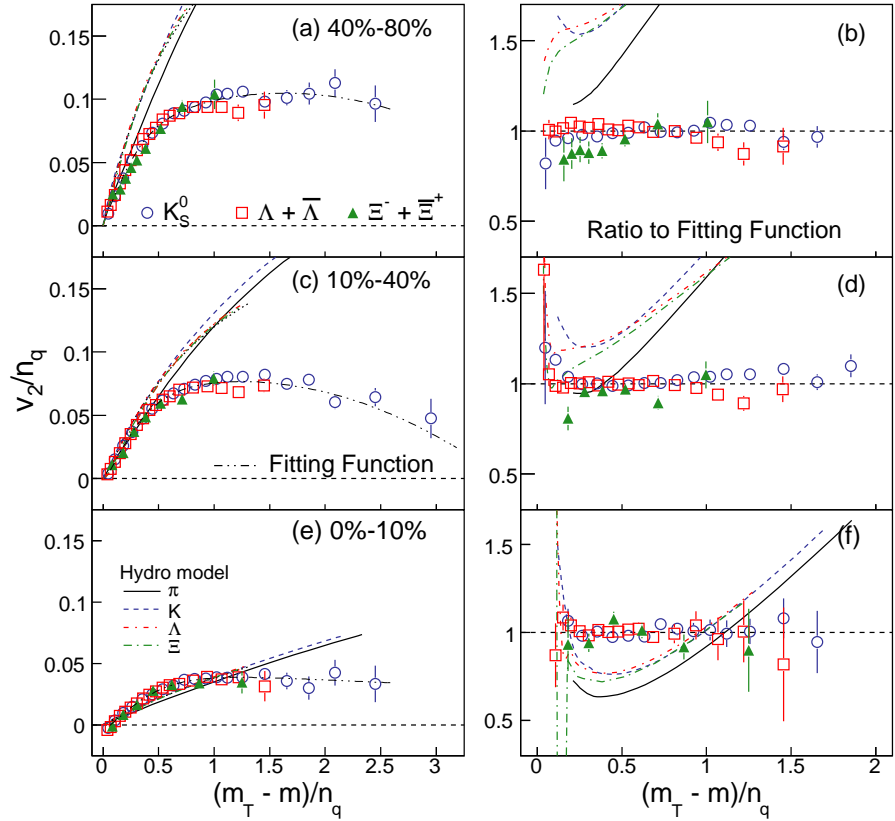}}
\caption{ A more detailed study of quark number scaling in Au+Au
  collisions. In the left panels (a, c, and e) $v_2/n$ is shown
  versus $(m_T-m_0)/n$ for three centrality classes. Hydrodynamic models
  are also shown for comparison. Data are fit to a single curve. In
  the right panels (b, d and f) the ratio of the data and hydro model
  to the fit function are shown. }
  \label{fig:ncqcent}
\end{center}
\end{figure}

Fig.~\ref{fig:ncqcent} shows $v_2/n$ versus $(m_T-m_0)/n$ for 200 GeV
Au+Au collisions in three different centrality
intervals\cite{Abelev:2008ed}. Data for $K_S^0$-mesons,
  $\Lambda$-baryons and $\Xi$-baryons are shown. The left panels show
  the data with a hydrodynamic calculation and a fitting function. The
  phenomenologically motivated function
\begin{equation}
v_2/n = \frac{a+bx+cx^2}{1+exp[\frac{-(x-d)}{e}]} - \frac{a}{2}
\label{eq:fit}
\end{equation}
with $x=(m_T-m_0)/n$, describes the data well for the three
centralities. The function captures the rise then saturation and
steady decline seen in the data. We ascribe no physical meaning to the
function or the five fit parameters but simply use it as a convenient
reference. The right panels show the ratio of the data and the hydro
model to the fit function. For reference, the fit parameters are shown
in Table~\ref{tab:parms}. The data is in good agreement with the fit
function for all centralities while this hydro model calculation does
not agree well with the data in any centrality.
\begin{table}[hbt]
\centering
\begin{tabular}{l|ccccc}
%\toprule
~ & \multicolumn{5}{c}{Fit parameters for Eq.~\ref{eq:fit}}\\
Centrality & a & b & c & d & e \\
\colrule
40\%-80\% & 20.0e-02 & 0.0 & 0.0 & -1.19e-02 & 2.37e-01 \\
10\%-40\% & 16.4e-02 & -4.53e-03 & 0.0 & 2.61e-02 & 2.40e-01 \\
0\%-10\%  & 8.96e-02 & -4.08e-03 & 0.0 & 6.52e-02  & 2.70e-01 
%\botrule
\end{tabular}
  \caption{The fit parameters describing the curves in Fig.~\ref{fig:ncqcent}.}
\label{tab:parms}
\end{table}

There is a systematic deviation from the ideal n scaling at
$(m_T-m_0)/n>0.8$~GeV/c$^2$ with $K_S^0$ mesons having slightly larger
$v_2/n$ values than $\Lambda$ baryons. This deviation from ideal
scaling was predicted based on the inclusion of higher fock states in
the hadrons or the inclusion of a finite width in the hadron wave
function\cite{Greco:2004ex,fock}. Deviations can also arise in a
hadronic phase when the hadronic cross sections are relevant. In the
case that hadronic cross-sections are an important factor, higher
statistics data for $\Omega$ baryons and $\phi$ mesons should deviate
from their respective groups. We also note that hadronic cascade
models also obtain approximate $v_2/n$ scaling due to the use of the
additive quark model for hadronic cross-sections\cite{Lu:2006qn}. On
the other hand, these models under-predict the integrated $v_2$ by a
factor of two. We also note that non-flow contributions can affect the
scaling observed in this range and the particle-type dependence of
non-flow sources is still being investigated.

\begin{figure}[ht]
\vspace{1.0cm}
\begin{minipage}[b]{1.0\linewidth}
\centering
\includegraphics[width=0.8\textwidth]{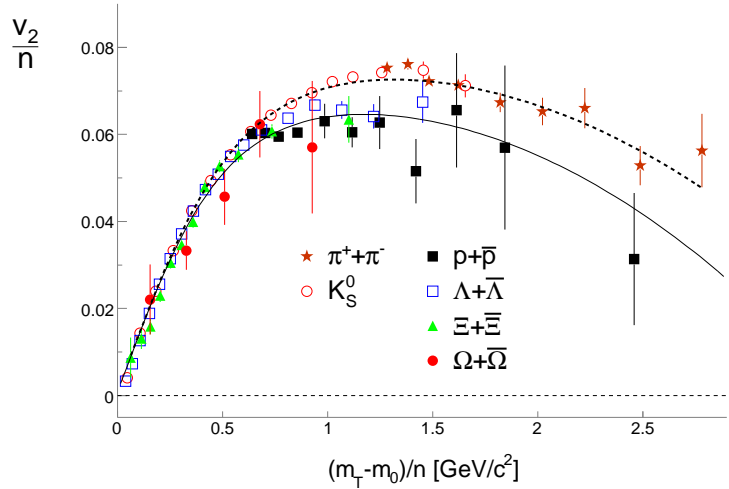}
\end{minipage}
%\begin{minipage}[b]{0.49\linewidth}
%\centering
%\includegraphics[width=1.\textwidth]{figures/nplus1.eps}
%\end{minipage}

%\begin{figure}[hbt]
%\begin{center}
%\centering\centerline{  \includegraphics[width=.8\textwidth]{figures/scaling_twofit.eps}}
%\caption{ Left Panel: $v_2/n$ versus $(m_T-m_0)/n$ for minimum bias Au+Au
%  collisions. The quark number scaling appears to be violated when
%  integrating over a wide centrality bin. Mesons and Baryons are fit
%  with the same functional form but with different parameters. Right
%  Panel: The same data as the left panel but scaled by n+1. This
%  indicates that the number of constituent quark scaling is violated
%  to the level that baryon $v_2$ is closer to four-thirds the meson
%  $v_2$ than three-halves.
\caption{ $v_2/n$ versus $(m_T-m_0)/n$ for minimum bias Au+Au
  collisions. The quark number scaling appears to be violated when
  integrating over a wide centrality bin. Mesons and Baryons are fit
  with the same functional form but with different parameters. The
  scaling is violated to the extent that $n+1$ would give a better
  agreement \textit{i.e.} baryon $v_2$ is closer to four-thirds the
  meson $v_2$ rather than three-halves.  }
  \label{fig:twofit}
%\end{center}
\end{figure}

In Fig.~\ref{fig:twofit} we investigate the breaking of ideal scaling
in more detail with data integrated over a larger centrality
interval. While this reduces the statistical uncertainty, it also
introduces uncertainties due to the large centrality bin width. In
particular, when particle yields have different centrality
dependencies, the average eccentricity of events producing a particle
can deviate from particle to particle. For example, the enhancement of
baryons in central collisions will mean that the average baryon comes
from a more central event than the average meson. Given the decrease
of $v_2$ with centrality, this can lead to a decrease of baryon $v_2$
simply due to the wide centrality bin. Although there are caveats and
systematic errors still to be quantified, we note that the baryons in
Fig.~\ref{fig:twofit} appear to lie systematically and significantly
below the mesons. Self-similar curves are fit to mesons and
baryons. The curves appear to describe the data. We note that the two
self-similar curves shown in Fig.~\ref{fig:twofit}
can be nearly unified if we replace n with n+1. This demonstrates that
the naive constituent quark scaling is violated to the extent that
baryon $v_2$ is actually closer to 4/3 the meson $v_2$ rather than
3/2. The connection of the baryon versus meson dependence and the
number of constituent quark scaling appears to not be as directly
connected to the number of constituent quarks as originally
conceived. Whether this is indicative of higher fock states, the
wave-function of the hadrons, an as yet un-accounted for experimental
systematic error, or something else is yet to be determined. The
systematic uncertainties based on the particle-type dependence of
non-flow are still being investigated.

\subsubsection{System-Size Scaling}

\begin{figure}[hbt]
\begin{center}
\centerline{  \includegraphics[width=1.\textwidth]{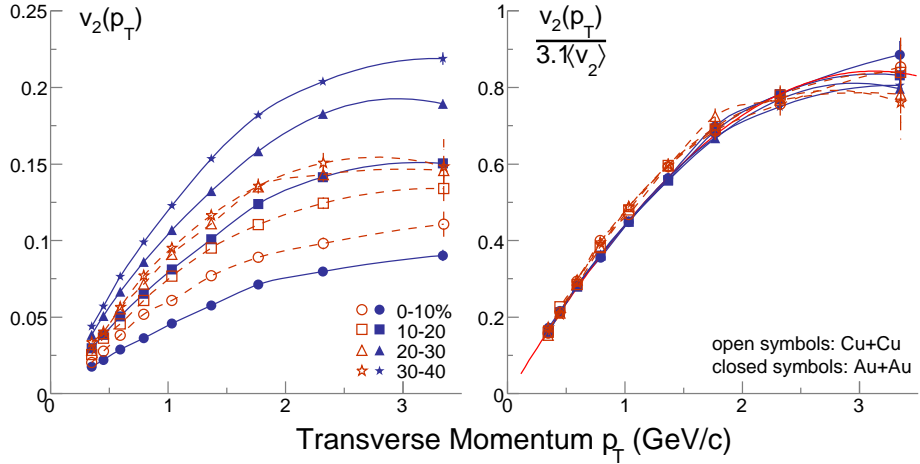}}
\caption{Left panel: $v_2$ versus $p_T$ in Au+Au and Cu+Cu collisions for
  four centrality intervals. Right panel: the same $v_2$ scaled by
  $3.1$ times the $p_T$ integrated $v_2$. The scaling demonstrates
  that the shape of $v_2(p_T)$ is approximately independent of
  centrality and system size. A function is fit to the Au+Au data and
  shown as a solid orange line. }
  \label{fig:cucuauau}
\end{center}
\end{figure}

The system-size dependence of $v_2$ can be studied by looking at the
centrality dependence of $v_2$ or by colliding smaller nuclei. Ideal
hydro predictions, having a zero mean-free-path assumption, should be
independent of the system-size. In this case, given the same
eccentricity, the $v_2$ should be independent of system size. One can
try to account for the change in eccentricity by dividing $v_2$ by
eccentricity from a model but this introduces a large amount of
uncertainty. Another approach is to study the shape of $v_2(p_T)$ to
see if that varies\cite{Teaney:2003kp}. The left panel of
Fig.~\ref{fig:cucuauau} shows $v_2$ measured in Au+Au and Cu+Cu
collisions for several centrality intervals\cite{Adare:2006ti}. In
the right panel, $v_2(p_T)$ is scaled by 3.1 times the mean $v_2$ for
that data set. $3.1\langle v_2\rangle$ was taken as a proxy for the
eccentricity of the collision system, and this proxy is not
inconsistent with models of eccentricity which are quite
uncertain. What is best demonstrated by this scaling, is that although
the magnitude of $v_2$ changes significantly for the different
centralities and systems, the shape of $v_2(p_T)$ is very similar.

\begin{figure}[hbt]
\begin{center}
\centerline{  \includegraphics[width=.8\textwidth]{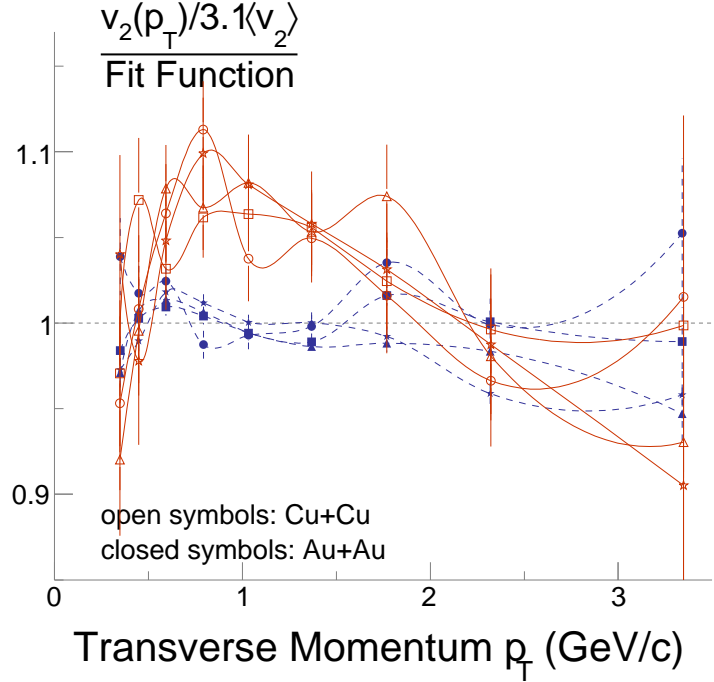}}
\caption{ The $v_{2}(p_T)/3.1\langle v_2\rangle$ data from the right
  panel of Fig.~\ref{fig:cucuauau} scaled by a function fit to the
  Au+Au data. This figure illustrates that their seems to be a
  significant difference between the shape of $v_2(p_T)$ in Au+Au and
  Cu+Cu collisions with the Cu+Cu collisions exhibiting a more abrupt
  turn-over -- \textit{i.e.} Cu+Cu data first rises faster with $p_T$
  then falls faster at $p_T>1$ GeV/c.}
  \label{fig:cucuauauratio}
\end{center}
\end{figure}

The invariance of $v_2(p_T)$ with system-size can be taken as an
indication that the viscosity of the expanding medium created in
heavy-ion collisions can not be large when $v_2$ is established; Large
viscous effects should introduce a system-size dependence to
$v_2(p_T)$ with viscosity causing $v_2$ to saturate at lower $p_T$
values in the smaller system\cite{Teaney:2003kp}. Hydrodynamic
calculations including viscosity confirm this
idea\cite{Romatschke:2007mq,Luzum:2008cw,Song:2007fn,Song:2007ux,Dusling:2007gi,Song:2008si}. To
look more carefully for a system-size dependence in the shape of $v_2$
we plot the ratio of the scaled data to a curve fit to the Au+Au
data. The results are shown in Fig.~\ref{fig:cucuauauratio}. The Cu+Cu
data systematically deviate from the Au+Au data. The $p_T$ dependence
of the ratio indicates that the Cu+Cu data begins to saturate before
the Au+Au data. This leads to a ratio that first rises then
falls. This would happen the other way around if the Au+Au data
saturated first. The uncertainties in the figure are large but the
shapes are still significantly different. The system-size dependence
of $v_2(p_T)$ may be a valuable tool for estimating the viscosity of
the matter created in heavy-ion collisions.

Although the data on $v_2$ includes many particle types, a wide
kinematic range in $p_T$ and $\eta$, a variety of system-sizes and a
wide range in center-of-mass energy, we've been able to identify
several regular features of the data. These include a nearly linear
rise of $v_2$ at mid-rapidity with $\log(\sqrt{s_{_{NN}}})$:
\begin{equation}
  v_2 = 0.008+0.0084\log(\sqrt{s_{_{NN}}})
\end{equation}
for 0\%-20\% central Au+Au or Pb+Pb collisions and a $p_T$, mass and
particle-type dependence that can be parametrized by
\begin{equation}
v_2/n = \frac{a+bx+cx^2}{1+exp[\frac{-(x-d)}{e}]} - \frac{a}{2},
\end{equation}
where $x=(m_T-m_0)/n$, while $v_2(\eta)$ to good approximation
decreases linearly from it's maximum at mid rapidity to beam
rapidity. This linear rise may be a trivial consequence of the
$\log(\sqrt{s_{_{nn}}})$ dependence of mid-rapidity $v_2$ or
vice-versa.

%% file: hydro.tex
\section{Confronting the Hydrodynamic Paradigm with RHIC Data}

We have discussed the hydrodynamic model extensively in this review as
a convenient reference for how well the matter produced in heavy-ion
collision converts spatial deformation into momentum space
anisotropy. Hydrodynamic models of heavy-ion collisions have many
uncertainties. These include, uncertain initial conditions, uncertain
thermalization times, and uncertain freeze-out conditions. A
successful description of data using a hydrodynamic model offers the
promise of not only establishing the attainment of local equilibrium
but also the promise of providing information on the Equation-of-State
of the matter and its transport properties. The uncertainty in the
models, however, are large and it has not yet been possible to extract
this desired information with satisfactory certainty. In addition, the
possibility that significant $v_2$ arises from initial-state
effects\cite{Krasnitz:2002ng,Boreskov:2008uy} could call into question
the applicability of hydrodynamics and the need for prolific
final-state rescattering. Measurements of two particle correlations,
which have often been interpreted as arising from
mini-jets\cite{Adams:2004pa,Trainor:2007ny}, need to be reconciled
with the idea of a locally thermalized matter with extensive
final-state rescattering. If the hydrodynamic models and data are
irreconcilable, the paradigm will, of course, have to be abandoned.

\begin{figure}[ht]
\vspace{0.5cm}
\begin{minipage}[b]{0.49\linewidth}
\centering
\includegraphics[width=1.\textwidth]{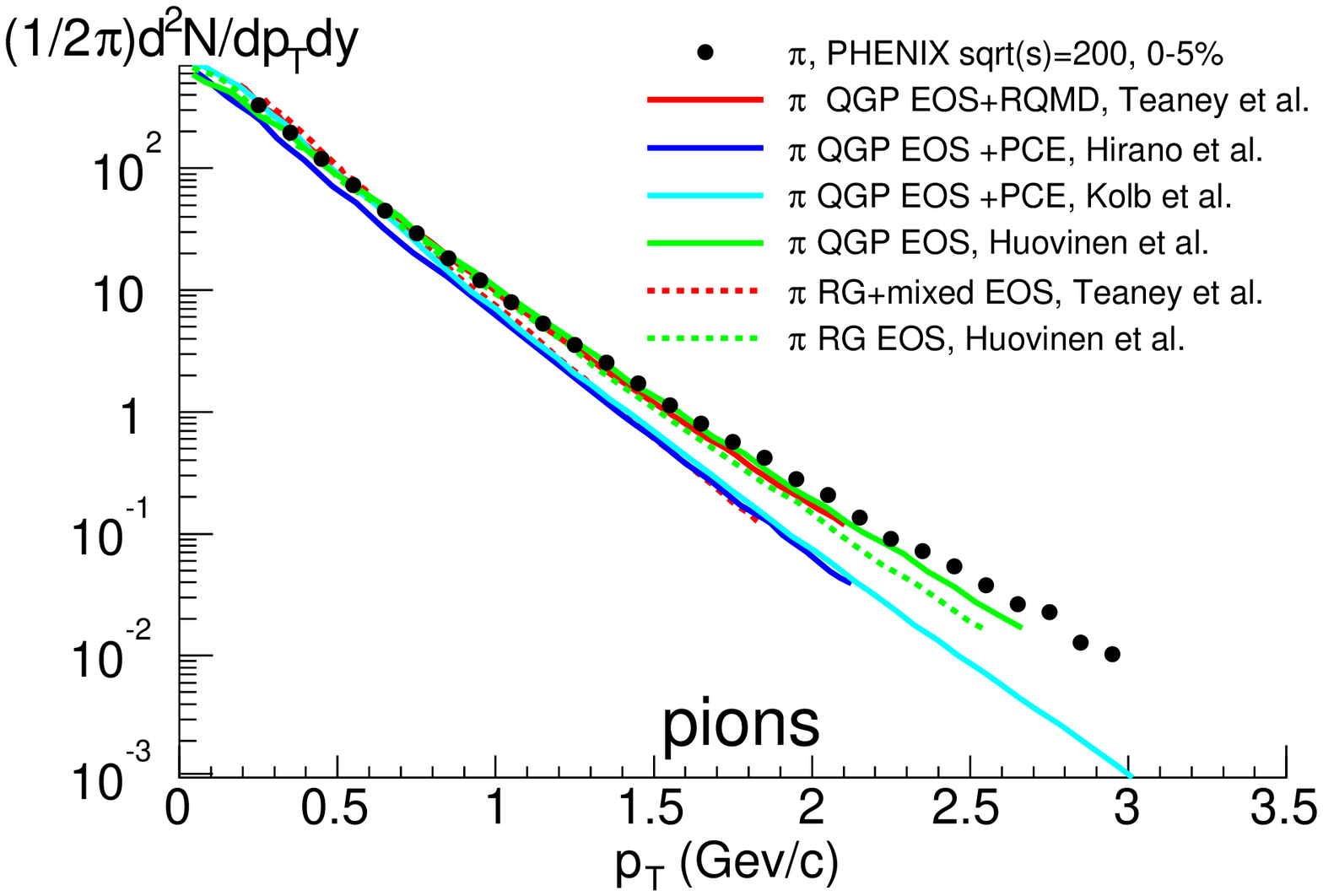}
\end{minipage}
\begin{minipage}[b]{0.49\linewidth}
\centering
\includegraphics[width=1.\textwidth]{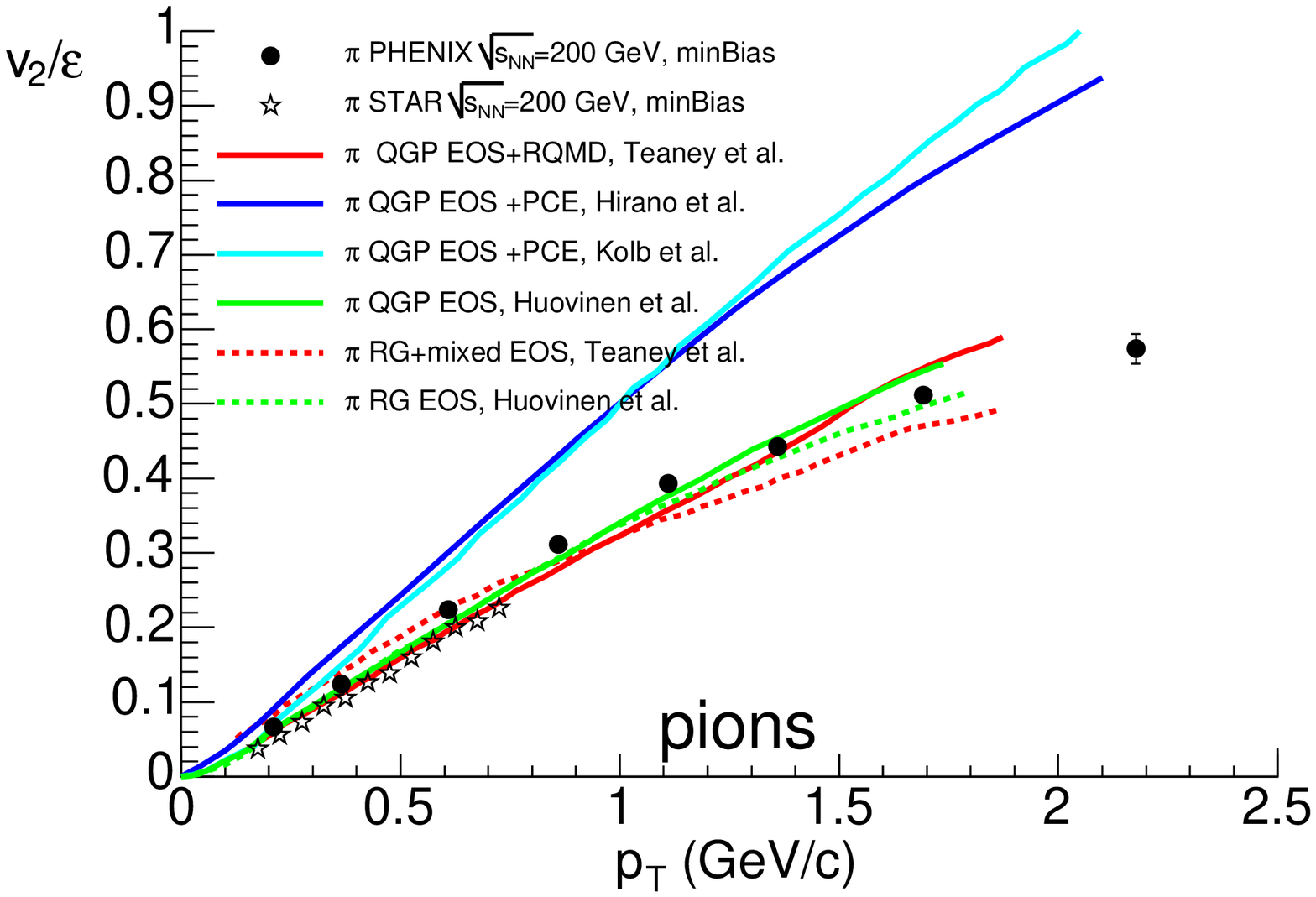}
\end{minipage}
\begin{minipage}[b]{0.49\linewidth}
\centering
\vspace{0.4cm}
\includegraphics[width=1.\textwidth]{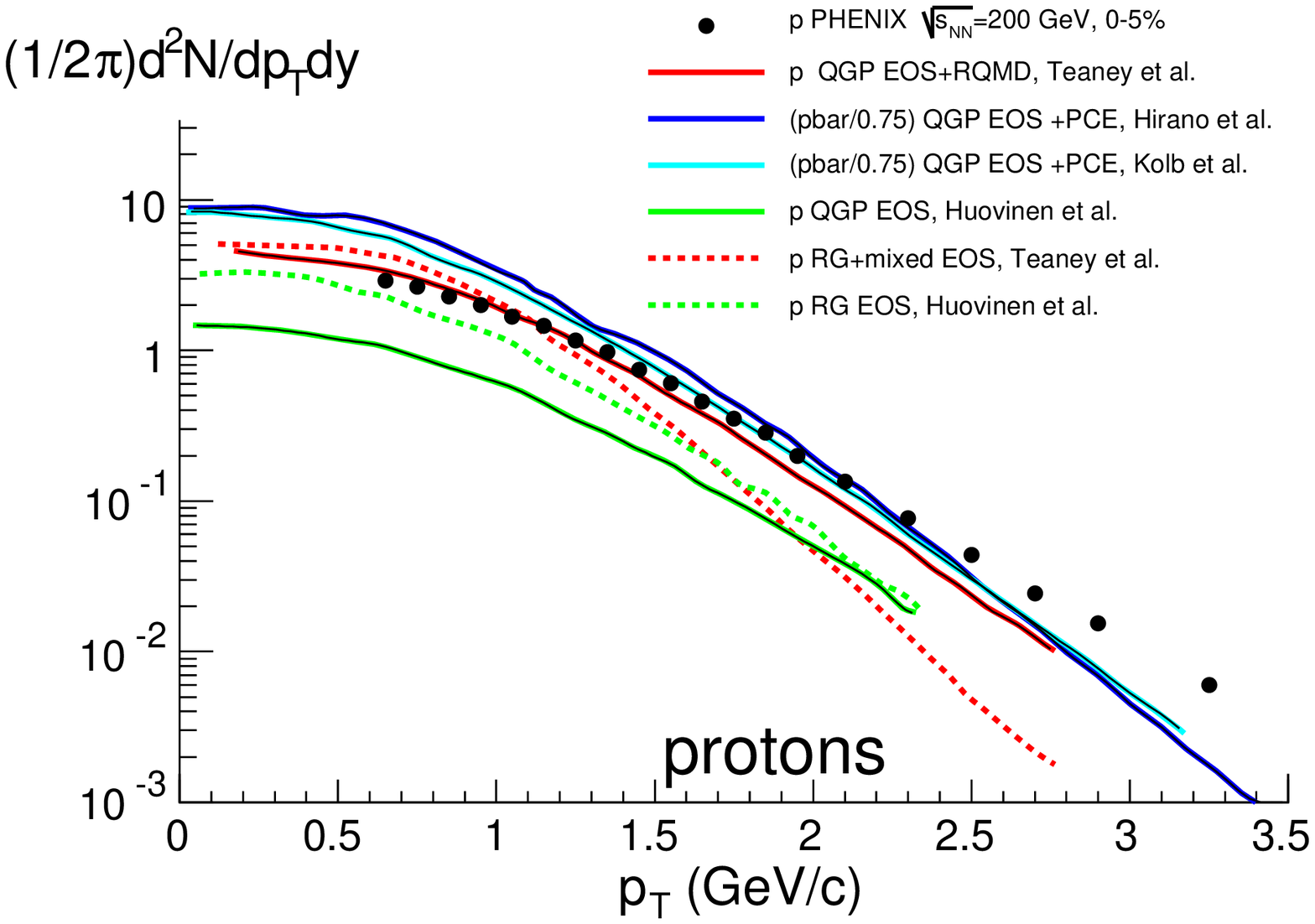}
\end{minipage}
\begin{minipage}[b]{0.49\linewidth}
\vspace{0.4cm}
\centering
\includegraphics[width=1.\textwidth]{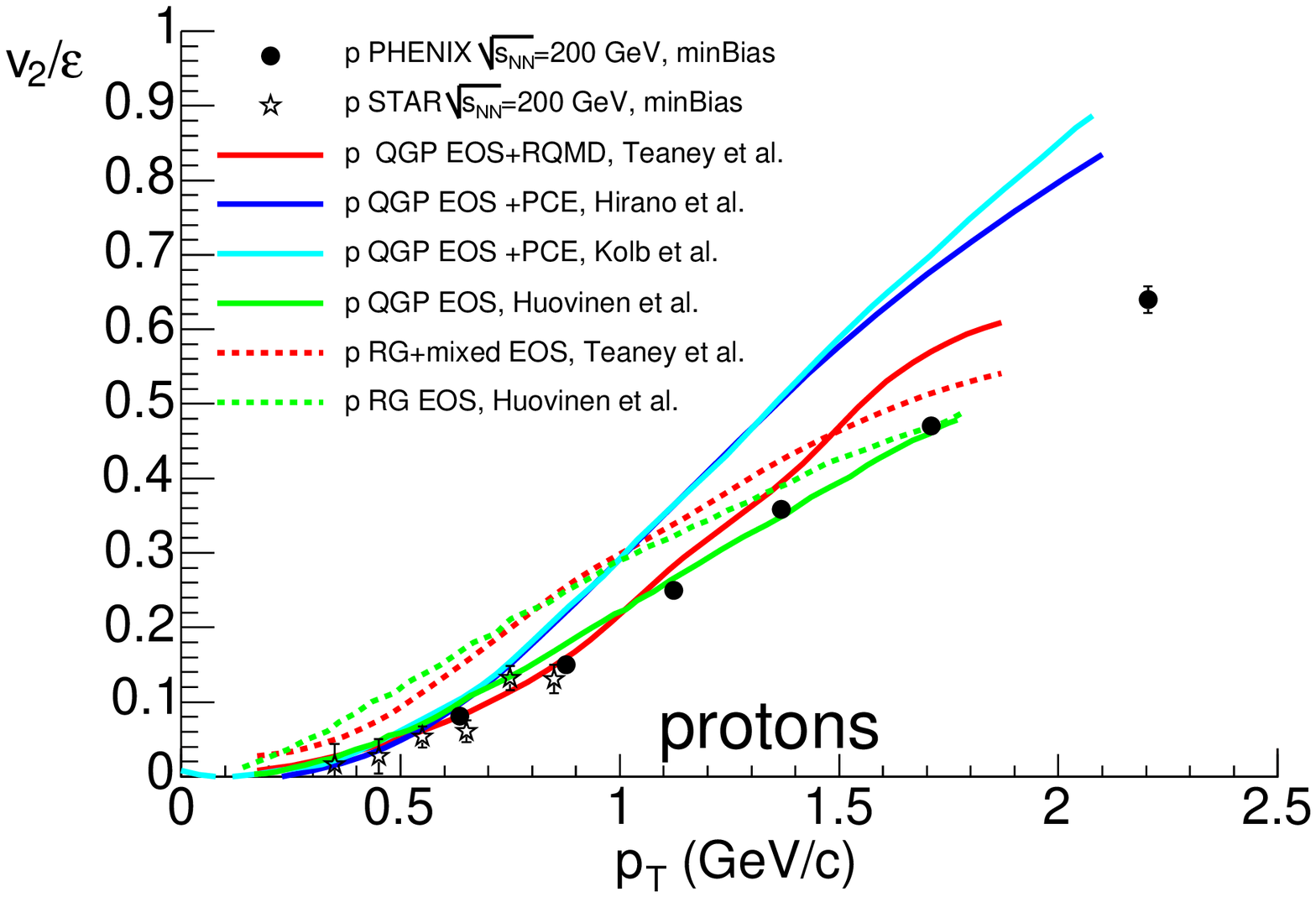}
\end{minipage}
\caption{ The four panels show $p_T$ spectra and $v_2$ (top and
  bottom) for pions and protons (left and right). Data from 200 GeV
  Au+Au collisions is compared to a variety of hydrodynamic
  models. Most models do not agree with $p_T$ spectra and $v_2$
  simultaneously.}
\label{fig:hydrocomp}
\end{figure}

To check for consistency with hydrodynamic
models\cite{Teaney:2000cw,Huovinen:2001cy,Kolb:2002ve,Hirano:2002ds,Hirano:2004en},
the PHENIX collaboration created a comprehensive comparison between
heavy-ion data on $p_T$ spectra and
$v_2/\varepsilon$\cite{whitepapers}. The inclusion of a comparison to
HBT data was hampered by the lack of predictions from some of the
models. The comparison to $p_T$ spectra and $v_2$ is shown in
Fig.~\ref{fig:hydrocomp}. The left panels show $p_T$ spectra with
pions in the top panel and protons in the bottom. The right panels
show $v_2$ for the same particles. The combination of data on $v_2$
and spectra provide a stringent test for the models as some models can
reproduce one quantity but only by adjusting parameters in such a way
that the agreement with other observables is spoiled.

The models shown in the figure differ in several ways. Models that
include a phase transition and a QGP phase are shown with solid lines
while models without a pure QGP phase are shown as dotted
lines. Including this phase transition acts to reduce the value of
$v_2$ since the equation of state is soft during the transition. This
means that the speed of sound drops (in these models to zero), so that
conversion of coordinate space eccentricity to momentum space
anisotropy is halted during the phase transition. In the case that the
models, do approximately match the pion spectra and $v_2$, the most
directly observable consequence of the lack of a phase transition is
on the proton spectra and proton $v_2$. The proton spectra end up
being too soft, and the splitting between proton and pion $v_2$ is
reduced with the proton $v_2$ becoming larger. This is somewhat
counter-intuitive but is a consequence of fixing the parameters to
match central data.
% Recall that mass splitting arises
%because more massive particles shift to larger $p_T$ and in the
%in-plane direction, where the pressure gradients are largest the low
%$p_T$ region is depopulated most.
%For models without a
%phase transition it is therefore found that, matching the $p_T$
%spectra leads to an over-prediction of $v_2$.

The models also differ in their treatments of the final hadronic
stage. The calculations from Teaney \textit{et al.} include a hybrid
model that uses a hadronic cascade (RQMD) for the final hadronic
evolution. Hirano and Kolb do not use such an afterburner but allow
the particle abundances to stop changing at a temperature above the
temperature at which they stop interacting; chemical freeze-out
happens before kinetic freeze-out. Huovinen on the other hand,
maintains chemical and kinetic equilibrium throughout the
expansion. These different treatments have very important consequences
for the particle-type dependence of the $p_T$ spectra and $v_2$.
Huovinen's treatment can reproduce the $v_2$ for pions and protons,
but only at the expense of under-predicting the number of protons; a
direct consequence of maintaining chemical equilibrium until the final
freeze-out at a relatively low temperature. 

\begin{figure}[hbt]
\begin{center}
\centerline{  \includegraphics[width=.8\textwidth]{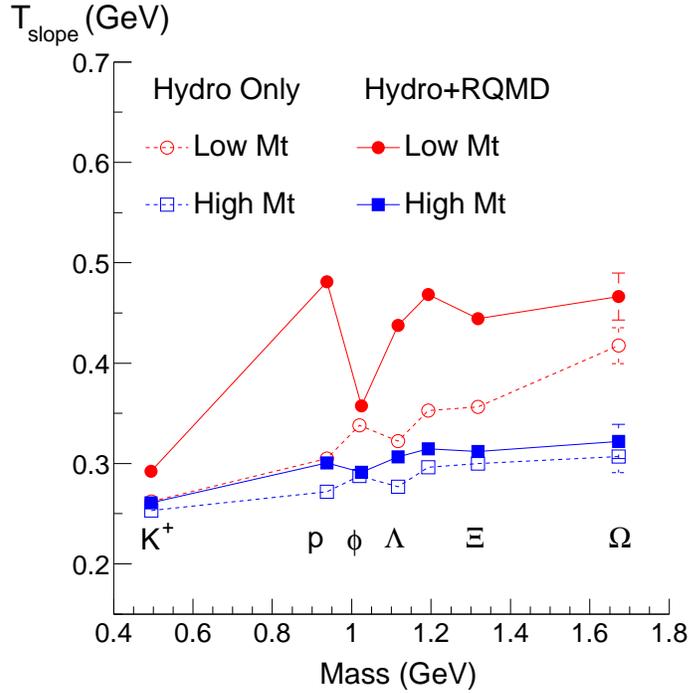}}
\caption{ The inverse slope parameter $T_{slope}$ for a variety of
  particle species when a heavy-ion collision is modeled with only a
  hydrodynamic model compared to a hybrid model which includes a stage
  modeled with a hadronic cascade model RQMD.}
  \label{fig:Tslope}
\end{center}
\end{figure}

The only model which compares well to all the data is Teaney's model
including a QGP phase, a phase transition, and a hadronic phase
modeled with RQMD. Such a hybrid model adds significantly to the
number of tunable parameters as compared for example to Huovinen's
model. On the other hand, the Teaney model shows that some particle
types are less affected by the hadronic phase and therefore less
sensitive to some of the uncertainty in freeze-out
prescription. Fig.~\ref{fig:Tslope} shows the Teaney calculation with
Hydro only versus Hydro+RQMD. The particle species least affected by
the inclusion of a hadronic afterburner, are the $\phi$-meson and the
$\Omega$-baryon. This arises presumably from the small hadronic
cross-section for these hadrons. This suggests high-statistics
measurements for these particles are a viable way to avoid
uncertainties in the effects of hadronic re-scattering.

\begin{figure}[hbt]
\begin{center}
\centerline{  \includegraphics[width=.8\textwidth]{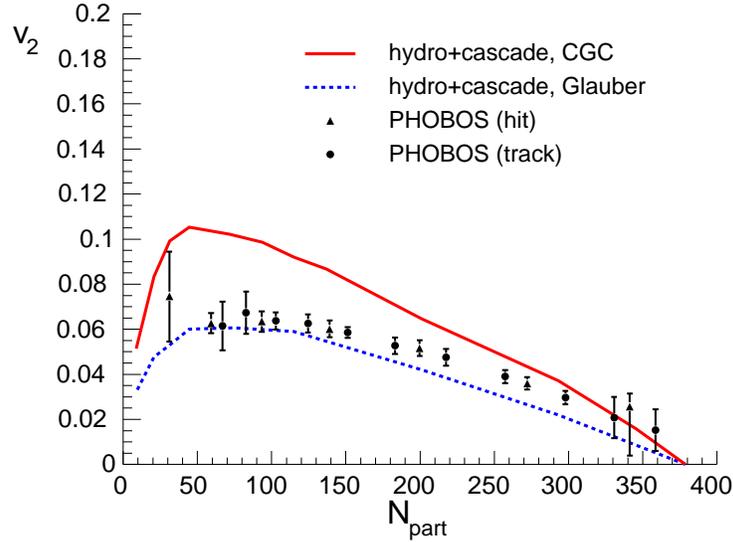}}
\caption{ A hybrid hydrodynamic model showing the uncertainty in the
  model calculations arising from two different models for the initial
  conditions. }
  \label{fig:v2cgc}
\end{center}
\end{figure}

Besides the uncertainty in the freeze-out prescription, there is
uncertainty on the eccentricity of the expanding fire-ball at the
start of the conjectured hydrodynamic evolution. Fig.~\ref{fig:v2cgc}
shows a hybrid hydro+cascade model compared to $v_2$
data\cite{Hirano:2005xf}. Two model curves are shown: one with a
Color-Glass-Condensate (CGC) initial
eccentricity\cite{Drescher:2006pi}, the other with a
Monte-Carlo-Glauber (MCG) eccentricity. As discussed previously the
CGC eccentricity is larger than the MCG eccentricity; this leads to an
over-prediction for $v_2$. On the other hand, this hybrid model does
not include viscous effects in the QGP phase so the difference between
the hybrid+CGC prediction could be related to viscosity. In fact,
since viscosity acts to reduce $v_2$, the hybrid+MCG curve shows that
there is no room for viscosity in this model. This violates the lower
bound on viscosity derived based on quantum mechanical
arguments\cite{Danielewicz:1984ww} and also later from string
theory\cite{Kovtun:2004de}. Clearly, to estimate the
viscosity allowed, or required by the data, the uncertainty on the
initial conditions must be reduced. As discussed previously, the
measured quantity $\sigma_{tot}^2=\delta_2 + 2\sigma_{v_2}^{2}$
provides a sensitive test of the models of the initial conditions and
needs to be carefully compared to the hydrodynamic model predictions
with various initial conditions.

%\begin{figure}[hbt]
%\begin{center}
%\centerline{  \includegraphics[width=.6\textwidth]{figures/v2ptyt.eps}}
%\caption{ $v_2/p_T$ versus transverse rapidity $y_T$. A hydrodynamic model
%  for pion $v_2$ is shown along with boost model described in the
%  text. }
%  \label{fig:v2ptyt}
%\end{center}
%\end{figure}

\subsubsection{Transport Model Fits}

\begin{figure}[hbt]
\begin{center}
\centerline{  \includegraphics[width=1.0\textwidth]{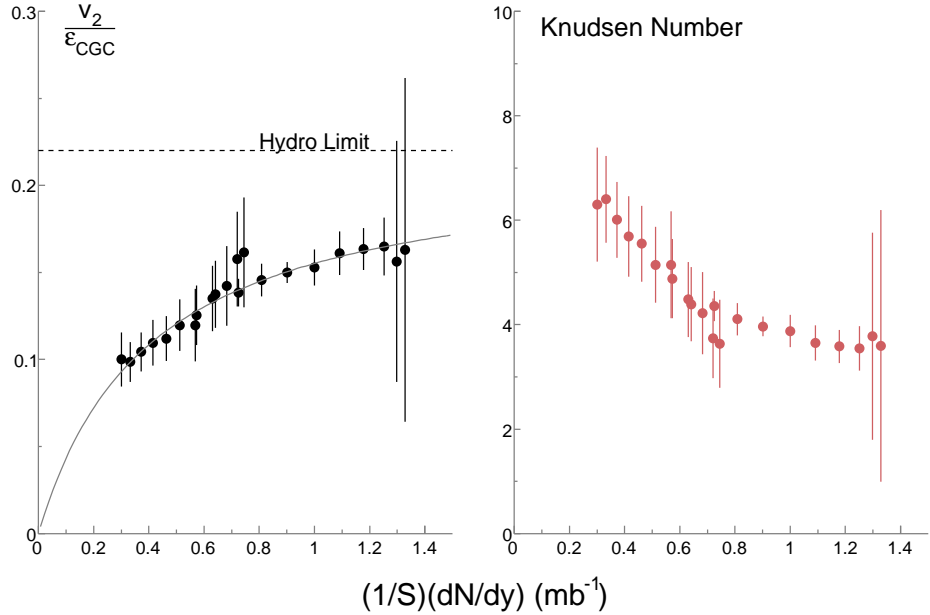}}
\caption{ Left panel: $v_2$ in Au+Au and Cu+Cu collisions scaled by
  eccentricity calculated in a CGC framework and plotted
  versus $(1/S)(dN/dy)$. The fit function and hydrodynamic limit are
  explained in the text. Right panel: The effective Knudsen extracted
  from the data and fit in the left panel.}
  \label{fig:knudsen}
\end{center}
\end{figure}

An approach to circumvent the uncertainties in the hydrodynamic models
has been outlined in Ref.\cite{Bhalerao:2005mm,Drescher:2007cd} where
$v_2/\varepsilon$ is fit as a function of
$\frac{1}{S}\frac{dN}{dy}$. The fit function is used to infer how
close the data come to a saturated value in the collisions with the
highest density achieved. The fit function is constrained by how
$v_2/\varepsilon$ should approach the high density and the low density
limits. One can construct different equations but in a transport code,
the following was found to represent the approach to the zero
mean-free-path limit well\cite{Gombeaud:2007ub}:
\begin{equation}
  \frac{v_2}{\varepsilon} = \frac{v_2^{sat}}{\varepsilon}\frac{1}{1+K/K_0}
\end{equation}
where $K$ is the Knudsen number and $K_0$ is a constant of order
one. Fig.~\ref{fig:knudsen} shows the data and fit in the left panel
and the inferred Knudsen number in the right panel. Based on this
procedure it is found that RHIC $v_2$ data are still some 20\% below
the saturation value anticipated within the fit function. This
conclusion however, not only depends on the assumptions built into the
transport model approach but also the centrality dependence of the
eccentricity. The Color Glass Condensate model for example predicts a
stronger centrality dependence for the eccentricity than the Monte
Carlo Glauber model. As a consequence, this fit implies that if the
initial conditions at RHIC are described by the CGC model, then the
$v_2$ data is closer to its saturation limit than if the MCG gives the
correct description. This is counter intuitive and opposite to the
conclusions reached based on real hydrodynamic calculations, which
indicate that the larger CGC initial eccentricity allows more room for
viscous effects in the QGP
phase\cite{Romatschke:2007mq,Song:2007fn}. The transport based fit
circumvents the actual solving of hydrodynamics but the conclusions
are dependent on the centrality dependence of the initial eccentricity
which is strongly model dependent. The fit also includes the speed of
sound as a free parameter. This effectively leads to an equation of
state which has no phase transition but which is allowed to vary in
the fit. A complimentary and perhaps better method for accessing the
Knudsen number and the viscosity is to study the shape change of
$v_2(p_T)$ for different system-sizes which avoids the uncertainty in
the eccentricity\cite{Teaney:2003kp}. This is a work currently in
progress.

\subsubsection{Viscous Hydrodynamics}

\begin{figure}[ht]
\vspace{0.5cm}
\begin{minipage}[b]{1.0\linewidth}
\centering
\includegraphics[width=0.7\textwidth]{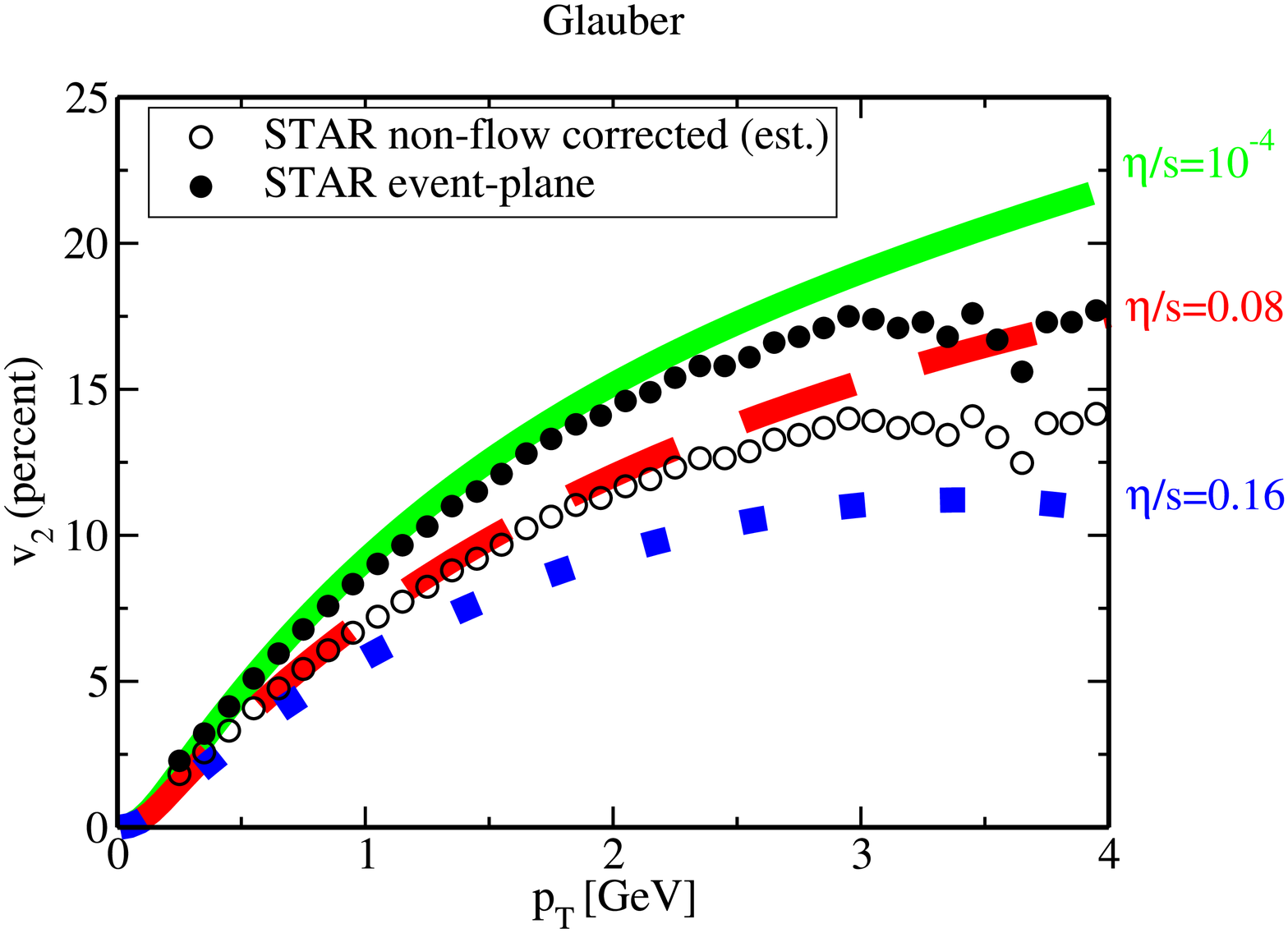}
\end{minipage}
\begin{minipage}[b]{1.0\linewidth}
\centering
\includegraphics[width=0.7\textwidth]{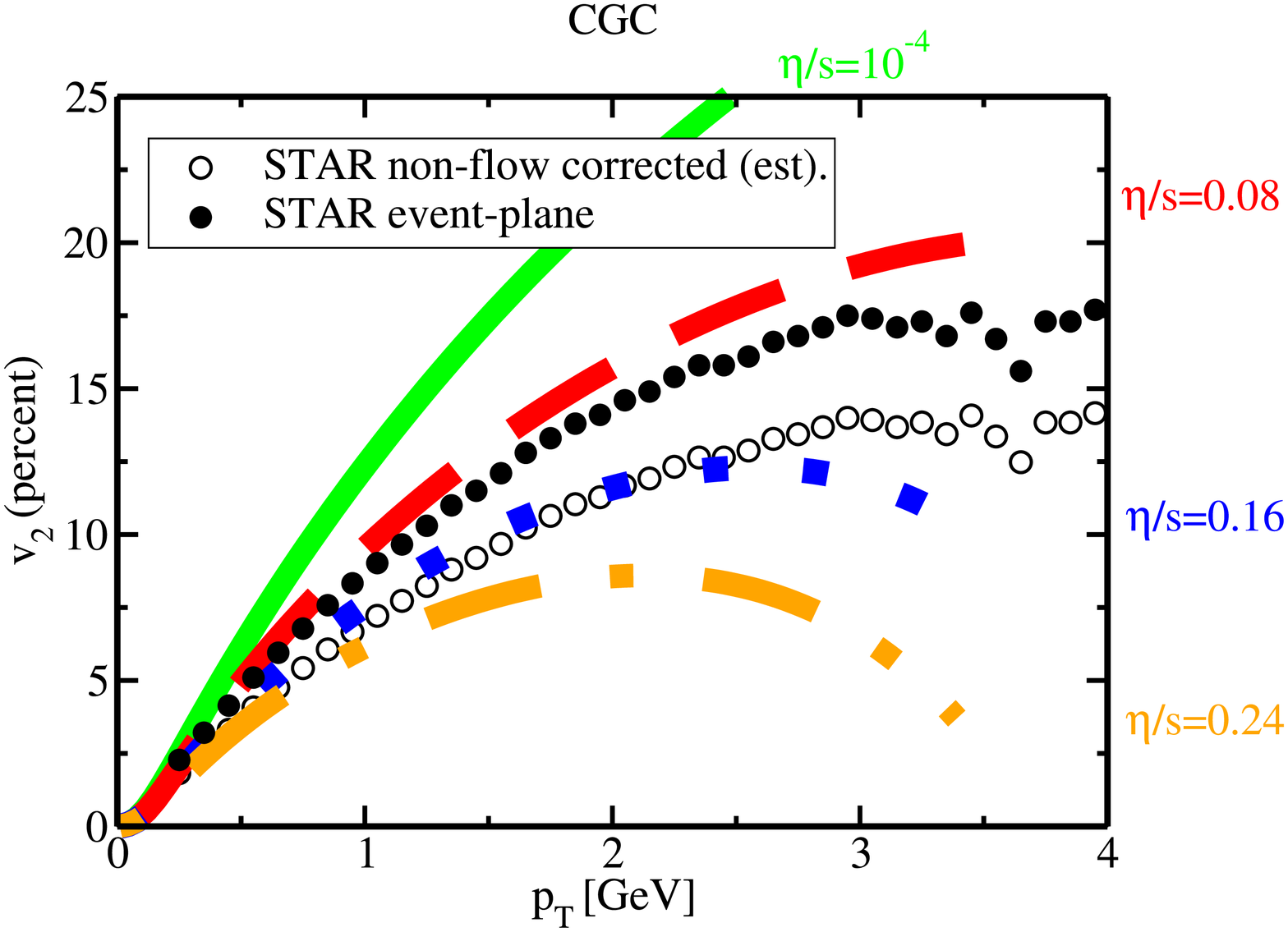}
\end{minipage}
\caption{ Hydrodynamic calculations including viscous effects. The
  top panel shows $v_2(p_T)$ for the case that the initial conditions
  are described with a Monte Carlo Glauber model. The bottom panel is
  based on Color Glass Condensate initial conditions. The curves show
  results for different values of $\eta/s$, the ratio of shear
  viscosity to entropy.}
\label{fig:viscous}
\end{figure}

The apparent success of ideal hydrodynamic models to describe the
gross features of RHIC data has led to the inference of small
viscosity and the claim of the discovery of the perfect liquid at
RHIC.  The perfect liquid announcement was listed as the top physics
story of 2005 by the American Institute of Physics and was widely
covered in the popular press. Much recent work has gone towards
including viscous effects in hydrodynamic calculations so that the
viscosity can be more accurately
estimated\cite{Romatschke:2007mq,Luzum:2008cw,Song:2007fn,Song:2007ux,Dusling:2007gi,Song:2008si}.
Fig.~\ref{fig:viscous} shows one such calculation. The top panel shows
results when the hydrodynamic evolution starts from Glauber initial
conditions while the bottom panel shows the case of CGC initial
conditions. The results that come closest to the STAR
data\cite{Abelev:2007qg} are given by $\eta/s=0.08$ for the Glauber
initial conditions and $\eta/s=0.16$ for the CGC initial
conditions. The STAR non-flow corrected data are from Figure 4 of
Ref.\cite{Abelev:2008ed}. The 10\%-40\% central data from that
reference was scaled to account for the difference between the
10\%-40\% centrality interval and the 0\%-80\% (minbias) centrality
interval. The larger $\eta/s$ inferred based on the CGC model arises
from the larger initial eccentricity which leaves more room for
viscous effects that tend to reduce the $v_2$. This contradicts the
conclusions drawn from the transport model inspired fit, which allows
the equation-of-state to change for the two different initial
conditions. The $p_T$ dependence of the data is also better captured
in the larger viscosity CGC scenario. The larger viscosity inferred
from the CGC initial conditions gives a more pronounced turn over of
$v_2(p_T)$ which better describes the $p_T$ dependence of $v_2$. The
comparison shown in Fig.~\ref{fig:viscous} shows that hydrodynamic
models including viscosity have a good chance of reproducing RHIC data
as long as the shear viscosity to entropy ratio $\eta/s$ is less than
$\frac{2}{4\pi}$ where $\frac{1}{4\pi}$ is the conjectured lower
limit.

\subsubsection{Fluctuating Initial Conditions}

\begin{figure}[ht]
\vspace{0.5cm}
\centering
\begin{minipage}[b]{0.60\linewidth}
\centering
\includegraphics[width=0.98\textwidth]{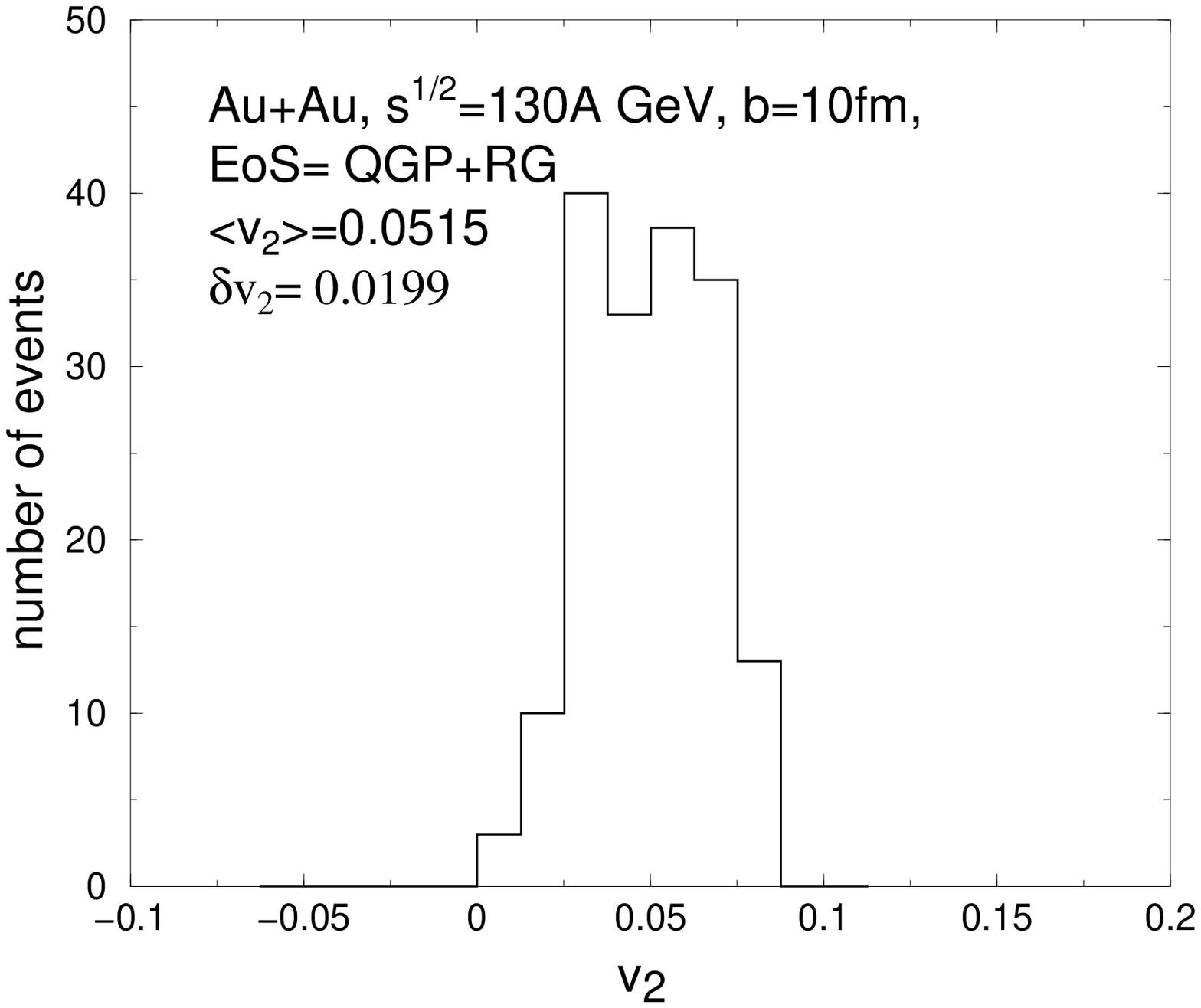}
\end{minipage}
\centering
\begin{minipage}[b]{0.75\linewidth}
\centering
\includegraphics[width=0.98\textwidth]{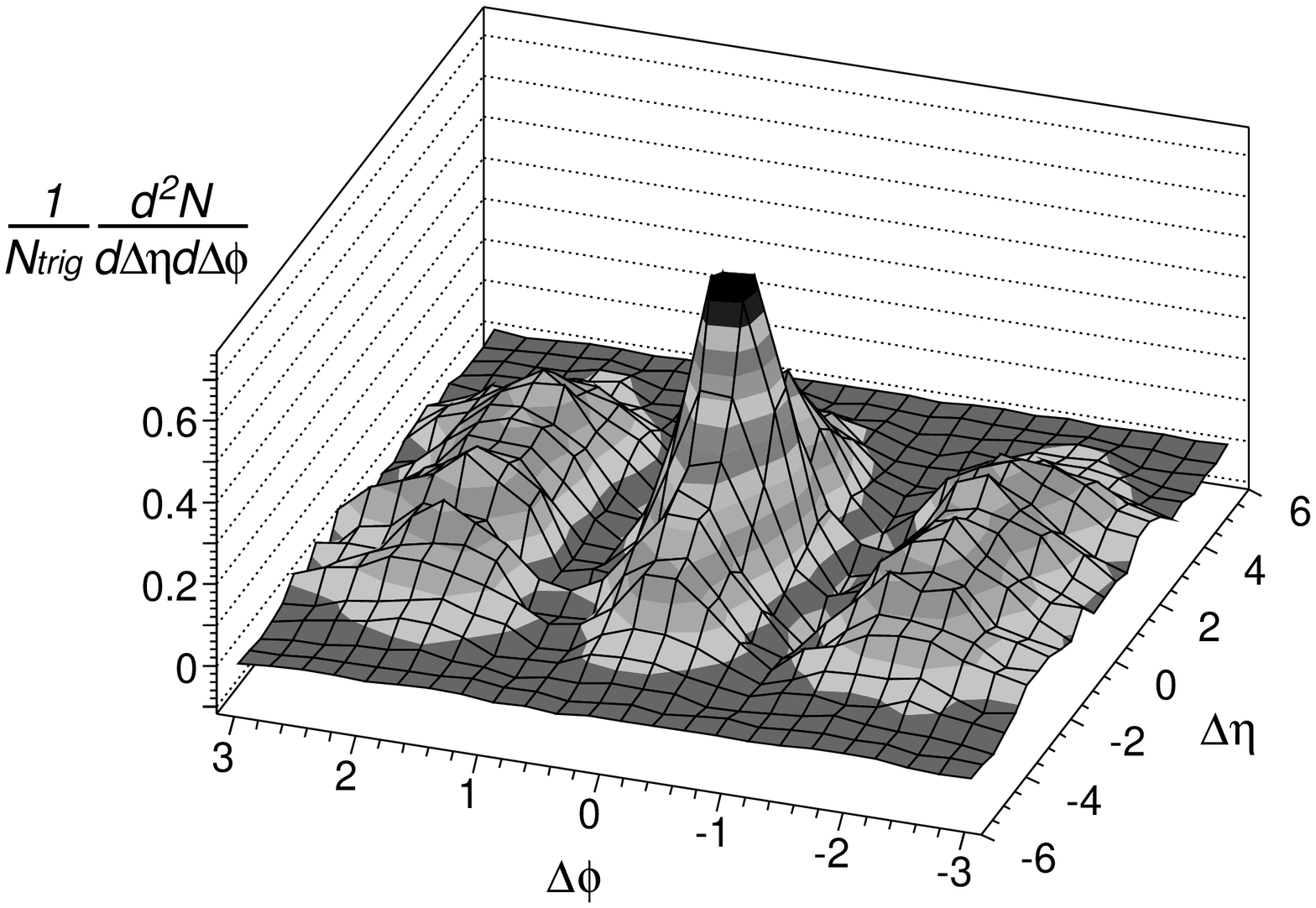}
\end{minipage}
\caption{ Hydrodynamic calculations including fluctuations in the
  initial conditions. The left panel shows the event-by-event $v_2$
  distribution. The right panel shows the two-particle correlations
  that arise due to the correlations in the initial energy density.}
\label{fig:nexspherio}
\end{figure}

The comparison of $v_2$ and other RHIC data to hydrodynamic models
seems to indicate that when viscous corrections are included, a
successful description of the data may be possible. There is
uncertainty in this comparison, however, related to uncertainties in
the initial conditions and in the freeze-out prescription. The
uncertainty in the initial conditions can be addressed experimentally
with measurements of $v_2$ fluctuations which in turn require an
understanding of non-flow correlations; The experimentally accessible
information appears to reduce to $v_2\{2\}^2$ and
$v_{2}\{2\}^2-v_{2}\{4\}^4=\delta_2 + 2\sigma_{v_{2}}^2$. An
alternative approach may be for hydrodynamic models to predict
$v_2\{2\}$ and $v_2\{4\}$ by including correlations and fluctuations
in the models. Progress has been made in this direction. Early work
relating to the effect of fluctuations in the initial conditions on
hydrodynamic calculations was carried out using the NeXSPheRIO
hydrodynamic model\cite{Aguiar:2001ac,Andrade:2006yh}. The initial
eccentricity fluctuations were indeed found to lead to $v_2$
fluctuations as shown in Fig.~\ref{fig:nexspherio}. Later it was
suggested that correlations in the initial conditions could lead to
$v_n$ fluctuations of even and odd orders of $n$ that would manifest
themselves as non-sinusoidal, apparently non-trivial, two-particle
correlations as seen in the RHIC
data\cite{Sorensen:2008dm,Sorensen:2008bf}.

Subsequent work following through on this idea shows that hydrodynamic
models with fluctuating initial conditions do lead to two-particle
correlations with structure beyond a simple $\cos(2\Delta\phi)$
shape\cite{Takahashi:2009na}. The correlation structure arising from
the fluctuations in the initial conditions is shown in the right panel
of Fig.~\ref{fig:nexspherio}. The model exhibits many of the features
seen in the data including a jet-like peak, a near-side ridge, and an
away-side ridge shifted away from $\Delta\phi=\pi$. All this structure
arises without the explicit inclusion of jets in the model. The
apparently exotic correlations do not appear in the model when a
smooth initial condition is used. This calculation illustrates the
importance of accounting for fluctuations in the initial conditions
when interpreting the correlation landscape. It also demonstrates that
complex interactions between jets and the medium, including
mach-cones, are not needed to explain the correlations data nor is the
concept of mini-jets necessarily required.

In light of the NeXSPheRIO calculations, the highly structured
correlation landscape at RHIC should not necessarily be taken as an
invalidation of the hydrodynamic models. The correlations may simply
reflect the need to abandon certain approximations, including the
approximation of infinitely smooth initial conditions.  Besides
comparing to two-particle correlation data, these models can be used
to calculate $v_2\{2\}$ and $v_2\{4\}$ to directly compare to data. It
will be interesting to see how the correlation landscape in this model
depends on the parameters of the model, in particular, the
thermalization time and the freeze-out time. The connection of $v_n$
fluctuations (related to two-particle correlations) to the lifetime of
the system was first pointed out by Mishra \textit{et
  al.}\cite{Mishra:2007tw}. In that reference the authors also
introduce the anaology between $\sqrt{\langle v_n^2\rangle}$
fluctuations and the power spectrum of the Cosmic Microwave Background
Radiation.

\subsubsection{Addressing Uncertainties}

Uncertainties in the freeze-out prescription and the effects of the
hadronic phase can be experimentally addressed through precise
measurements of $\phi$-mesons and $\Omega$-baryons. Models indicate
that due to their small hadronic cross-sections, these hadrons are
minimally influenced by the hadronic phase and reflect well the QGP
phase. In addition, heavy flavor hadrons may help determine or provide
a cross-check for the transport properties of the QGP. Another
approach to extracting the viscosity is by studying the shape of
$v_2(p_T)$ versus system size. This approach does not rely on a model
for the initial eccentricity. Uncertainties in the eccentricity and
the initial conditions can be reduced through measurements of $v_2$
fluctuations and two-particle correlations. These studies are
ongoing. One can also measure $v_n$ fluctuations for arbitrary $n$
value. These are of course related to the two-particle correlation
landscape which has already been extensively studied at RHIC. It will
be of great interest to see how the correlation landscape predicted in
hydrodynamic models with fluctuating initial conditions changes
depending on the model parameters. The correlations data may help
constrain quanties like the lifetime of the system. The studies listed
above, along with a beam-energy scan at RHIC and the first data from
LHC, will allow for more progress in understanding the matter created
in heavy-ion collisions and its subsequent evolution.